\documentclass[prb,amsfonts,amssymb,showpacs,twocolumn,floatfix]{revtex4}
\usepackage{graphics,bm,natbib}
\begin{document}
\widetext
\title{First-principle Wannier functions and
effective
lattice fermion models for narrow-band compounds
}
\author{I. V. Solovyev}
\email[Electronic address: ]{solovyev.igor@nims.go.jp}
\altaffiliation[Present address: ]
{Computational Materials Science Center (CMSC),
National Institute for Materials Science (NIMS),
1-2-1 Sengen, Tsukuba, Ibaraki 305-0047, Japan}
\affiliation{
PRESTO, Japan Science and Technology Agency, \\
Institute for Solid State Physics, University of Tokyo, \\
Kashiwanoha, Kashiwa, Chiba, 277-8581, Japan
}
\date{\today}

\widetext
\begin{abstract}
We propose a systematic procedure for constructing effective
lattice fermion models for narrow-band compounds on the basis
of first-principles electronic-structure calculations.
The method is illustrated for the series of transition-metal (TM)
oxides: SrVO$_3$, YTiO$_3$, V$_2$O$_3$, and Y$_2$Mo$_2$O$_7$,
whose low-energy properties are linked
exclusively to the electronic structure of
an isolated
$t_{2g}$ band. The method consists of three
parts, starting from the electronic structure in the
local-density approximation (LDA). (i)
construction of the kinetic-energy Hamiltonian
using
formal downfolding
method. It allows to describe the band structure
close to the Fermi level in terms of a limited number
of (unknown yet)
Wannier functions (WFs), and eliminate the rest of the basis states.
(ii) solution of an inverse problem and
construction of WFs for the
given kinetic-energy Hamiltonian. Here, we closely follow
the construction of the basis functions in
the liner-muffin-tin-orbital (LMTO) method, and enforce the orthogonality
of WFs to other band.
In this approach,
one can easily control the
contributions of the kinetic energy to the WFs.
(iii)
calculation of screened Coulomb interactions in the basis of \textit{auxiliary} WFs.
The latter are defined as the WFs
for which the kinetic-energy term is set to be zero.
Meanwhile, the hybridization between TM $d$ and other atomic states is well
preserved by the orthogonality condition to other bands.
The use of auxiliary WFs is necessary in order to
avoid the double counting of the
kinetic-energy term, which is included explicitly in the model
Hamiltonian.
In order to calculate the screened Coulomb interactions we employed a
hybrid approach. First, we evaluate the screening caused by
the change of occupation numbers and the
relaxation of
the LMTO basis functions, using the conventional constraint-LDA approach, where
all matrix elements of hybridization connecting the TM $d$ orbitals
and other orbitals are set to be zero.
Then, we switch on the hybridization and evaluate the screening
of on-site Coulomb interactions
associated with the
change of this hybridization in the random-phase approximation.
The second channel of screening appears to be very important, and results in
relatively small value of the
effective
Coulomb interaction for isolated
$t_{2g}$ bands (about 2-3 eV, depending on the material).
We
discuss details of this screening and consider its
band-filling dependence, frequency dependence, influence of the lattice
distortion, proximity of other bands, as well as the effect of dimensionality
of the model Hamiltonian.
\end{abstract}



\maketitle


\onecolumngrid

\begin{center}
\section{\label{sec:Intro}Introduction}
\end{center}

  Many successes of modern solid-state physics and chemistry
are related with the development of the Hohenberg-Kohn-Sham
density-functional theory (DFT),\cite{HohenbergKohn,KohnSham}
which is designed for the ground state and based on the minimization
of the total energy $E[\rho]$ with respect to the electron
density $\rho$.
For practical applications, DFT resorts to iterative
solution of single-particle Kohn-Sham (KS) equations,
\begin{equation}
\left( -\frac{\hbar^2 }{2m} \nabla^2 + V_{\rm H} + V_{\rm XC} + V_{\rm ext} \right) \psi_i = \varepsilon_i \psi_i,
\label{eqn:KS}
\end{equation}
together with the equation for the electron density:
\begin{equation}
\rho = \sum_i n_i |\psi_i|^2,
\label{eqn:rho}
\end{equation}
defined in terms of
eigenfunctions ($\psi_i$), eigenvalues ($\varepsilon_i$), and the
occupation numbers ($n_i$) of KS quasiparticles.
Different terms in Eq.(\ref{eqn:KS}) are correspondingly the kinetic-energy operator,
the Hartree potential, the exchange-correlation potential, and the external potential.
In the following we will also reserve the notation
$H_{\rm KS}$$=$$- (\hbar^2/2m) \nabla^2$$+$$V_{\rm H}$$+$$V_{\rm XC}$$+$$V_{\rm ext}$ for the total
KS Hamiltonian in the real (${\bf r}$) space.

  The exchange-correlation potential is typically treated in the local-density
approximation (LDA). It employs an analytical expression for $V_{\rm XC}[\rho]$
borrowed from the theory of
homogeneous electron gas in which the density of the electron gas
is replaced by the local density of the real system.
LDA is far from being perfect and
there are many examples of so-called strongly-correlated materials
for which the conventional LDA appears to be insufficient
both for the excited-state
and ground-state properties.\cite{AZA,LDAUreview,IFT}

\begin{figure}[h!]
\begin{center}
\resizebox{7cm}{!}{\includegraphics{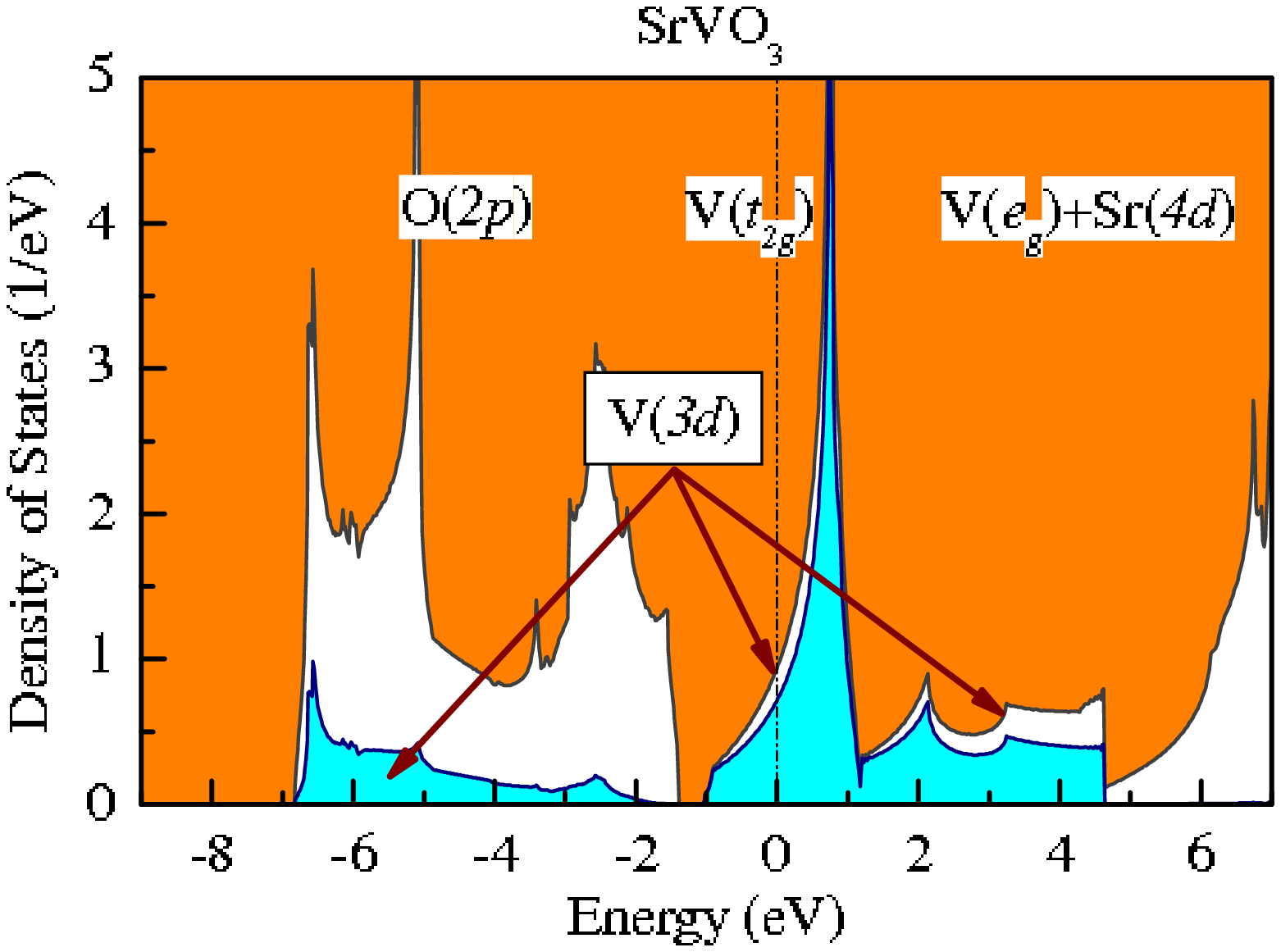}}
\resizebox{7cm}{!}{\includegraphics{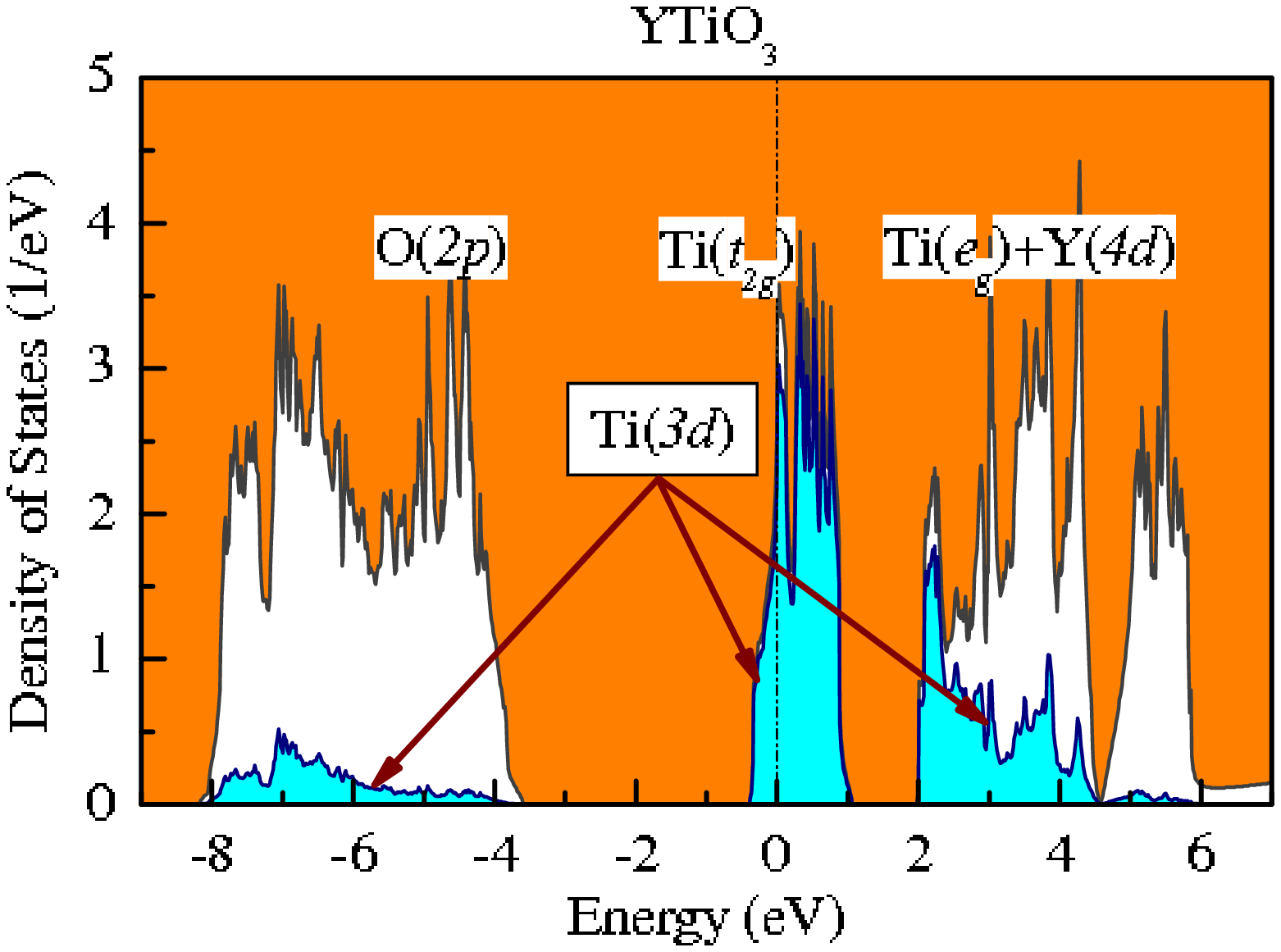}}
\end{center}
\begin{center}
\resizebox{7cm}{!}{\includegraphics{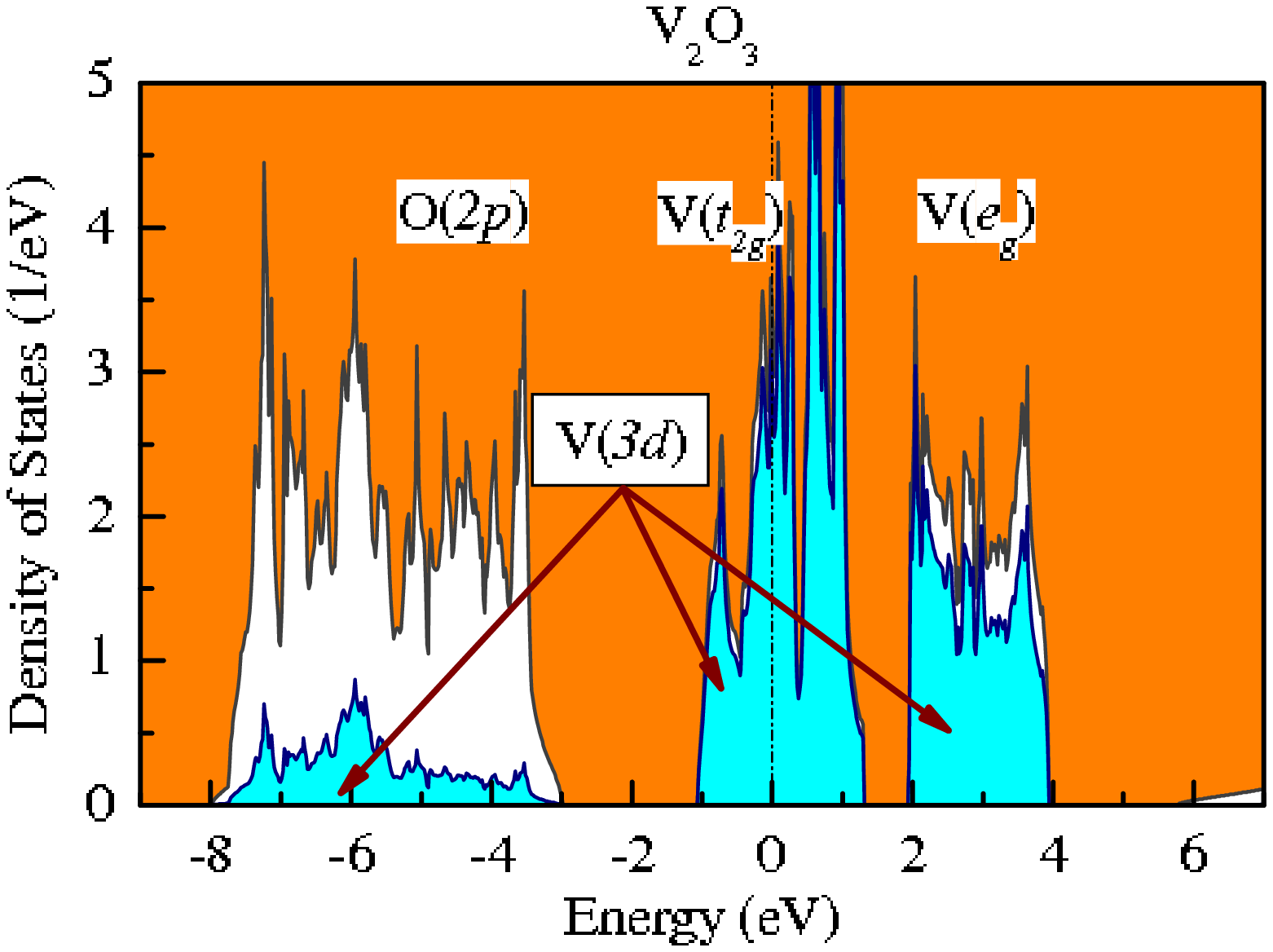}}
\resizebox{7cm}{!}{\includegraphics{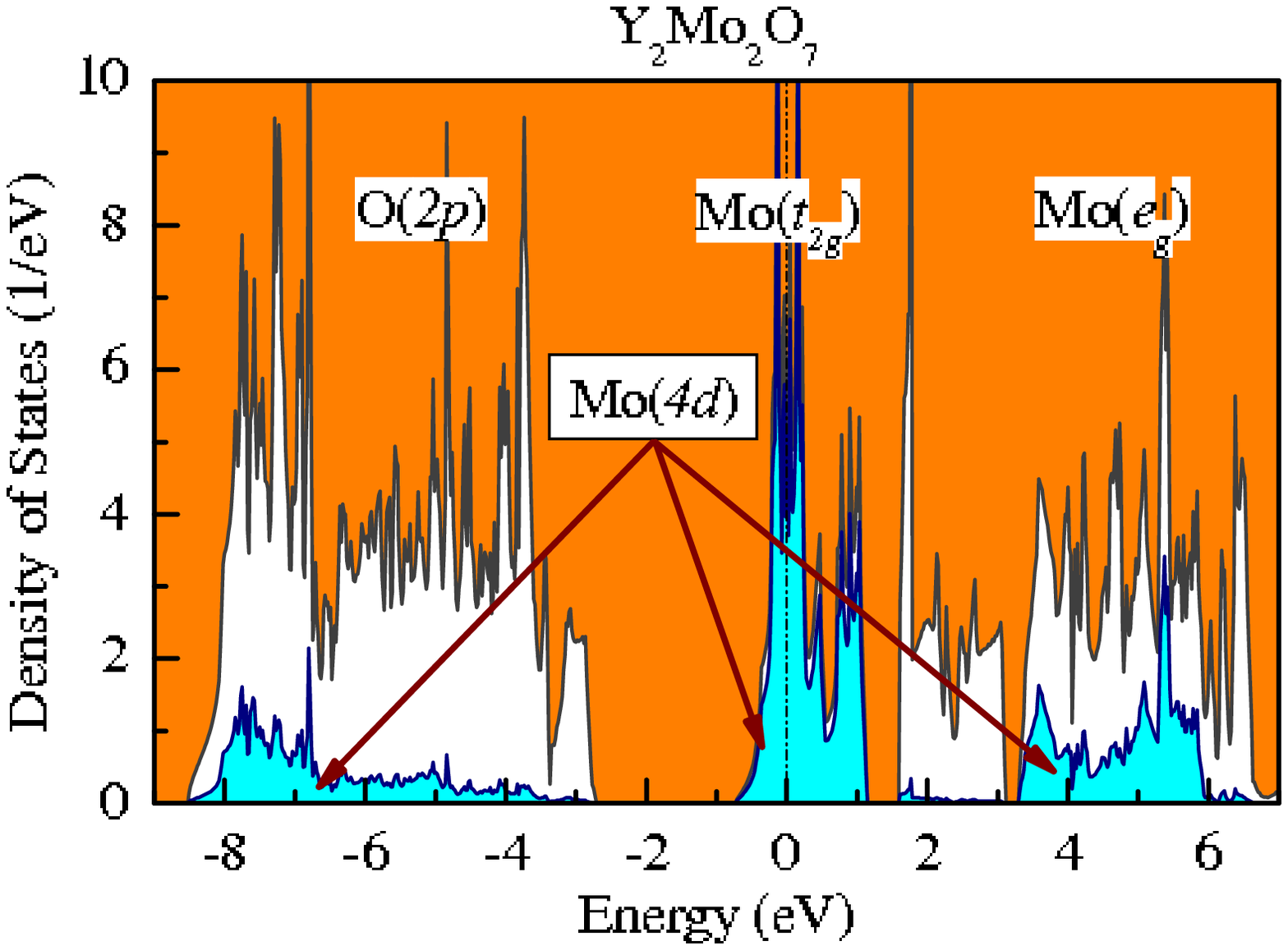}}
\end{center}
\caption{\label{fig.DOSsummary} Total and partial densities of states
for SrVO$_3$, YTiO$_3$, V$_2$O$_3$, and Y$_2$Mo$_2$O$_7$ in the local-density approximation.
The shaded area shows the contributions of transition-metal $d$-states.
Other symbols show positions of the main bands.
The Fermi level is at zero energy.}
\end{figure}

  A typical situation realized in transition-metal (TM) oxides
is shown in Fig.~\ref{fig.DOSsummary}.
We would like to emphasize two points.\\
(i) The common feature of many TM oxides is the existence of the well isolated
narrow band
(or the group of bands) located near the Fermi level and well isolated from
the rest of electronic stares.
For compounds shown in Fig.~\ref{fig.DOSsummary}, this is the TM $t_{2g}$ band,
which is sandwiched between O($2p$) band (from below)
and a group of bands (from above), which have an appreciable
weight of the TM $e_g$ states
(the meaning of many notations will become clear in Sec.~\ref{sec:KineticApplications},
where we will discuss details of the crystal and electronic structure for the
considered oxide compounds).
Electronic and magnetic properties of these compounds are
predetermined mainly by the behavior of this $t_{2g}$ band.
The effect of other bands can be included indirectly, through the renormalization
of interaction parameters in the $t_{2g}$ band. \\
(ii) The LDA description appears to be especially bad for the
$t_{2g}$ states located
near the Fermi level. It often fails to reproduce the insulating
behavior of these compounds, as well as the correct magnetic ground state,
which is directly related with the existence of the band gap.\cite{TRN}
The source of the problem is know to be the on-site Coulomb correlations,
whose form is greatly oversimplified
in the model of homogeneous electron gas. Therefore, the basic strategy
which was intensively pursued already for more than decade was to
incorporate the physics of on-site Coulomb correlations in LDA and
to solve this problem using modern many-body techniques.
This way of thinking gave rise to such directions as
LDA$+$$U$ (e.g., Refs.~\onlinecite{AZA}, \onlinecite{LDAUreview}, and
\onlinecite{PRB94}) and LDA$+$DMFT
(dynamical mean-field theory, Ref.~\onlinecite{LSDADMFT}).

  Taking into account two above arguments, we believe that the most logical way
to approach the problem of Coulomb correlations in narrow-band compounds is
to divide it in two part: \\
(i) mapping of conventional electronic structure calculations onto the
multi-orbital Hubbard model,
and derivation of the parameters of this model from the first principles, for example
starting from the simplest electronic structure in LDA; \\
(ii) solution of this multi-orbital Hubbard model using modern many-body methods.\cite{Sr2VO4preprint}

  In this paper we will discuss the first part of this project and show
how results of conventional LDA calculations
for the $t_{2g}$ bands can be mapped onto the multi-orbital Hubbard model:
\begin{equation}
\hat{\cal{H}}= \sum_{{\bf R}{\bf R}'} \sum_{\alpha \beta}
h_{{\bf R}{\bf R}'}^{\alpha \beta}\hat{c}^\dagger_{{\bf R}\alpha}\hat{c}^{\phantom{\dagger}}_{{\bf R}'\beta} +
\frac{1}{2} \sum_{\bf R}  \sum_{\alpha \beta \gamma \delta}
U_{\alpha \beta \gamma \delta}
\hat{c}^\dagger_{{\bf R}\alpha} \hat{c}^\dagger_{{\bf R}\gamma}
\hat{c}^{\phantom{\dagger}}_{{\bf R}\beta} \hat{c}^{\phantom{\dagger}}_{{\bf R}\delta},
\label{eqn:Hmanybody}
\end{equation}
where $\hat{c}^\dagger_{{\bf R}\alpha}$ ($\hat{c}_{{\bf R}\alpha}$)
creates (annihilates) an electron in the Wannier orbital
$\tilde{W}_{{\bf R} \alpha}$ of the site ${\bf R}$, and
$\alpha$ is a joint index, incorporating all remaining (spin and orbital)
degrees of freedom.
The matrix $\| h_{{\bf R}{\bf R}'}^{\alpha \beta} \|$ parameterizes the
kinetic energy of electrons.
The matrix elements $h_{{\bf R}{\bf R}'}^{\alpha \beta}$
have the following meaning:
the
site-diagonal part (${\bf R}$$=$${\bf R}'$)
describes the local level-splitting,
caused by the crystal field and (or) the spin-orbit interaction, while
the off-diagonal part
(${\bf R}$$\neq$${\bf R}'$) stands for the transfer integrals
(or the transfer interactions).
$U_{\alpha \beta \gamma \delta}
$$=$$
\int d{\bf r} \int d{\bf r}' \tilde{W}_\alpha^\dagger({\bf r}) \tilde{W}_\beta({\bf r})
v_{\rm scr}({\bf r}$$-$${\bf r}') \tilde{W}_\gamma^\dagger({\bf r}') \tilde{W}_\delta({\bf r}')$
are the matrix elements of
\textit{screened} Coulomb interaction $v_{\rm scr}({\bf r}$$-$${\bf r}')$, which are supposed to be diagonal
with respect to the site indices.
In principle, the off-diagonal elements can be also included into the model.
However, we do not consider them in the present work.
In Sec.~\ref{sec:Summary} we will discuss several open questions
related with the definition of the intersite Coulomb interactions
in LDA.

  The first part of this paper
will be devoted to derivation of the parameters of the kinetic energy.
Then, we will explain how to construct the Wannier functions (WFs), which generate
these parameters after applying to the KS Hamiltonian in the
real space.
The next part will be devoted to calculations of
screened Coulomb interactions, using the WF formalism.

\section{\label{sec:LMTO}The LMTO Method}

  In this section we will briefly review the main ideas of
linear-muffin-tin-orbital (LMTO) method, as they will be widely used in
the subsequent sections for the construction of transfer interactions and
the Wannier orbitals.
The method is designed for the solution of KS equations in LDA.
For the details and recent developments, the reader
is referred to the activity of the O.~K.~Andersen group at the
Max-Plank Institute in Stuttgart.\cite{LMTO}

  Majority of modern electronic structure methods use
some basis. The basis functions of the LMTO method, $\{ | \chi
\rangle \}$ (the so-called muffin-tin orbitals -- MTOs) have many
similarities with orthogonalized atomic orbitals.
As we will see below, the LMTO method is very convenient
for constructing the WFs, and for certain applications,
the basis function of the LMTO method, from the very beginning, can be chosen
as a Wannier function.

\begin{figure}[h!]
\begin{center}
\resizebox{5cm}{!}{\includegraphics{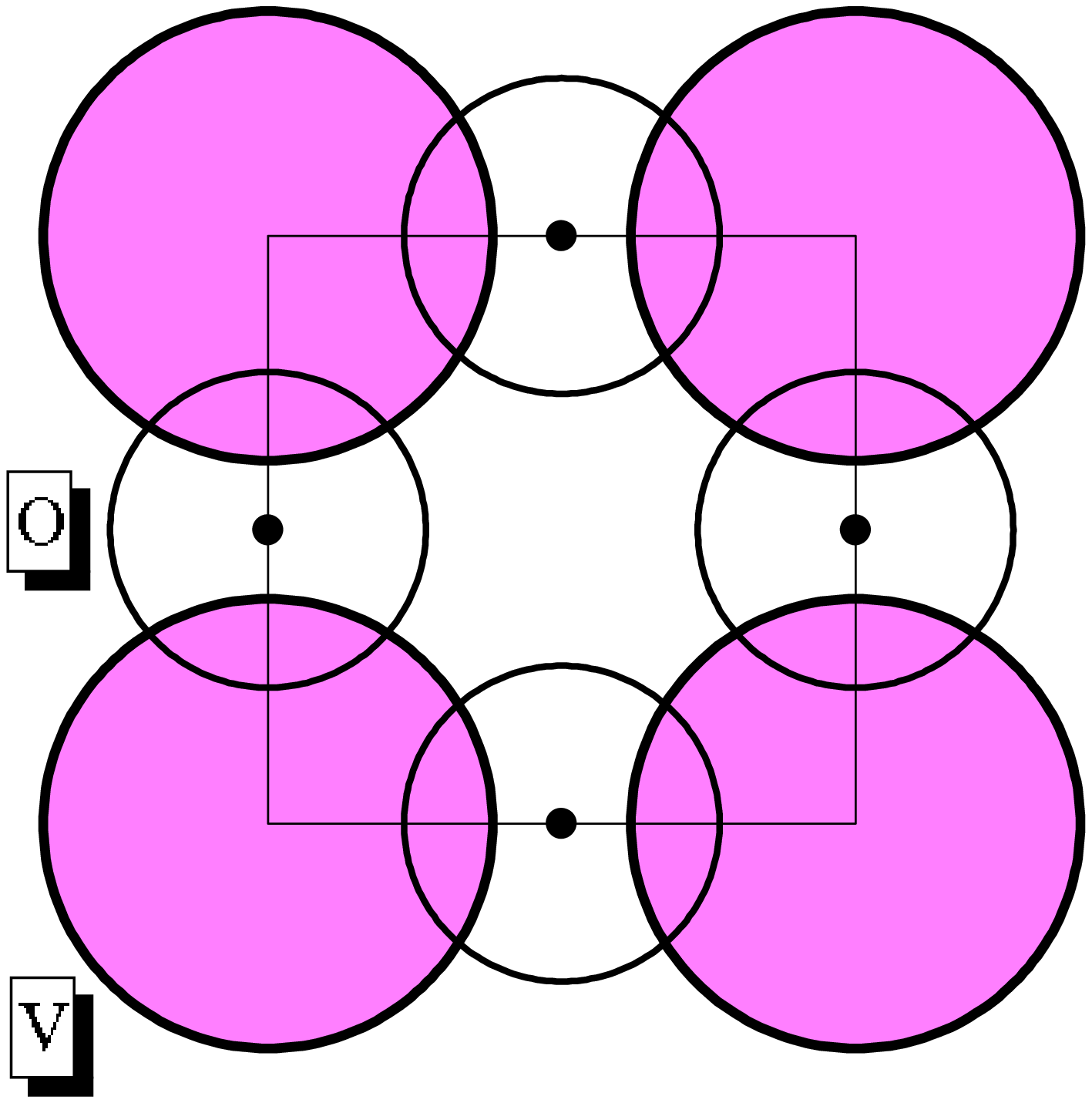}}
\resizebox{8cm}{!}{\includegraphics{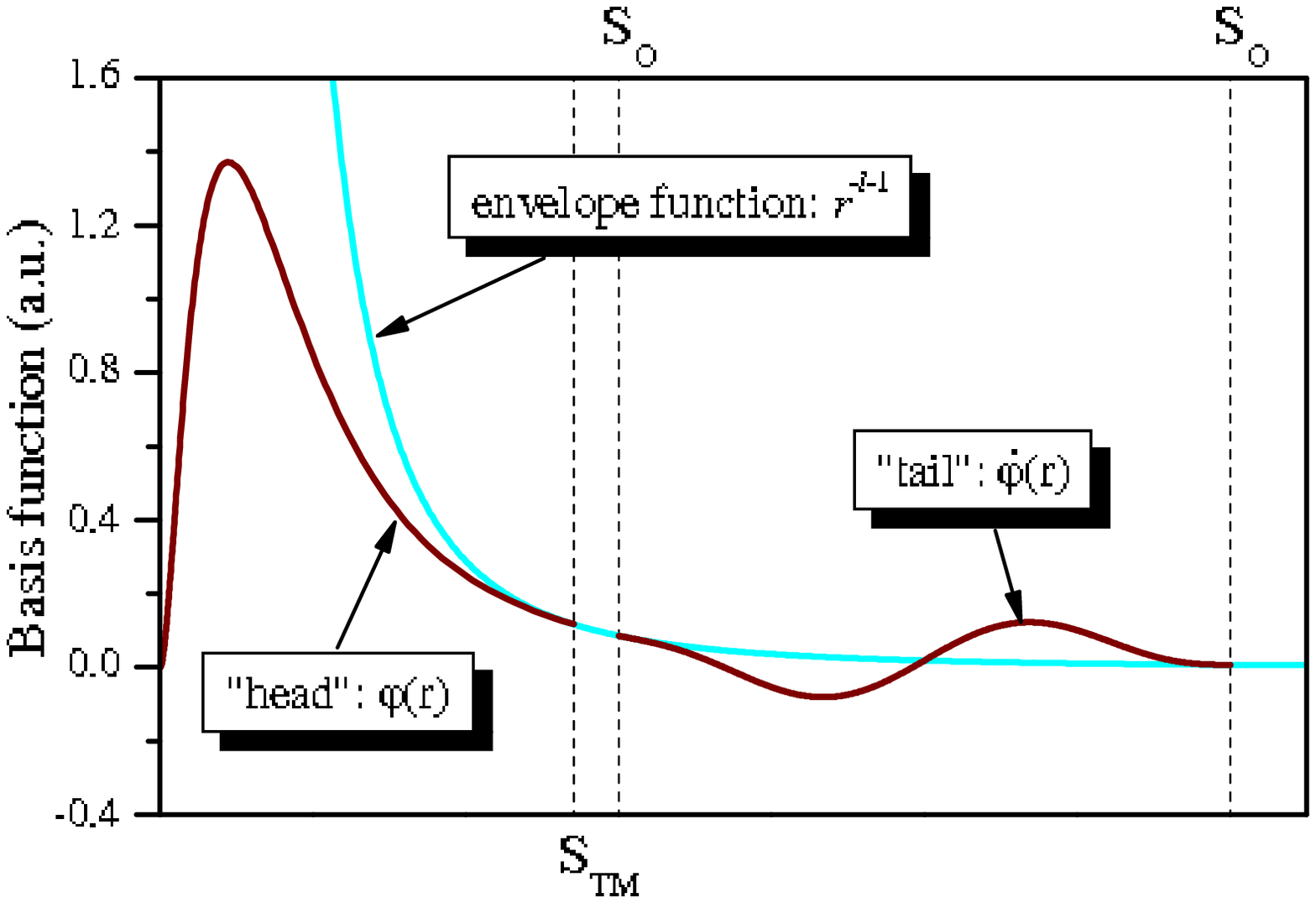}}
\end{center}
\caption{\label{fig.MTO} Construction of basis functions of the LMTO method.
It is assumed that the whole space of the crystal is filled by the atomic spheres.
The left panel shows a concrete example of such filling in the VO$_2$ plane
of SrVO$_3$.
The central part of the basis function (the ``head'') is constructed as the
solution of Kohn-Sham equation inside atomic sphere ${\bf R}$
($\phi_{{\bf R}L}$), which is
calculated in the center of gravity of the band and whose angular
character is denoted by $L$.
The ''tail'' of the basis
function at the neighboring site ${\bf R}'$ is constructed from the energy-derivatives
of $\phi_{{\bf R}'L'}$ (denoted as $\dot{\phi}_{{\bf R}'L'}$).
At the atomic spheres boundaries (shown by ${\rm S_{TM}}$ and ${\rm S_O}$ for the
central transition-metal site and neighboring oxygen sites, respectively),
$\phi_{{\bf R}L}$ and $\dot{\phi}_{{\bf R}'L'}$ match
continuously and differenciably onto certain envelop function
(typically, an irregular solution of the Laplace equation).}
\end{figure}

  The conventional LMTO approach employs the atomic-spheres-approximation (ASA),
which assumes that the whole space of the crystal can be filled
by overlapping atomic spheres (Fig.~\ref{fig.MTO}), so that the overlap
between the spheres as well as the empty spaces, which are not encircled
by any spheres, can be neglected.
The MTOs are
constructed from solutions of KS equations inside
atomic spheres (the partial waves), calculated at some energies $E_{\nu L}$
(typically, the center of gravity of the occupied band or of the
entire band), $\phi_{{\bf R} L}$, and their energy derivatives $\dot{\phi}_{{\bf R} L}$.
In each atomic sphere, the KS potential is spherically averaged.
Therefore, the solutions are proportional to the
angular harmonics, which are specified by the indices $L$$\equiv$$(\ell,m)$
(correspondingly, orbital and azimuthal quantum numbers) .
At the atomic
sphere boundaries, $\{ \phi_{{\bf R} L} \}$ and $\{ \dot{\phi}_{{\bf R} L} \}$
match continuously and differenciably onto certain
envelop functions. The latter are typically
constructed from irregular solutions of Laplace
equation, which rapidly decay in the real space (Fig.~\ref{fig.MTO}).

  It is easy to verify that the functions
$\phi_{{\bf R} L}$ and $\dot{\phi}_{{\bf R} L}$
obey the following ``LMTO algebra'':
\begin{equation}
\langle \phi_{{\bf R} L} | \phi_{{\bf R} L} \rangle = 1,
\label{eqn:LMTOalgebra1}
\end{equation}
\begin{equation}
\langle \dot{\phi}_{{\bf R} L} | \phi_{{\bf R} L} \rangle =
\langle \phi_{{\bf R} L} | \dot{\phi}_{{\bf R} L} \rangle = 0,
\label{eqn:LMTOalgebra2}
\end{equation}
\begin{equation}
\langle \dot{\phi}_{{\bf R} L} | \dot{\phi}_{{\bf R} L} \rangle = p_{{\bf R} L},
\label{eqn:LMTOalgebra3}
\end{equation}
\begin{equation}
\left( H_{\rm KS} - E_{\nu L} \right) | \phi_{{\bf R} L} \rangle = 0,
\label{eqn:LMTOalgebra4}
\end{equation}
and
\begin{equation}
\left( H_{\rm KS} - E_{\nu L} \right) | \dot{\phi}_{{\bf R} L} \rangle = | \phi_{{\bf R} L} \rangle,
\label{eqn:LMTOalgebra5}
\end{equation}
where $H_{\rm KS}$$=$$H_{\rm KS}({\bf r})$ is the Kohn-Sham Hamiltonian
in the real space. Then, one possible choice of the muffin-tin orbitals is
\begin{equation}
| \chi \rangle = | \phi \rangle + | \dot{\phi} \rangle ( \hat{\cal H} - \hat{E}_{\nu} ).
\label{eqn:LMTObasis1}
\end{equation}
In this paper we use the shorthanded notations by Andersen \textit{et al.}.
For instance, Eq.~(\ref{eqn:LMTObasis1}) should be read as follows:
$$
| \chi_{{\bf R} L} \rangle = | \phi_{{\bf R} L} \rangle +
\sum_{{\bf R}' L'} | \dot{\phi}_{{\bf R}' L'} \rangle ( {\cal H}^{L'L}_{{\bf R}'{\bf R}} -
\delta_{{\bf R}'{\bf R}} \delta_{L'L} E_{\nu L} ).
$$
The first and second terms in the right-hand side of this equation are sometimes
called, correspondingly, the ``head'' and the ``tail'' of MTO.
The angular character of MTO at the central site ${\bf R}$ is specified by that of
partial wave $\phi_{{\bf R}L}$. The matrix elements of $H_{\rm KS}$ in the MTOs basis
(\ref{eqn:LMTObasis1}) can be immediately derived by using the properties
(\ref{eqn:LMTOalgebra1})-(\ref{eqn:LMTOalgebra5}) of $\phi_{{\bf R}L}$ and $\dot{\phi}_{{\bf R}L}$:
$$
\langle \chi | H_{\rm KS} | \chi \rangle = \hat{\cal H} +
(\hat{\cal H}-\hat{E}_{\nu}) \hat{E}_{\nu} \hat{p} (\hat{\cal H}-\hat{E}_{\nu}),
$$
where $\hat{E}_{\nu}$ and $\hat{p}$ are the diagonal matrices constructed from
$\{ \hat{E}_{\nu L} \}$ and $\{ p_{{\bf R}L} \}$, and
$\langle \chi | H_{\rm KS} | \chi \rangle$ is the shorthanded notation for the matrix
$\| \langle \chi_{{\bf R}L} | H_{\rm KS} | \chi_{{\bf R}'L'} \rangle \|$.
The corresponding overlap matrix, $\| \langle \chi_{{\bf R}L} | \chi_{{\bf R}'L'} \rangle \|$, is
$$
\langle \chi | \chi \rangle = \hat{1} +
(\hat{\cal H}-\hat{E}_{\nu}) \hat{p} (\hat{\cal H}-\hat{E}_{\nu}).
$$
Since the second term in the right-hand side of $\langle \chi | \chi \rangle$ is
typically small, the basis functions (\ref{eqn:LMTObasis1}) are said to form a
nearly orthogonal representation of the LMTO method, and $\hat{\cal H}$ is the LMTO
Hamiltonian in the second order of $(\hat{\cal H}$$-$$\hat{E}_{\nu})$.
The basis (\ref{eqn:LMTObasis1}) can be orthonormalized numerically, by applying
the transformation
$$
| \chi \rangle \rightarrow | \tilde{\chi} \rangle = | \chi \rangle
\langle \chi | \chi \rangle^{-1/2}.
$$
The corresponding LMTO Hamiltonian,
$\hat{H}$$=$$\langle \tilde{\chi} | H_{\rm KS} | \tilde{\chi} \rangle$,
which is formally valid in all orders of $(\hat{\cal H}$$-$$\hat{E}_{\nu})$,
is given by
$$
\hat{H} = \langle \chi | \chi \rangle^{-1/2}
\langle \chi | H_{\rm KS} | \chi \rangle
\langle \chi | \chi \rangle^{-1/2}.
$$
This Hamiltonian will be used as the starting point in the next section, for the
definition of transfer interactions between certain Wannier orbitals.
We start with the formal description of the downfolding method. The construction
of the Wannier basis functions, underlying this approach, will be considered in
Sec.~\ref{sec:Wannier}, where we will use again all merits of the LMTO method
and show that proper WFs can be constructed by retaining the
``heads'' of MTOs and attaching to them different ``tails''.

  For periodic crystals, it is convenient to work in the reciprocal (${\bf k}$)
space. Therefore, if it is not specified otherwise, we assume that the LMTO Hamiltonian
is already constructed in the reciprocal space, after the Fourier transformation of MTOs:
$$
| \chi_{{\bf k}L} \rangle = \frac{1}{\sqrt{N}} \sum_{\bf R}
e^{-i {\bf kR}} | \chi_{{\bf R}L} \rangle,
$$
where $N$ is the number of sites.

  For all considered compounds, results of our
ASA-LMTO calculations are in a good agreement
the ones obtained using more accurate full-potential methods.
In our definition of the crystal-field splitting we go beyond the conventional
ASA and take into account
nonsphericity of the electron-ion interactions (see Sec.~\ref{sec:nsphCoulomb}).

\section{\label{sec:hoppings}Downfolding Method for the Kinetic-Energy part}

  Parameters of the kinetic energy are obtained using the downfolding
method, starting from the electronic structure in LDA.
In order to describe properly the electronic structure of the TM oxides
in the valent part of the spectrum
using the LMTO method, it is typically required several tens or even hundreds basis functions
(including the ones associated with empty
spheres, which are added in order to improve the atomic spheres approximation for
loosely packed atomic structures).
Several examples of such bases will given in Secs.~\ref{sec:tSrVO3}-\ref{sec:tV2O3}.

  What we want to do next is to describe some part of this electronic structure
by certain tight-binding (TB) Hamiltonian $\hat{h}$, which, contrary $\hat{H}$,
is formulated in the basis of a very limited number of orthogonal
atomic-like orbitals.
For example, in order to reproduce the $t_{2g}$ bands located near the Fermi level,
one would like to use only three
$t_{2g}$ orbitals centered at each TM site.
These orbitals have a meaning of Wannier orbitals, which will be
considered in Sec.~\ref{sec:Wannier}.

  We start with the identity by noting that any eigenstate  of the LMTO Hamiltonian
$\hat{H}$ can be presented as the sum $|\psi \rangle$$=$$|\psi_t \rangle$$+$$|\psi_r \rangle$,
where $|\psi_t \rangle$ is expanded over the LMTO basis function of the $t_{2g}$-type,
$\{ | \tilde{\chi}_t \rangle \}$ (here, the character of the basis function
is specified by its ``head''), and $|\psi_r \rangle$ is expanded over the rest of the basis
functions $\{ | \tilde{\chi}_r \rangle \}$.
Then,
the matrix equations for LMTO eigenstates can be rearranged identically as
\begin{eqnarray}
( \hat{H}^{tt}-\omega ) | \psi_t \rangle  +  \hat{H}^{tr} | \psi_r \rangle & = & 0, \label{eqn:seceq1}\\
\hat{H}^{rt} | \psi_t \rangle  +  ( \hat{H}^{rr}-\omega ) | \psi_r \rangle & = & 0. \label{eqn:seceq2}
\end{eqnarray}
By eliminating $| \psi_r \rangle$ from Eq.~(\ref{eqn:seceq2}) one obtains the effective
$\omega$-dependent Hamiltonian in the basis of $t_{2g}$-states
$$
\hat{H}^{tt}_{\rm eff}(\omega) = \hat{H}^{tt} - \hat{H}^{tr}
(\hat{H}^{rr} - \omega)^{-1}\hat{H}^{rt}
$$
and the ``overlap'' matrix
$$
\hat{S}(\omega)=1+\hat{H}^{tr}
(\hat{H}^{rr}-\omega)^{-2}\hat{H}^{rt},
$$
satisfying the condition $\langle \psi_t | \hat{S} | \psi_t \rangle$$=$$1$.
Then, the required TB Hamiltonian, $\hat{h}$, is obtained after the orthonormalization
of the vectors $| \psi_t \rangle$$\rightarrow$$| \tilde{\psi}_t \rangle$$=$$\hat{S}^{1/2}| \psi_t \rangle$
and fixing the energy $\omega$ in the center of gravity of the $t_{2g}$ band ($\omega_0$):
\begin{equation}
\hat{h} = \hat{S}^{-1/2}(\omega_0) \hat{H}^{tt}_{\rm eff}(\omega_0) \hat{S}^{-1/2}(\omega_0).
\label{eqn:TB}
\end{equation}
Typically, the downfolding is performed in the reciprocal space, and the
parameter $\omega_0$ may also depend on ${\bf k}$. The
Hamiltonian $\hat{h}_{\bf k}$ can be Fourier transformed back to the real space:
$$
\hat{h}_{{\bf RR}'} = \sum_{\bf k} e^{i{\bf k}({\bf R}-{\bf R}')}\hat{h}_{\bf k}.
$$
The site-diagonal part of
$\hat{h}_{{\bf RR}'}$ shall describe the crystal-field (CF) splitting
caused by
the lattice distortion and associated with the transfer interactions between
$t$- and $r$-orbitals, which are eliminated in the downfolding method,
while the off-diagonal elements
have a meaning of transfer interactions.
The first application of this approach has been considered in Ref.~\onlinecite{PRB04}.
In Sec.~\ref{sec:KineticApplications} we will
illustrate abilities of this method for several types
of $t_{2g}$ compounds.

  The crystal-field splitting may have another origin, which
is related with nonsphericity of the electron-ion interactions.\cite{MochizukiImada}
This contribution will be considered in Sec.~\ref{sec:nsphCoulomb}.

\section{\label{sec:Wannier}Wannier functions}
  In the previous section we have shown that there is a TB Hamiltonian, $\hat{h}$,
formulated in some basis of Wannier orbitals $\{  \tilde{W}  \}$.
It allows to generate
the electronic structure
of isolated $t_{2g}$ bands, which
is practically identical
to the electronic structure obtained after the
diagonalization of the total LMTO Hamiltonian $\hat{H}$.

  In this section we will solve an inverse problem and construct the basis
$\{  \tilde{W}  \}$, which
after applying to the original
KS Hamiltonian, generates the matrix $\hat{h}$:
\begin{equation}
\hat{h} = \langle \tilde{W} | H_{\rm KS} | \tilde{W} \rangle.
\label{eqn:HWannierME1}
\end{equation}

  In a close analogy with the LMTO method, we will first introduce the
orbitals $\{  W  \}$, which are related with $\{ \tilde{W}  \}$
by the orthonormalization transformation
\begin{equation}
| \tilde{W} \rangle = | W \rangle \langle W | W \rangle^{-1/2},
\label{eqn:WF1}
\end{equation}
and search $|  W \rangle$ in the form:
\begin{equation}
|  W \rangle = | W_t \rangle + \sum_{r=1}^{N_r} \Gamma_r | W_r \rangle,
\label{eqn:WF2}
\end{equation}
where $ W_t $ is constructed entirely from the $t_{2g}$-type solutions of KS
equations inside atomic spheres and their energy-derivatives $\{ \phi_t, \dot{\phi}_t \}$:
i.e. both $\phi_t$ and $\dot{\phi}_t$ belong to the TM sites, and `t'
stands for the $3d$-$t_{2g}$ partial waves. Each $ W_r $ is constructed
from the rest of the partial waves $\{ \phi_r, \dot{\phi}_r \}$.
Then, $ W_t $ and $ W_r $ can be found from the following
conditions: \\
1. We request only the $t$-part of $| W \rangle$ to contribute to
the matrix elements of the KS Hamiltonian, and search it in the form of MTO:
\begin{equation}
|W_t \rangle = | \phi_t \rangle + | \dot{\phi}_t \rangle (\hat{\mathfrak{h}} - \hat{E}_{\nu t}).
\label{eqn:WFt}
\end{equation}
In this definition, $|W_t \rangle$ is a function
of (yet unknown) matrix $\hat{\mathfrak{h}}$, which will be found later.
The matrix elements of $H_{\rm KS}$ in the basis of these Wannier orbitals are given by:
\begin{equation}
\langle W | H_{\rm KS} | W \rangle = \hat{\mathfrak{h}} +
(\hat{\mathfrak{h}} - \hat{E}_{\nu t}) \hat{E}_{\nu t} \hat{p}_t
(\hat{\mathfrak{h}} - \hat{E}_{\nu t}).
\label{eqn:HWannierME2}
\end{equation}
2. The $r$-parts of the WF, $\{ W_r \}$,
do not contribute to the matrix elements (\ref{eqn:HWannierME2}).
They are introduced only in order to make
the WFs (\ref{eqn:WF2}) orthogonal to the rest of the eigenstates of
the Hamiltonian $\hat{H}$. Therefore, we search $| W_r \rangle$ in the form:
$$
|W_r \rangle = | \phi_r \rangle + \alpha_r | \dot{\phi}_r \rangle,
$$
where
$$
\alpha_r = \frac{-2 E_{\nu r}}{1+ \sqrt{1- 4 E_{\nu r}^2 p_r}}
$$
is obtained from the condition $\langle W_r | H_{\rm KS} | W_r \rangle$$=$$0$. \\
3. The coefficients $\Gamma_r$ are found from the orthogonality condition of
$| W \rangle$ to $N_r$ eigenstates $\{ | \psi_i \rangle \}$
of the original LMTO Hamiltonian $\hat{H}$:
\begin{equation}
\sum_{r=1}^{N_r} \langle \psi_i | W_r \rangle \Gamma_r =
-\langle \psi_i | W_t \rangle, ~~ i=1,~\dots,~N_r.
\label{eqn:orthogonality}
\end{equation}
This allows to include the $r$-components of the WFs in a systematic
way. For example, by taking into consideration the $2p$-partial waves
inside oxygen spheres ($N_r$$=$$9$ wavefunctions for cubic perovskites),
the WFs can be orthogonalized to $N_r$$=$$9$ O($2p$) bands, etc.
For a given $\hat{\mathfrak{h}}$, the problem is reduced to the solution of the
system of linear equations (\ref{eqn:orthogonality}).

  Since $| W_t \rangle$ contributes to Eq.~(\ref{eqn:orthogonality}),
the coefficients $\{ \Gamma_r \}$ will also depend on $\hat{\mathfrak{h}}$.
Therefore, the total WF is an implicit function of the
matrix $\hat{\mathfrak{h}}$: $| W \rangle$$\equiv$$| W(\hat{\mathfrak{h}}) \rangle$. \\
4. The last step is the orthonormalization (\ref{eqn:WF1}), which after substitution
into Eq.~(\ref{eqn:HWannierME1}) yields the following equation for the
matrix $\hat{\mathfrak{h}}$:
$$
\hat{\mathfrak{h}} = \langle W(\hat{\mathfrak{h}}) | W(\hat{\mathfrak{h}}) \rangle^{1/2}
~\hat{h}~\langle W(\hat{\mathfrak{h}}) | W(\hat{\mathfrak{h}}) \rangle^{1/2} -
(\hat{\mathfrak{h}} - \hat{E}_{\nu t}) \hat{E}_{\nu t} \hat{p}_t
(\hat{\mathfrak{h}} - \hat{E}_{\nu t}).
$$
This equation is solved iteratively with respect to $\hat{\mathfrak{h}}$.

\subsection{\label{sec:WannierExtension}Spacial extension of Wannier functions}

  The choice of the WFs as well as their extension in the real space is not
uniquely defined. For many practical applications one would like to have
``maximally localized'' orbitals,\cite{MarzariVanderbilt}
although in the context of the WFs,
the term ``maximally localized'' itself bears certain arbitrariness and is merely a
mathematical construction, because depending on the considered physical property
one can introduce different criteria of the ``maximal localization''.

  Although we do not explicitly employ here any procedure which would pick up the
``most localized'' representation for the WFs, our method
well suits this general strategy and the obtained WFs are
expected to be well localized around the central TM sites.

  There are several ways of controlling the spacial extension of the WFs in the LMTO method. \\
1. By using different envelop functions one can, in principle, change the spacial extension
of MTOs (Fig.~\ref{fig.MTO}), which controls the decay
of the original LMTO Hamiltonian in the real space.
For example, instead of irregular solutions of Laplace equation, one can use
Hankel functions of the complex argument.
However, any choice should satisfy certain criteria of the completeness of the
basis set. From this point of view, the use of the Hankel functions is not well
justified
as it typically deteriorates the accuracy of LMTO
calculations. Therefore, in the present work we leave this problem as it is
and fix the LMTO basis set. \\
2. Once the LMTO basis is fixed, the relative weight of the TM
$d$-states and other atomic states which contribute to the $t_{2g}$ band cannot be
changed (see Fig.~\ref{fig.DOSsummary}). For example, the contribution of the
oxygen $2p$-states cannot be replaced by muffin-tin orbitals
centered at the TM sites and vise versa. The same proportion of
atomic orbitals should be preserved in the WFs, constructed for
this $t_{2g}$ band. Then, the only
parameter which can be controlled is how many WFs, centered at different
sites of the lattice, contribute to the density of $d$-states at the given
TM site. Then, the definition ``localized orbital'' mean that it is
mainly centered around given TM site. Conversely, the ``delocalized orbital''
may have a long tail spreading over other TM sites.
Then, it is easy to see that our procedure corresponds to the former choice.
Indeed, in the first order of
$(\hat{\mathfrak{h}}$$-$$\hat{E}_{\nu t})$ and neglecting for a while the
nonorthogonality to the rest of the electronic states, the norm of the WF
can be obtained from Eq.~(\ref{eqn:WFt}) as
$\langle \tilde{W} | \tilde{W} \rangle$$=$$\langle \phi_t | \phi_t \rangle$, meaning that
the WF is fully localized at the central TM site.
Then, it holds $\hat{\mathfrak{h}}$$=$$\hat{h}$, which is valid in the second order
of $(\hat{\mathfrak{h}}$$-$$\hat{E}_{\nu t})$.\cite{LMTO} Therefore, the leading
correction to the above approximation, which define the actual weight of the WF
at the neighboring TM sites is controlled by the parameters
of the kinetic energy $\hat{h}$, and is of the order of
$\hat{h} \hat{p}_t \hat{h}$. As we will see below, the latter is small.
The conclusion is rather generic and well anticipated for the strongly-correlated
systems for which the kinetic-energy terms is generally small. \\
3. The angular character of the WF at the central TM site
should be consistent with the one extracted from the local density of states in the region of
$t_{2g}$ bands (in the other words, the local density of states at the TM sites
should be well represented by atomic orbitals $\{ | \tilde{\chi}_t \rangle \}$
used in the downfolding method).
Therefore, we choose $\{ | \tilde{\chi}_t \rangle \}$ as the set of atomic orbitals
which mainly contribute to the local density of states in the region of $t_{2g}$ bands.
For these purposes, at each TM site we calculate the density matrix in the basis of
five $d$ orbitals $\{ | \tilde{\chi}_d \rangle \}$:
\begin{equation}
\hat{\cal N} = \sum_{i \in t_{2g}} \langle \tilde{\chi}_d | \psi_i \rangle
\langle \psi_i | \tilde{\chi}_d \rangle,
\label{eqn:DensityMatrix}
\end{equation}
and sum up the contributions of all $t_{2g}$ bands
(here, $i$ is an joint index, which incorporates the band index and
the coordinates of the ${\bf k}$-point in the first Brillouin zone).
Then, we diagonalize $\hat{\cal N}$, and assign three most populated orbitals, obtained
after the diagonalization to $\{ | \tilde{\chi}_t \rangle \}$.

\begin{center}
\section{\label{sec:nsphCoulomb}Crystal-field splitting caused by nonsphericity
of electron-ion Interactions}
\end{center}

  The contribution of Coulomb interactions to the crystal-field splitting is a
tricky issue. Despite an apparent simplicity of the problem, one should clearly
distinguish different contribution and not to include them twice, in the kinetic
and Coulomb parts of the model Hamiltonian (\ref{eqn:Hmanybody}). The use of
full-potential techniques does not automatically guarantee the right answer.
However, the atomic-spheres-approximation, which typically supplements the
LMTO method, will also require additional corrections for the crystal-field
splitting. In this section we would like to make two comments on this problem. \\
1. The nonsphericity of \textit{on-site} Coulomb interactions is already
included in the second part of the model Hamiltonian (\ref{eqn:Hmanybody}).
The problem will be discussed in details in Sec.~\ref{sec:screenedU}.
Therefore, in order to avoid the double counting,
the corresponding contribution to the
kinetic-energy part should be subtracted. From this point of view the use of the
spherically averaged KS potential in ASA is well
justified.

  The same is true for the intersite Coulomb interactions, if they are
explicitly included to the model Hamiltonian (\ref{eqn:Hmanybody}). \\
2. All remaining interactions should generally contribute to the
crystal-field splitting. In ASA, the proper correction at the site ${\bf R}$ can be
found by considering the matrix elements of the Coulomb potential produced by
all other atomic spheres (or ions) at the site ${\bf R}$,
$$
\Delta \hat{h}_{\bf RR} = \sum_{{\bf R}'\neq{\bf R}}
\langle \tilde{W}_{\bf R} |
\frac{- Z^{*}_{{\bf R}'} e^2}{|{\bf R}+{\bf r}-{\bf R}'|}
| \tilde{W}_{\bf R} \rangle,
$$
where $\tilde{W}_{\bf R}$$\equiv$$\tilde{W}_{\bf R}({\bf r})$ is the WF centered at the
site ${\bf R}$, and $Z^{*}_{{\bf R}'}$ is the total charge associated with the
sphere ${\bf R}'$: namely, the nuclear charge minus the electronic charge encircled by the
atomic sphere. The nonspherical part of this integral can be easily calculated in the
real space by using the multipole expansion for
$|{\bf R}+{\bf r}-{\bf R}'|^{-1}$.

  In all forthcoming discussions, unless it is specified otherwise, the matrix
elements of the crystal-field splitting will incorporate the correction
$\Delta \hat{h}_{\bf RR}$ associated with nonsphericity of the electron-ion interactions.

\section{\label{sec:KineticApplications}Applications to transition-metal oxides}

\subsection{\label{sec:tSrVO3}Cubic Perovskites: SrVO$_3$}
  SrVO$_3$ is a rare example of perovskite compounds, which crystallizes in the
ideal cubic structure. It attracted a considerable attention in the connection
with the bandwidth control of the metal-insulator transition.\cite{IFT}

  For the cubic compounds, the separation of the LMTO basis functions into
$\{  | \tilde{\chi}_t \rangle \}$ and $\{ | \tilde{\chi}_r \rangle \}$ used in the downfolding method
is rather straightforward: three $t_{2g}$ orbitals centered at each V site form the
subspace of $\{ | \tilde{\chi}_t \rangle \}$ orbitals, and the rest of the basis functions
are associated with $\{ | \tilde{\chi}_r \rangle \}$.

\begin{figure}[h!]
\begin{center}
\resizebox{6cm}{!}{\includegraphics{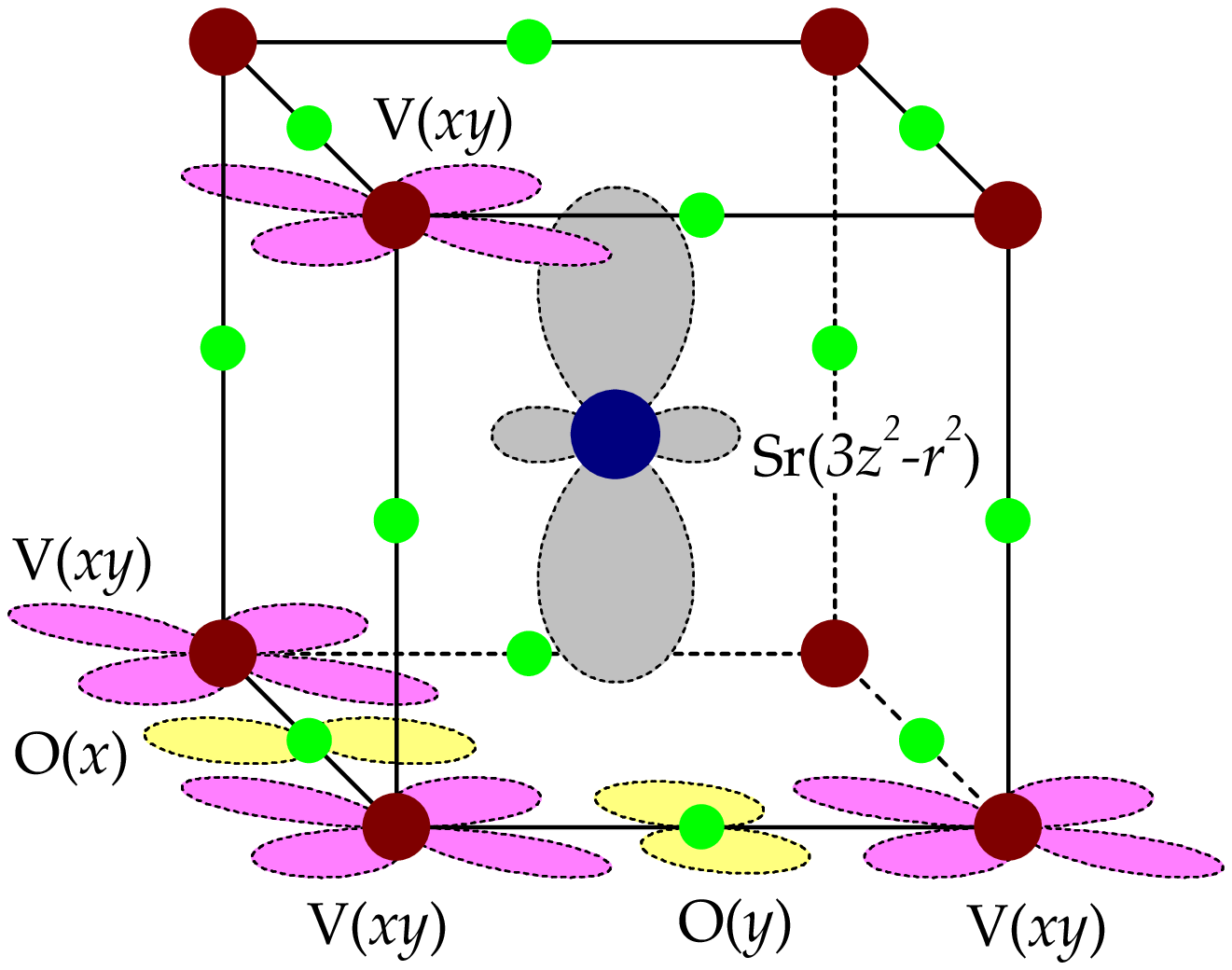}}
\resizebox{9cm}{!}{\includegraphics{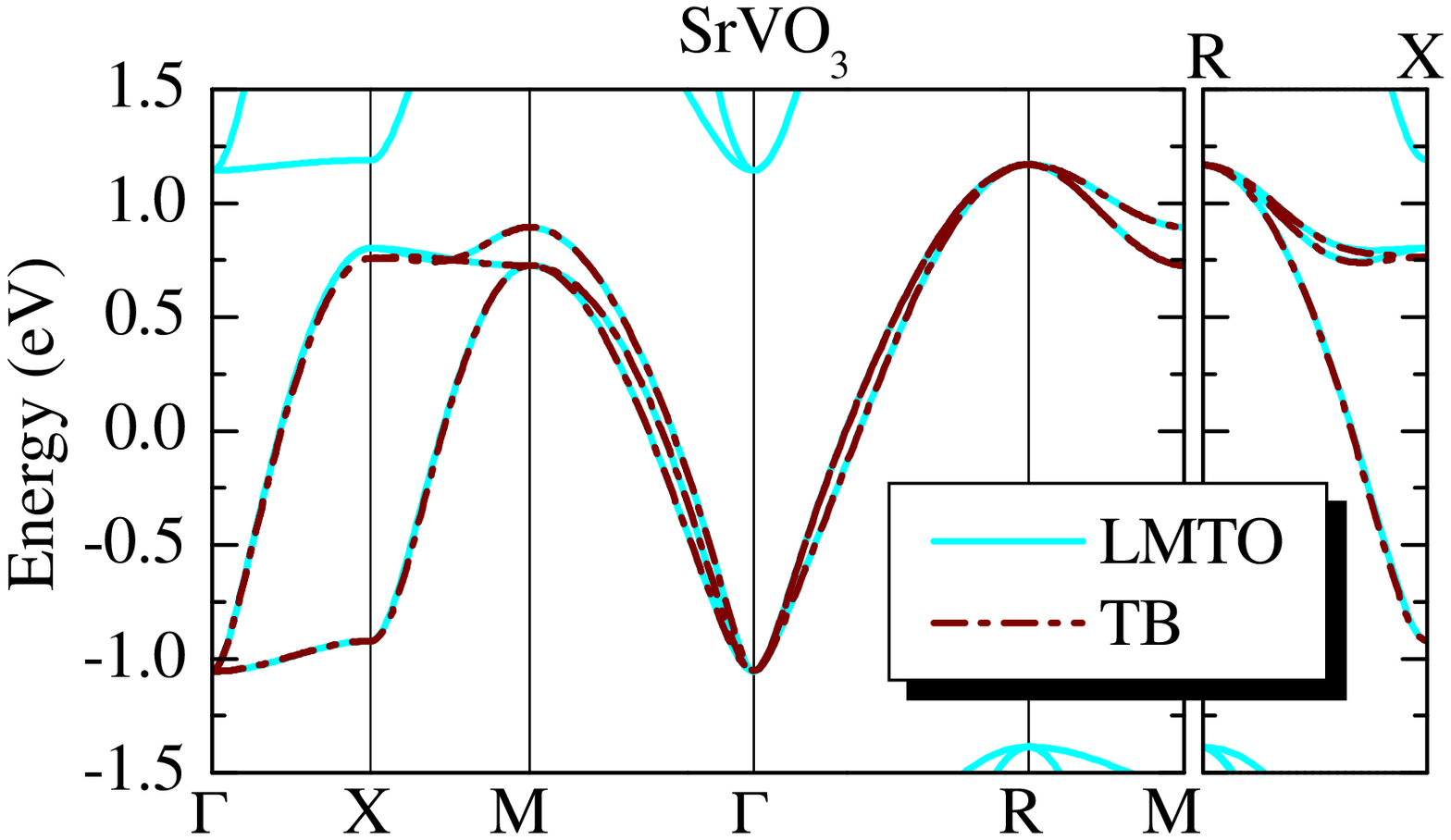}}
\end{center}
\caption{\label{fig.SrVO3bands}
Left panel: Crystal structure of cubic perovskites and atomic wavefunctions
mediating transfer interactions between V($t_{2g}$) orbitals.
The standard V($xy$)-O($y$)-V($xy$) and V($xy$)-O($x$)-V($xy$) interactions
operate in the $x$- and $y$-directions, respectively. The
V($xy$)-Sr($3z^2$-$r^2$)-V($xy$) interaction operate in the ``forbidden''
$z$-direction.
Right panel: LDA energy bands for SrVO$_3$ obtained in LMTO calculations and
after the tight-binding (TB) parametrization using the downfolding method.
Notations of the high-symmetry points of the Brillouin zone are taken from
Ref.~\protect\onlinecite{BradlayCracknell}.}
\end{figure}

  Parameters of LMTO calculations for SrVO$_3$ are given in Table~\ref{tab.SrVO3LMTOparameters}.
\begin{table}[h!]
\caption{\label{tab.SrVO3LMTOparameters} Atomic positions
(in units of cubic lattice parameter $a$$=$$3.842$\AA),
atomic radii (in \AA) and basis functions included in LMTO calculations for
cubic SrVO$_3$.}
\begin{ruledtabular}
\begin{tabular}{ccccc}
type of atom  & position       & atomic radius & LMTO basis & number of atoms \\
\colrule
Sr  & $(0.5,0.5,0.5)$          &  $1.919$      & $5s5p4d4f$ &  $1$   \\
V   & $(0,0,0)$                &  $1.470$      & $4s4p3d$   &  $1$   \\
O   & $(0.5,0,0)$              &  $1.032$      & $2s2p$     &  $3$   \\
\end{tabular}
\end{ruledtabular}
\end{table}
The corresponding electronic structure is shown in Fig.~\ref{fig.SrVO3bands}.
The downfolding procedure is nearly perfect and well reproduces the behavior
of three $t_{2g}$ bands.
As expected for cubic perovskite compounds,\cite{SlaterKoster}
the transfer interactions between different $t_{2g}$ orbitals are small.
The dispersion of $t_{2g}$ bands is well described in terms of three interaction
parameters
$t_1$, $t'_1$, and $t_2$:
$$
\varepsilon_{xy}({\bf k})= 2t_1 (\cos a k_x + \cos a k_y) + 2t_1'\cos a k_z +4t_2 \cos a k_x \cos a k_y
$$
($a$ being the cubic lattice parameter; similar expressions for the
$yz$ and $zx$ bands are obtained by cyclic permutations of the indices $x$, $y$, and $z$).
The parameters $t_1$, $t'_1$, and $t_2$, obtained after the Fourier transformation,
are listed in Table~\ref{tab.SrVO3TBrealspace}.
\begin{table}[h!]
\caption{\label{tab.SrVO3TBrealspace}
Parameters of transfer interactions for SrVO$_3$ (in eV).}
\begin{ruledtabular}
\begin{tabular}{ccc}
  $t_1$   &  $t'_1$ &   $t_2$                 \\
\colrule
$-$$0.209$             & $-$$0.023$          & $-$$0.084$         \\
\end{tabular}
\end{ruledtabular}
\end{table}
As expected, the nearest-neighbor (NN) $dd\pi$-interaction $t_1$ mediated by the oxygen
$2p$-states is the strongest. For the $xy$-orbitals, it operates in the $x$ and
$y$ directions. However, there is also an appreciable $dd\delta$-interaction
$t'_1$ operating in the ``forbidden'' direction (for example, the direction
$z$ in the case of $xy$ orbitals). These interactions are mediated by the
Sr($4d$) states and strongly depend on the proximity of the latter to the
Fermi level. Therefore, it is not quite right to say
that the transfer interactions between $t_{2g}$ orbitals are strictly
two-dimensional in the cubic lattice.\cite{HarrisPRL03}
Since the
La($5d$) states are located even lower in energy than the Sr($4d$) ones,
the interaction $t'_1$ is expected to be even stronger in LaTiO$_3$.
However, in the case of LaTiO$_3$ we have an additional complication associated with
the orthorhombic distortion. As we will see below, it changes the conventional form of transfer interactions
expected for the simplified cubic perovskite structure dramatically.

\begin{figure}[h!]
\begin{center}
\resizebox{6cm}{!}{\includegraphics{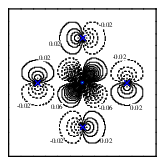}}
\resizebox{6cm}{!}{\includegraphics{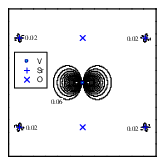}}
\end{center}
\caption{\label{fig.SrVO3WF}
Contour plot of the xy-Wannier orbital of SrVO$_3$,
in the $(001)$ (left) and $(\overline{1}10)$ (right) planes.
The solid and dashed line correspond to the positive and negative values of the Wannier function.
Atomic positions are shown by symbols.
Around each site, the Wannier function increases/decreses with the step $0.04$
from the values indicated on the graph.}
\end{figure}

  The corresponding WF is shown in Fig.~\ref{fig.SrVO3WF}.
In this case, the $t$-part of the WF was constructed from
the $3d$-$t_{2g}$ partial waves inside V spheres. The partial waves
of the Sr($4d5s$), V($3d$-$e_g$), and O($2p$) types were included into the
$r$-part, in order to enforce the orthogonality of the WF
to the bands of the
aforementioned type.

\begin{figure}[h!]
\begin{center}
\resizebox{10cm}{!}{\includegraphics{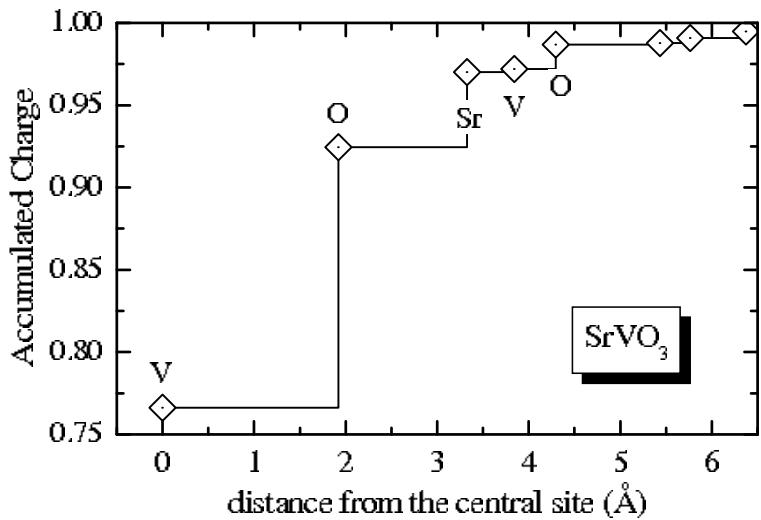}}
\end{center}
\caption{\label{fig.SrVO3WFextension}
Spacial extension of $t_{2g}$-Wannier function in SrVO$_3$: the electronic charge
accumulated around the central V site after adding every new sphere of
neighboring atomic sites.}
\end{figure}

  Since the $t_{2g}$ band is an \textit{antibonding} combination of the atomic
V($3d$-$t_{2g}$) and O($2p$) orbitals,
the WF has nodes located between V and O sites.
Fig.~\ref{fig.SrVO3WFextension} illustrates the spacial extension of the WF.
It shows the electronic charge accumulated around the central V site after adding
every new sphere of the neighboring atomic sites.
Since the WF is normalized, the total charge should be equal to one.
In the case of SrVO$_3$, 77\%
of the this charge belongs to the central V site, 16\% is distributed
over four neighboring oxygen sites, about 5\% belongs to the next eight Sr sites,
and 1\% -- to the eight oxygen sites located in the fourth coordination sphere.
Other contributions are small.

  It is also instructive to calculate the expectation value of the square of the
position operator, $\langle {\bf r}^2 \rangle$$=$$\langle W | {\bf r}^2 | W \rangle$,
which characterizes the spread of the WF in the method of Marzari
and Vanderbilt.\cite{MarzariVanderbilt} They proposed to define the
``maximally localized'' Wannier orbitals as the ones which minimize
$\langle {\bf r}^2 \rangle$. Using the WF shown in Fig.~\ref{fig.SrVO3WF},
we obtain $\langle {\bf r}^2 \rangle$$=$$2.37$ \AA$^2$. Unfortunately, at present all
applications of the method by Marzari and Vanderbilt to the TM oxides
are limited by MnO.\cite{Posternak} Therefore, we can make only indirect comparison
between two different compounds. The values of $\langle {\bf r}^2 \rangle$ reported
in Ref.~\onlinecite{Posternak} for individual WFs centered at the Mn
and O sites were of the order of 0.6-0.8 \AA$^2$, that is considerably smaller than
2.37 \AA$^2$ obtained in our work for SrVO$_3$. However, such a difference is not
surprising. \\
1. Our scheme of constructing the WFs is not based on the minimization
of $\langle {\bf r}^2 \rangle$. Therefore, our values of $\langle {\bf r}^2 \rangle$
should be generally larger. \\
2. More importantly, the spacial extension of the WFs depends on the
dimensionality of the Hilbert space, which is used in the construction of the
Hubbard model. For example, we will show in Sec.~\ref{sec:tV2O3} that by treating
explicitly the $e_g^\sigma$ states in V$_2$O$_3$ one can easily find more compact
representation for the Wannier orbitals. This characteristic can be further improved
by including the O($2p$) states explicitly into the Wannier basis,\cite{comment.2}
like in Ref.~\onlinecite{Posternak} for MnO.
However, there is a very high price to pay for this extra localization.
This is the dimensionality of the Hilbert space, which becomes crucial in the
many-body methods for the numerical solution of the
model Hamiltonian (\ref{eqn:Hmanybody}).

  Thus, we believe that our WFs constructed for \textit{isolated $t_{2g}$ band}
are indeed well localized.

  For cubic perovskites, there are several ways of extracting
parameters of transfer interactions from the first-principles
electronic structure calculations. For example, one can simply fit the LDA
band structure in terms of a small number of Slater-Koster interactions.\cite{SlaterKoster}
However, the situation becomes increasingly complicated in materials with the lower
crystal symmetry, like the orthorhombically distorted perovskite oxides, corundum, or
pyrochlore compounds.
First, the number of possible Slater-Koster interactions increases dramatically.
Second, the form of these interactions becomes very complicated and differs substantially
from the cubic perovskite compounds (one example is the
mixing of $t_{2g}$ and $e_g$ orbitals by the orthorhombic distortion, which does not
occur in cubic perovskites).
Therefore, it seems that for complex compounds
the only way to proceed is to use
the downfolding method.
In the next sections we will consider several examples along this line.

\subsection{\label{sec:tYTiO3}Orthorhombically Distorted Perovskites: YTiO$_3$}

  YTiO$_3$ is a \textit{ferromagnetic insulator}. The resent interest to
this compound has been
spurred by the behavior of orbital polarization, which
is closely related with the origin of the ferromagnetic ground state.
YTiO$_3$ is typically considered in combination with LaTiO$_3$, which is an
\textit{antiferromagnetic insulator}. The magnetic behavior of these two,
formally isoelectronic materials, is not fully understood.\cite{PRB04}

\begin{figure}[h!]
\begin{center}
\resizebox{6cm}{!}{\includegraphics{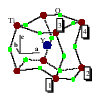}}
\resizebox{9cm}{!}{\includegraphics{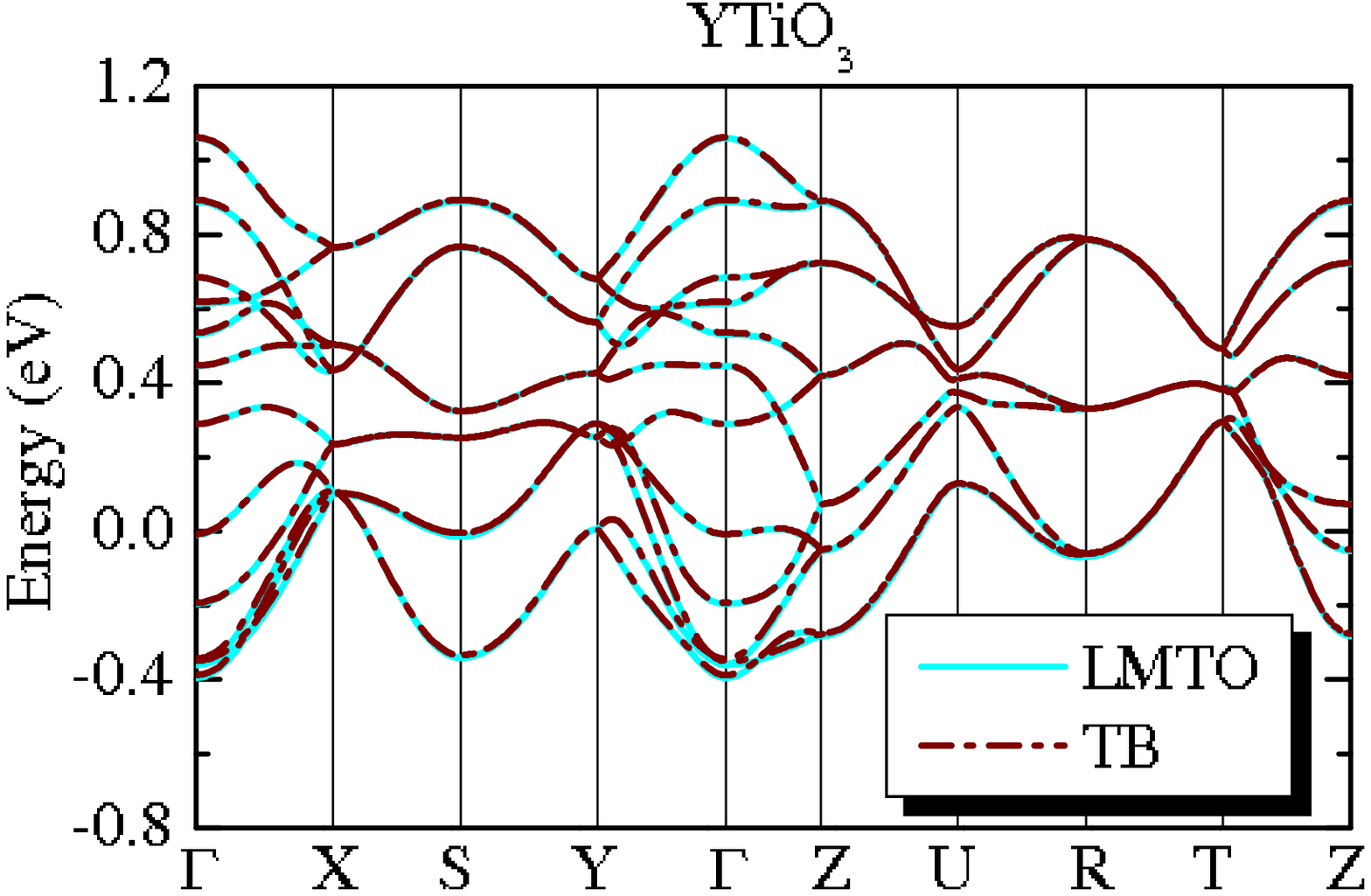}}
\end{center}
\caption{\label{fig.YTiO3bands}
Left panel: Fragment of crystal structure of YTiO$_3$ (a ``distorted cube'').
The inset shows the directions of orthorhombic axes.
Right panel: LDA energy bands for YTiO$_3$ obtained in LMTO calculations and
after tight-binding (TB) parametrization using the downfolding method.
Notations of the high-symmetry points of the Brillouin zone are taken from
Ref.~\protect\onlinecite{BradlayCracknell}.
}
\end{figure}

  Contrary to SrVO$_3$, both YTiO$_3$ and LaTiO$_3$ crystallize in the strongly distorted
orthorhombic structure (shown in Fig.~\ref{fig.YTiO3bands} for YTiO$_3$, the space group
No.~62 in the International Tables; the Sch\"{o}nflies notation
is $D^{16}_{2h}$). In this
section we will illustrate abilities of the downfolding method for distorted
perovskite compounds, using YTiO$_3$ as an example.
Parameters of LMTO calculations for YTiO$_3$ are given in Table~\ref{tab.YTiO3LMTOparameters}.

\begin{table}[h!]
\caption{\label{tab.YTiO3LMTOparameters} Atomic positions
(in units of orthorhombic lattice parameters $a$$=$$5.316$,
$b$$=$$5.679$, and $c$$=$$7.611$\AA),
atomic radii (in \AA) and basis functions included in LMTO calculations for
orthorhombically distorted YTiO$_3$.
The symbol `Em' stands for the empty spheres.}
\begin{ruledtabular}
\begin{tabular}{ccccc}
type of atom  & position       & atomic radius & LMTO basis & number of atoms \\
\colrule
Y   & $(0.521,0.073,0.250)$    &  $1.758$      & $5s5p4d4f$ &  $4$   \\
Ti  & $(0,0,0)$                &  $1.507$      & $4s4p3d$   &  $4$   \\
O   & $(0.379,0.458,0.250)$    &  $1.052$      & $2s2p$     &  $4$   \\
O   & $(0.309,0.191,0.558)$    &  $1.043$      & $2s2p$     &  $8$   \\
Em  & $(0.192,0.268,0.116)$    &  $0.595$      & $1s$       &  $8$   \\
Em  & $(0.438,0.329,0.750)$    &  $0.757$      & $1s2p$     &  $4$   \\
Em  & $(0.149,0.368,0.027)$    &  $0.556$      & $1s$       &  $8$   \\
Em  & $(0.106,0.208,0.199)$    &  $0.479$      & $1s$       &  $8$   \\
\end{tabular}
\end{ruledtabular}
\end{table}

  A new problem we have to address here is how to separate the basis functions of the LMTO
method onto the $\{ | \tilde{\chi}_t \rangle \}$ and  $\{ | \tilde{\chi}_r \rangle \}$
orbitals.
Note that although the $t_{2g}$ band is well separated from the rest of the electronic
structure also in the case of YTiO$_3$, the \textit{atomic} $t_{2g}$ and $e_g$ orbitals are
strongly mixed by the crystal-field effects and the transfer interactions in the distorted
perovskite structure. Therefore, the conventional separation into atomic $t_{2g}$
orbitals and the rest of the basis functions does not apply here, and in order to
generate $\{ | \tilde{\chi}_t \rangle \}$ we use eigenvectors of the density matrix
(see Sec.~\ref{sec:WannierExtension}).

 For the site 1, shown in
Fig.~\ref{fig.YTiO3bands}, these three ``$t_{2g}$ orbitals''
have the following form (in the basis of
$| xy \rangle$, $| yz \rangle$,
$| z^2 \rangle$, $| zx \rangle$, and $| x^2$$-$$y^2 \rangle$ orbitals,
in the orthorhombic coordinate frame):
\begin{eqnarray}
| \tilde{\chi}_1 \rangle & = & (\phantom{-}0.13,          -0.60,-0.24,
\phantom{-}0.34,          -0.67), \nonumber\\
| \tilde{\chi}_2 \rangle & = & (\phantom{-}0.17,\phantom{-}0.50,-0.35,
\phantom{-}0.77,\phantom{-}0.11),  \label{eqn:YTiO3localframe}\\
| \tilde{\chi}_3 \rangle & = & (          -0.43,          -0.54, -0.29,
\phantom{-}0.22,\phantom{-}0.62). \nonumber
\end{eqnarray}
At the sites 2, 3, and 4 similar orbitals can be generated from the ones at the
site 1
using the symmetry properties of the $D^{16}_{2h}$ group and
applying
the 180$^\circ$ rotations around the orthorhombic axes ${\bf a}$,
${\bf c}$, and ${\bf b}$, respectively. These orbitals define the local basis
(or the local coordinate frame) around each Ti site.

  Then, the rest of the basis functions $\{ \tilde{\chi}_r \}$ can be eliminated
using the downfolding method. The corresponding electronic structure for the $t_{2g}$ bands
is shown in Fig.~\ref{fig.YTiO3bands}, which reveals an excellent agreement between
results of LMTO calculations and their tight-binding parametrization
using the down-folding method.

  Parameters obtained after the transformation to
the real space are listed in Table~\ref{tab.YTiO3TBrealspace},
in the local coordinate frame.
\begin{table}[h!]
\caption{\label{tab.YTiO3TBrealspace}
Parameters of the kinetic energy for YTiO$_3$ (in eV), in the local
coordinate frame.
the atomic positions are shown in Fig.~\protect\ref{fig.YTiO3bands}.
The basis functions at the site 1 are given by
Eqs.~(\protect\ref{eqn:YTiO3localframe}).
The basis functions at the sites 2 and 3 are obtained
by the 180$^\circ$ rotations of the site 1 around the orthorhombic
axes ${\bf a}$ and ${\bf c}$, respectively. The matrix $\hat{h}_{11}$ describes the
crystal-field splitting at the site 1
(nonsphericity of electron-ion interactions
has been added in
$\hat{h}_{11}$) The matrices $\hat{h}_{12}$ and $\hat{h}_{13}$
stand for the transfer integrals in the bonds 1-2 and 1-3, respectively. }
\begin{ruledtabular}
\begin{tabular}{cc}
  ${\bf i}$-${\bf j}$   &  $\hat{h}_{\bf ij}$              \\
\colrule
1-1 &
$
\begin{array}{rrr}
 -0.060 & -0.039           & -0.014             \\
 -0.039 & \phantom{-}0.028 & -0.001             \\
 -0.014 & -0.001           & \phantom{-}0.032   \\
\end{array}
$ \\
\colrule
1-2 &
$
\begin{array}{rrr}
 -0.080            & -0.036            &  \phantom{-}0.072 \\
  \phantom{-}0.162 &  \phantom{-}0.013 &  \phantom{-}0.073 \\
  \phantom{-}0.063 &  \phantom{-}0.031 & -0.038            \\
\end{array}
$ \\
\colrule
1-3 &
$
\begin{array}{rrr}
  \phantom{-}0.000 & -0.080            &  \phantom{-}0.055 \\
 -0.080            &  \phantom{-}0.082 & -0.007            \\
  \phantom{-}0.055 & -0.007            &  0.073            \\
\end{array}
$ \\
\end{tabular}
\end{ruledtabular}
\end{table}
We note
a substantial crystal-field splitting associated with the orthorhombic
distortion. After the diagonalization of the site-diagonal part of the TB Hamiltonian
$\hat{h}$, we obtain the following (``one-down, two-up'')
splitting of $t_{2g}$ levels:
$-$$0.076$, $0.032$, and $0.046$ eV. Some implications of the crystal-field splitting
to the orbital polarization and the magnetic ground state of YTiO$_3$ and
LaTiO$_3$ have been
discussed in Refs.~\onlinecite{PRB04} and \onlinecite{MochizukiImada}.\cite{comment.1}
The form of transfer interactions becomes extremely
complicated, and differs dramatically from many naive expectations
based on the analogy with the cubic perovskites.
Generally, the transfer interactions are three-dimensional and operate
between different $t_{2g}$ orbitals.

\begin{figure}[h!]
\begin{center}
\resizebox{5cm}{!}{\includegraphics{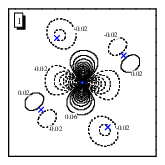}}
\resizebox{5cm}{!}{\includegraphics{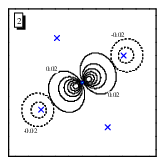}}
\resizebox{5cm}{!}{\includegraphics{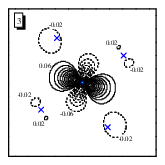}}
\end{center}
\caption{\label{fig.YTiO3WF}
Contour plot of three $t_{2g}$-Wannier orbitals
in the $(001)$ plane of YTiO$_3$. The orbitals are centered at the site 1
shown in Fig.~\protect\ref{fig.YTiO3bands}.
Projections of the vanadium and oxygen sites onto the plane are shown
by circles and crosses, respectively. Other notations are the same as
in Fig.~\protect\ref{fig.SrVO3WF}.}
\end{figure}

\begin{figure}[h!]
\begin{center}
\resizebox{10cm}{!}{\includegraphics{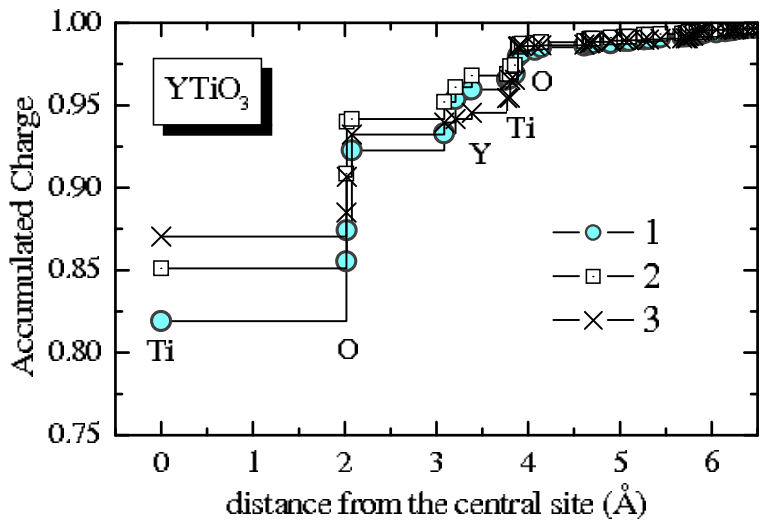}}
\end{center}
\caption{\label{fig.YTiO3WFextension}
Spacial extension of $t_{2g}$-Wannier functions in YTiO$_3$.}
\end{figure}

  The WFs are shown in Fig.~\ref{fig.YTiO3WF}, and their spacial extension
in the real space is illustrated in Fig.~\ref{fig.YTiO3WFextension}.
In these calculations,
the WFs have been orthogonalized to the O($2p$), Ti($3d$-$e_g$), and
Y($4d$) bands.
The orbitals appear to be more localized than in SrVO$_3$:
82-87\% of the total charge belongs the central Ti site, and
only 6 to 10\% is distributed over neighboring oxygen sites.
This is because of the large orthorhombic distortion, which suppresses
all interatomic interactions mediated by the oxygen states.
Another reason is the larger energy distance between O($2p$) and
$t_{2g}$ bands in YTiO$_3$ ($3.2$ eV against $0.3$ eV in SrVO$_3$ -- see
Fig.~\ref{fig.DOSsummary}), which explains smaller weight of the
atomic oxygen states in the $t_{2g}$ band and the WFs of YTiO$_3$.

  Another interesting feature is that the degree of localization is
pretty different for three orbitals. For example, we obtain
$\langle {\bf r}^2 \rangle$$=$ 2.28, 1.90, and 2.05 \AA$^2$, correspondingly
for $W_1$, $W_2$, and $W_3$ shown in Fig.~\ref{fig.YTiO3WF}.
One can paraphrase it in a different way: the degree of hybridization can be
different for different $t_{2g}$ orbitals, unless they are related with each
other by symmetry operations.

\subsection{\label{sec:tY2Mo2O7}Pyrochlores: Y$_2$Mo$_2$O$_7$}

  The pyrochlore compounds exhibit a variety of interesting properties.
Many of them are not fully understood.

  Y$_2$Mo$_2$O$_7$ is a canonical example of geometrically frustrated systems.
In this compound, the magnetic atoms form the networks of
corner-sharing tetrahedra. Then, the antiferromagnetic coupling
between NN Mo spins leads to the frustration.
The origin of this antiferromagnetic coupling can be
understood on the basis of semi-empirical
Hartree-Fock calculations.\cite{PRB03}
A remaining question, which is not fully understood, is the origin of
the spin-glass state realized in Y$_2$Mo$_2$O$_7$ below
20 K.\cite{Y2Mo2O7exp}
For comparison,
Nd$_2$Mo$_2$O$_7$ is a ferromagnet, revealing a large anomalous Hall effect.\cite{PRB03,Taguchi}

  Another interesting group is superconducting $\beta$-pyrochlores with the chemical formula
$A$Os$_2$O$_6$ ($A$$=$ K, Rb, and Cs).\cite{Hiroi}

  In this section we will derive parameters of the kinetic energy for the $t_{2g}$
band of Y$_2$Mo$_2$O$_7$. Very similar strategy can be applied for other pyrochlores.
Parameters of LMTO calculations for Y$_2$Mo$_2$O$_7$ are listed in
Table~\ref{tab.Y2Mo2O7}.
\begin{table}[h!]
\caption{\label{tab.Y2Mo2O7} Atomic positions
(in units of cubic lattice parameter $a$$=$$10.21$ \AA),
atomic radii (in \AA) and basis functions included in LMTO calculations for
pyrochlore Y$_2$Mo$_2$O$_7$.}
\begin{ruledtabular}
\begin{tabular}{ccccc}
type of atom  & position       & atomic radius & LMTO basis & number of atoms \\
\colrule
Y   & $(0.5,0.5,0.5)$          &  $1.746$      & $5s5p4d$   &  $4$   \\
Mo  & $(0,0,0)$                &  $1.587$      & $5s5p4d$   &  $4$   \\
O   & $(0.375,0.375,0.375)$    &  $1.032$      & $2s2p$     &  $2$   \\
O   & $(0.338,0.125,0.125)$    &  $1.032$      & $2s2p$     &  $12$  \\
Em  & $(0.125,0.125,0.125)$    &  $0.952$      & $1s2p3d$   &  $2$   \\
Em  & $(0.25,0,0)$             &  $1.044$      & $1s2p3d$   &  $8$   \\
\end{tabular}
\end{ruledtabular}
\end{table}

\begin{figure}[h!]
\begin{center}
\resizebox{6cm}{!}{\includegraphics{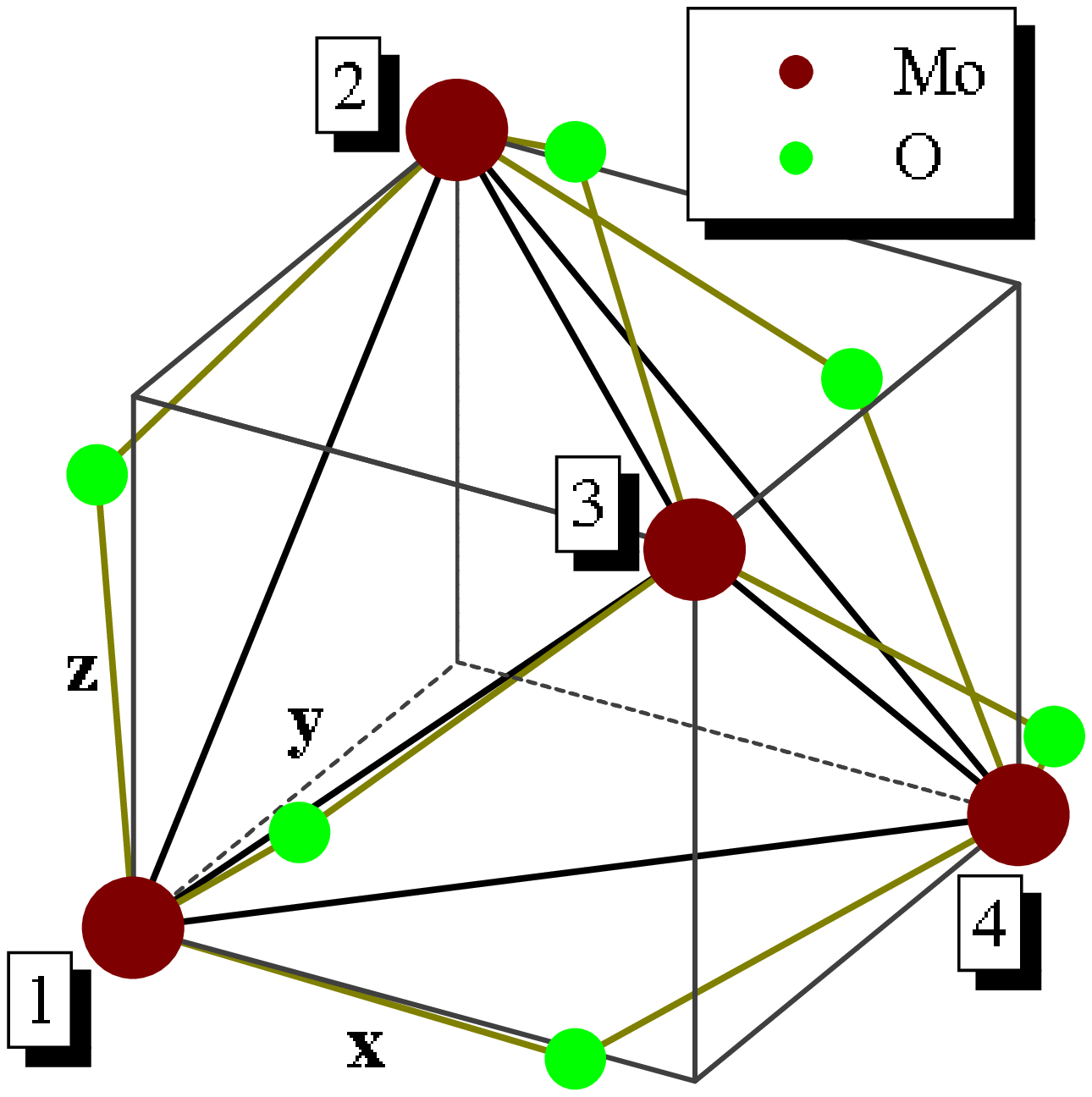}}
\resizebox{9cm}{!}{\includegraphics{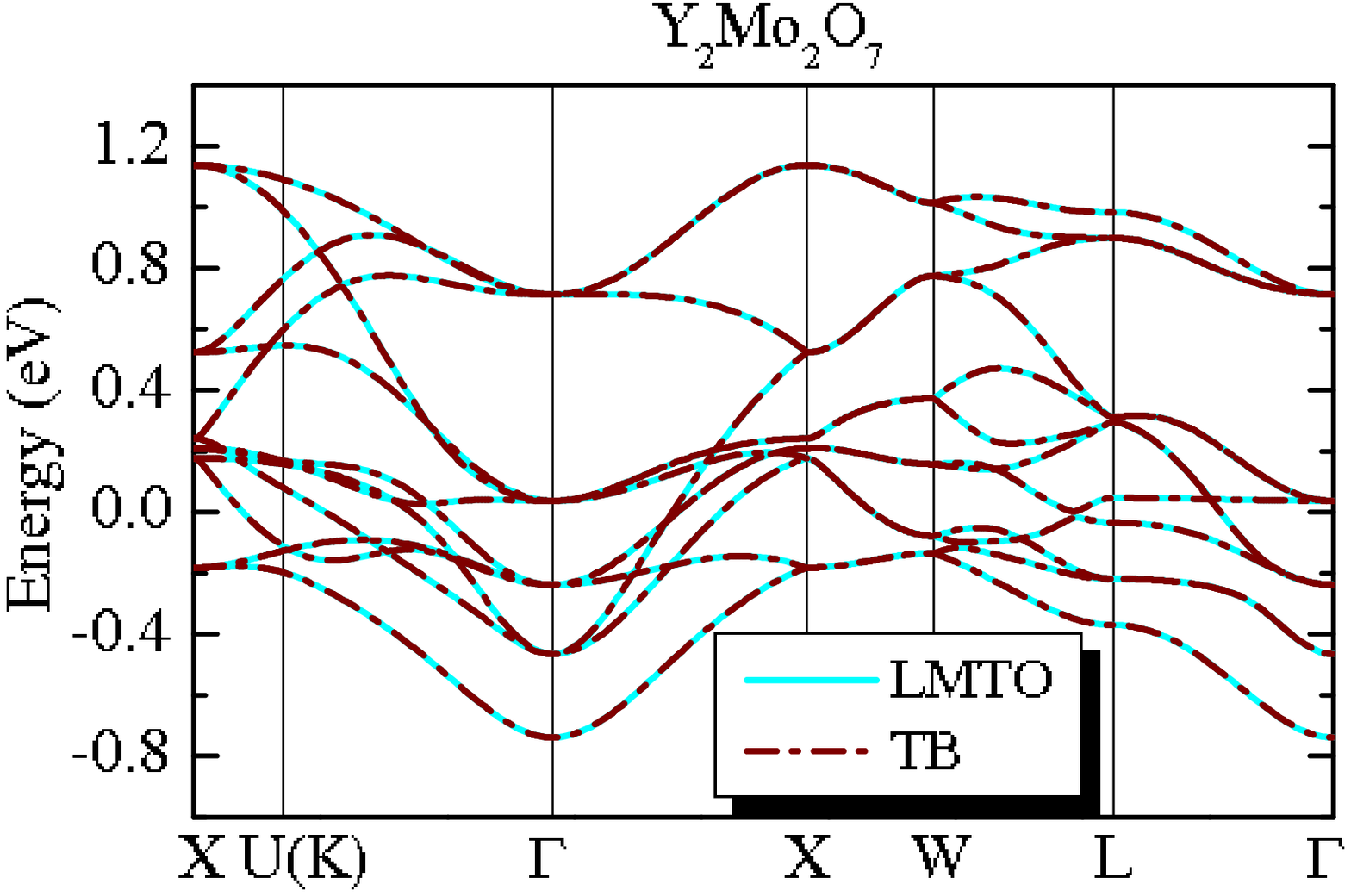}}
\end{center}
\caption{\label{fig.Y2Mo2O7bands}
Left panel shows the fragment of crystal structure of Y$_2$Mo$_2$O$_7$: a single Mo-tetrahedron
surrounded by the oxygen sites. In Y$_2$Mo$_2$O$_7$, the Y and Mo ions from two
sublattices of corner-sharing tetrahedra. The sublattices interpenetrate in a way, in
which each Mo is located in the center of Y hexagon (not shown here).
Right panel: LDA energy bands for Y$_2$Mo$_2$O$_7$ obtained in LMTO calculations and
after tight-binding (TB) parametrization using the downfolding method.
Notations of the high-symmetry points of the Brillouin zone are taken from
Ref.~\protect\onlinecite{BradlayCracknell}.
}
\end{figure}

  In the pyrochlore lattice, each Mo site is located in the trigonal environment (Fig.~\ref{fig.Y2Mo2O7bands}).
Therefore, the atomic Mo($t_{2g}$) levels will be split into one-dimensional
$a_{1g}$ and two-dimensional $e_g^\pi$ representations.
The latter states can mix with the
Mo($e_g^\sigma$) states, which belong to the same representation.
Therefore, the basis functions $\{ | \tilde{\chi}_t \rangle \}$ can be constructed in the
same manner as for YTiO$_3$, by diagonalizing the site-diagonal part of the density matrix
for the $t_{2g}$ bands. For the site 1 shown in Fig.~\ref{fig.Y2Mo2O7bands}, this yields the
following atomic orbitals (in the basis of $| xy \rangle$, $| yz \rangle$,
$| z^2 \rangle$, $| zx \rangle$, and $| x^2$$-$$y^2 \rangle$
orbitals):
\begin{eqnarray}
| \tilde{\chi}_1 \rangle & = & (\phantom{-}0.58,\phantom{-}0.58,\phantom{-}0\phantom{.00},
\phantom{-}0.58,\phantom{-}0\phantom{.00}), \nonumber \\
| \tilde{\chi}_2 \rangle & = & (          -0.06,\phantom{-}0.18,-0.29,
          -0.13,\phantom{-}0.93),  \label{eqn:Y2Mo2O7localframe}\\
| \tilde{\chi}_3 \rangle & = & (\phantom{-}0.18,          -0.04,\phantom{-}0.93,
          -0.14,\phantom{-}0.29). \nonumber
\end{eqnarray}
In these notations, the first orbital correspond to the $a_{1g}$ representation, and
two other -- to the $e_g$ representation.
Similar orbitals at the sites 2, 3, and 4 can be generated from the ones at the site 1 using
the symmetry operations of the $O_h^7$ group (No.~227 in the
International Tables): namely, the 180$^\circ$ rotations
around the cubic axes ${\bf x}$, ${\bf y}$, and ${\bf z}$, respectively. The rest of the
basis functions form the subspace $\{ | \tilde{\chi}_r \rangle \}$.

  The electronic structure obtained after the elimination of the $\{ | \tilde{\chi}_r \rangle \}$
orbitals is shown in Fig.~\ref{fig.Y2Mo2O7bands}. Again, we note an excellent agreement
with the results of the original LMTO calculations. The parameters of the kinetic energy in the real
space are listed in Table~\ref{tab.Y2Mo2O7TBrealspace}.
\begin{table}[h!]
\caption{\label{tab.Y2Mo2O7TBrealspace}
Parameters of the kinetic energy for Y$_2$Mo$_2$O$_7$ (in eV), in the local
coordinate frame.
The atomic positions are shown in Fig.~\protect\ref{fig.Y2Mo2O7bands}.
The basis orbitals at the site 1 are given by
Eqs.~(\protect\ref{eqn:Y2Mo2O7localframe}).
The basis orbitals at the sites 2, 3, and 4 are obtained
by the 180$^\circ$ rotations
of the site 1
around the cubic
axes ${\bf x}$, ${\bf y}$, and ${\bf z}$, respectively. The matrix $\hat{h}_{11}$ describes the
crystal-field splitting at the site 1 (nonsphericity of
electron-ion interactions has been added in $\hat{h}_{11}$).
The matrix $\hat{h}_{14}$
stands for transfer integrals between atoms 1 and 4. }
\begin{ruledtabular}
\begin{tabular}{cc}
  ${\bf i}$-${\bf j}$   &  $\hat{h}_{\bf ij}$            \\
\colrule
1-1 &
$
\begin{array}{rrr}
             -0.177 &   \phantom{-}0.000 &   \phantom{-}0.000 \\
   \phantom{-}0.000 &   \phantom{-}0.088 &   \phantom{-}0.000 \\
   \phantom{-}0.000 &   \phantom{-}0.000 &   \phantom{-}0.088 \\
\end{array}
$ \\
\colrule
1-4 &
$
\begin{array}{rrr}
            -0.064 & -0.001 &   \phantom{-}0.002 \\
            -0.001 & -0.186 &             -0.070 \\
  \phantom{-}0.002 & -0.070 &   \phantom{-}0.022 \\
\end{array}
$  \\
\end{tabular}
\end{ruledtabular}
\end{table}
We note an appreciable ($\sim$$0.25$ eV)
crystal-field splitting between the $a_{1g}$ and $e_g$ orbitals.\cite{PRB03}
The NN transfer interactions in the bonds other than 1-4 can be obtained
using the symmetry operations of the $O_h^7$ group. The transfer
interactions beyond the nearest neighbors are considerably smaller.

\begin{figure}[h!]
\begin{center}
\resizebox{5cm}{!}{\includegraphics{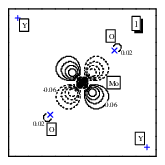}}
\resizebox{5cm}{!}{\includegraphics{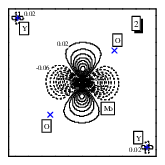}}
\resizebox{5cm}{!}{\includegraphics{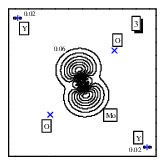}}
\end{center}
\caption{\label{fig.Y2Mo2O7WF}
Contour plot of three $t_{2g}$-Wannier orbitals
in the $(001)$ plane of Y$_2$Mo$_2$O$_7$.
The orbitals correspond to the site 1 in Fig.~\protect\ref{fig.Y2Mo2O7bands}.
The orbital $1$ corresponds to the $a_{1g}$ representation,
the orbitals $2$ and $3$ correspond to the $e_g$ representation.
Other notations are the same is
in Fig.~\protect\ref{fig.SrVO3WF}.}
\end{figure}

\begin{figure}[h!]
\begin{center}
\resizebox{10cm}{!}{\includegraphics{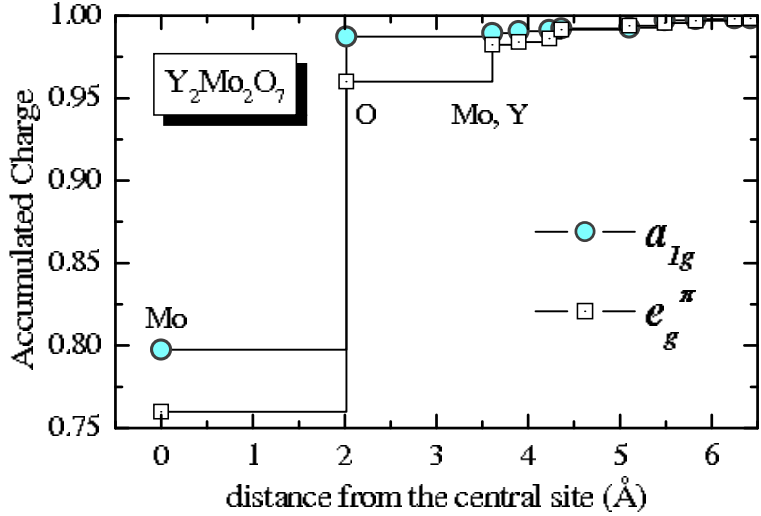}}
\end{center}
\caption{\label{fig.Y2Mo2O7WFextension}
Spacial extension of $t_{2g}$-Wannier functions in Y$_2$Mo$_2$O$_7$.}
\end{figure}

  The corresponding WFs are shown in Fig.~\ref{fig.Y2Mo2O7WF}, and their
spacial extension is depicted in Fig.~\ref{fig.Y2Mo2O7WFextension}.
The WFs have been orthogonalized to the neighboring Y($4d$-$e_g$) and
O($2p$) bands. Since the $4d$-wavefunctions are typically more extended
in comparison with the $3d$ ones, the WFs are less localized.
In the case of Y$_2$Mo$_2$O$_7$, 75-80\% of the total charge is located at the
central site, and about 20\% is distributed over neighboring oxygen sites.
The degree of localization also depends on the symmetry of WFs.
So, the $a_{1g}$ orbital is well localized within the MoO$_6$ cluster,
whereas the $e_g$ orbitals have a noticeable weight ($\sim$2.5\% of the
total charge) at the Y and Mo sites belonging to the next coordination sphere.

\subsection{\label{sec:tV2O3} Corundum-type V$_2$O$_3$}

   V$_2$O$_3$ is regarded as the canonical Mott-Hubbard system, where the Coulomb
interaction between conduction electrons leads to a breakdown of the conventional
one-electron band theory.\cite{McWhanRiceRemeika}
It was and continues to be the subject of vast research activity, which has been
summarized in many review articles (for instance, Ref.~\onlinecite{IFT}).

\begin{figure}[h!]
\begin{center}
\resizebox{6cm}{!}{\includegraphics{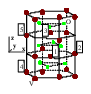}}
\resizebox{9cm}{!}{\includegraphics{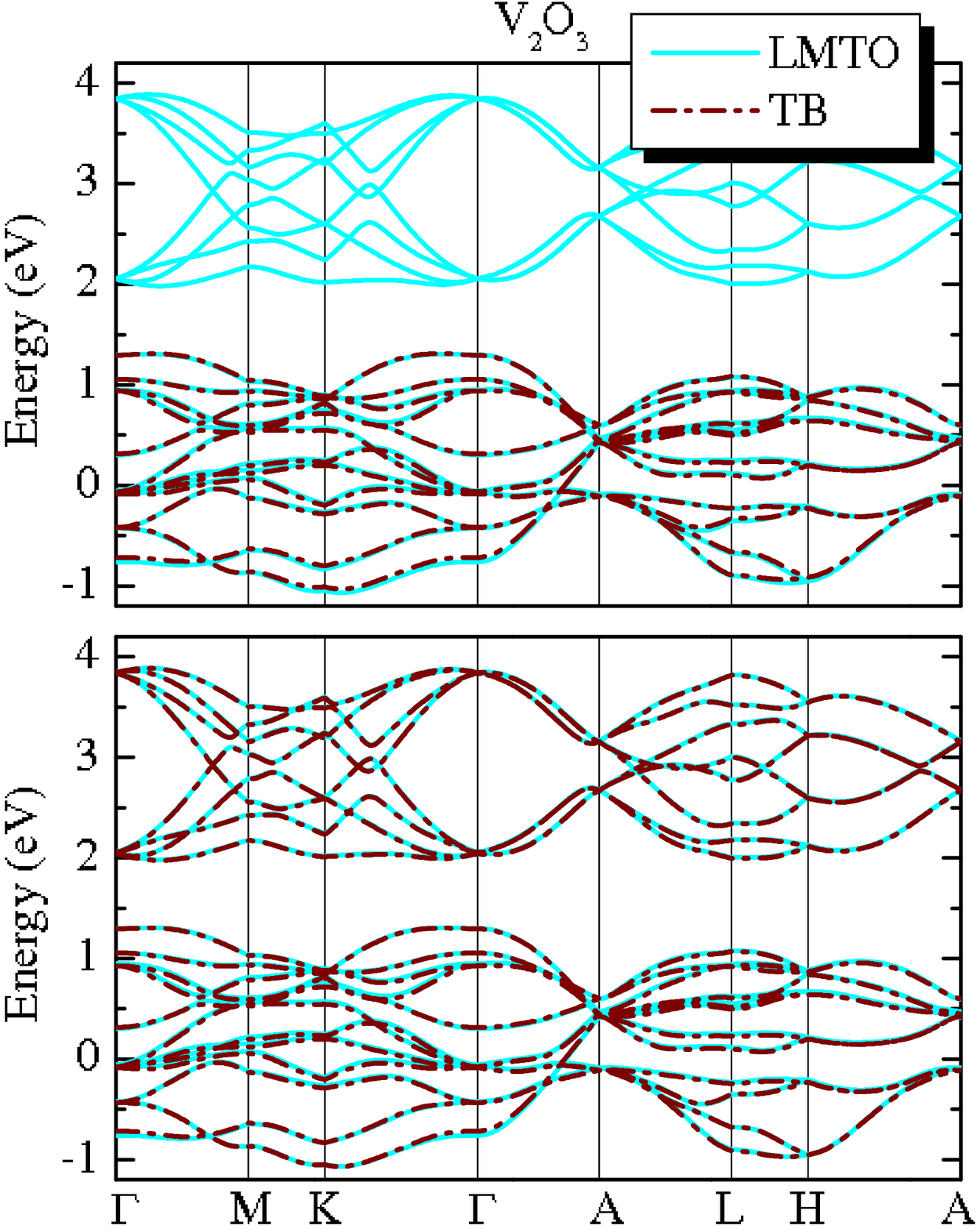}}
\end{center}
\caption{\label{fig.V2O3bands}
Left panel: Crystal structure of V$_2$O$_3$ with the notation of atomic sites.
The sites 3 and 4 are equivalent and can be transformed to each other by
primitive translations (both atomic positions
are required in order to specify the form of transfer integrals
in Table~\protect\ref{tab.V2O3TBrealspace}).
Right panel: LDA energy bands for V$_2$O$_3$ obtained in LMTO calculations and
after tight-binding (TB) parametrization using the downfolding method for the
three-orbital (top) and five-orbital (bottom) models.
Notations of the high-symmetry points of the Brillouin zone are taken from
Ref.~\protect\onlinecite{BradlayCracknell}.
}
\end{figure}

  V$_2$O$_3$ crystallizes in the corundum structure with two formula units per
rhombohedral cell (the space group is $D_{3d}^6$,
No.~167 in the International Tables). The local environment of the
V sites is trigonal (Fig.~\ref{fig.V2O3bands}), in which the $t_{2g}$ levels are
split into one-dimensional $a_{1g}$ and two dimensional $e^{\pi}_g$ representations.

  Parameters of LMTO calculations for V$_2$O$_3$ are given in Table~\ref{tab.V2O3}.
\begin{table}[h!]
\caption{\label{tab.V2O3} Atomic positions (in units of rhombohedral
lattice parameters $a$$=$$2.859$ and $c$$=$$4.667$ \AA),
atomic radii (in \AA) and basis functions included in LMTO calculations for
V$_2$O$_3$.}
\begin{ruledtabular}
\begin{tabular}{ccccc}
type of atom  & position       & atomic radius & LMTO basis & number of atoms \\
\colrule
V   & $(0,0,  0.5)$           &  $1.317$      & $4s4p3d$   &  $4$   \\
O   & $(-0.5,0.326,0.25)$     &  $1.009$      & $2s2p3d$   &  $6$   \\
Em  & $(0,0,0)$               &  $1.223$      & $1s2p3d$   &  $2$   \\
Em  & $(0.5,0.267,-0.25)$     &  $0.921$      & $1s2p$     &  $6$   \\
\end{tabular}
\end{ruledtabular}
\end{table}

  In the present work, our main interest in V$_2$O$_3$ will be purely academic.
As we can see in Fig.~\ref{fig.DOSsummary}, V$_2$O$_3$ has two well
separated bands, which are mainly formed by the V($3d$) states.
One is the $t_{2g}$ band, which in LDA is crossed by the Fermi level.
Another one is the $e_g$ band, which is located around 3 eV, and composed mainly
of the $e_g^\sigma$ states. The latter can mix with the $e_g^\pi$ ones.
Therefore, for V$_2$O$_3$ (and related corundum-type oxides) one can introduce two
different models.

  The first one is more general and explicitly treats all V($3d$) bands. In the following
we will call it as the ``five-orbital'' model, according to the number of
basis orbitals $\{ | \tilde{\chi}_t \rangle \}$ per one V site. The oxygen degrees of
freedom will be eliminated using the downfolding method. The corresponding TB Hamiltonian
will be denoted as $\hat{h}^{(5)}$.

  The second one is the minimal model, which can be derived from the previous one
by eliminating the $e^{\sigma}_g$ states. We will call it the ``three-orbital model''.
The corresponding TB Hamiltonian will be denoted as $\hat{h}^{(3)}$.

  The basic difference between these two models is that the first one treats the $e^{\sigma}_g$
states explicitly, while in the second case the effect of these states is included
implicitly, through the renormalization of interaction parameters between $a_{1g}$ and
$e_g^\pi$ orbitals.

  For both models, the local orbitals at each V site were obtained from the diagonalization
of the density matrix, which sums up the
contributions over twelve $t_{2g}$ bands. For the three-orbital model,
such choice of the basis functions is very important, as it controls the accuracy of the
downfolding method. For the five-orbital model, one can use any unitary transformation
of the five $3d$ orbitals. Obviously, the final result will not depend on this transformation.
However, for a better comparison between two models, we use the same basis in both
cases.
In principle, the $e_g^\pi$ and $e_g^\sigma$ states will be mixed in the density
matrix, as they belong to the same representation. However, we will continue
to call the lower- and upper-lying $e_g$ states as $e_g^\pi$ and $e_g^\sigma$,
despite the fact that each of them may have an admixture of another type.

  Then, the basis orbitals at the sites 1 and 2 (see Fig.~\ref{fig.V2O3bands}) have the
following form (in the basis of atomic
$| xy \rangle$, $| yz \rangle$,
$| z^2 \rangle$, $| zx \rangle$, and $| x^2$$-$$y^2 \rangle$ orbitals):
\begin{eqnarray}
| \tilde{\chi}_1 \rangle & = & (\phantom{-}0.77,-0.29,\phantom{-}0\phantom{.00},
\phantom{-}0.48,-0.30), \nonumber \\
| \tilde{\chi}_2 \rangle & = & (\phantom{-}0.30,\phantom{-}0.48,\phantom{-}0\phantom{.00},
\phantom{-}0.29,\phantom{-}0.77), \nonumber \\
| \tilde{\chi}_3 \rangle & = & (\phantom{-}0\phantom{.00},\phantom{-}0\phantom{.00},\phantom{-}1\phantom{.00},
\phantom{-}0\phantom{.00},\phantom{-}0\phantom{.00}),  \label{eqn:V2O3localframe}\\
| \tilde{\chi}_4 \rangle & = & (\phantom{-}0.49,\phantom{-}0.51,\phantom{-}0\phantom{.00},
-0.65,-0.27), \nonumber \\
| \tilde{\chi}_5 \rangle & = & (\phantom{-}0.27,-0.65,\phantom{-}0\phantom{.00},
-0.51,\phantom{-}0.49). \nonumber
\end{eqnarray}
In these notations, $| \tilde{\chi}_1 \rangle$ and $| \tilde{\chi}_2 \rangle$ are the $e^{\pi}_g$ orbitals,
$| \tilde{\chi}_3 \rangle$ is the $a_{1g}$ orbital, and $| \tilde{\chi}_4 \rangle$ and $| \tilde{\chi}_5 \rangle$
are the $e^{\sigma}_g$ orbitals.
The basis orbitals at the sites 3, 4, and 5 are generated
by the mirror-reflection ${\bf y}$$\rightarrow$$-$${\bf y}$
in Eq.~(\ref{eqn:V2O3localframe}). In the three-orbital model, first three orbitals
constitute the subspace $\{ | \tilde{\chi}_t \rangle \}$, while two remaining
orbitals are included in $\{ | \tilde{\chi}_r \rangle \}$.
In the five-orbital model, all five orbitals are included in $\{ | \tilde{\chi}_t \rangle \}$.

   The electronic structure obtained after the downfolding is shown in Fig.~\ref{fig.V2O3bands}.
The three-orbital model well reproduces the
behavior of twelve $t_{2g}$ bands, while the five-orbital model allows to reproduce both $t_{2g}$
and $e_g$ bands.

  The corresponding parameters in the real space are given in Table~\ref{tab.V2O3TBrealspace}.
In the five-orbital model, one can see a noticeable hybridization between
$t_{2g}$ and $e^{\sigma}_g$ states. Therefore, the elimination of the $e^{\sigma}_g$ states in the
three-orbital model should lead to an additional renormalization of the parameters
of the crystal-field splitting and the transfer interactions. Generally, the matrix elements
of the kinetic-energy part in the subspace of $t_{2g}$
orbitals
\textit{are not the same} for two considered models.
\begin{table}[h!]
\caption{\label{tab.V2O3TBrealspace}
Parameters of the kinetic energy (in eV)
in the local
coordinate frame
for three- and
five-orbital  models of V$_2$O$_3$ (denoted as $\hat{h}^{(3)}$ and $\hat{h}^{(5)}$, respectively).
The atomic positions are shown in Fig.~\protect\ref{fig.V2O3bands}.
The basis functions at the sites 1 and 2 are given by
Eqs.~(\protect\ref{eqn:V2O3localframe}). The basis functions at the sites 3, 4, and 5
are obtained 1 by the refection
${\bf y}$$\rightarrow$$-$${\bf y}$ of the site 1.
The order of orbitals is $e^{\pi}_g$, $a_{1g}$, and $e^{\sigma}_g$.
The $e^{\sigma}_g$ orbitals
are eliminated in the three-orbital model.
}
\begin{ruledtabular}
\begin{tabular}{ccc}
  ${\bf i}$-${\bf j}$   &  $\hat{h}^{(3)}_{\bf ij}$    &  $\hat{h}^{(5)}_{\bf ij}$          \\
\colrule
1-1 &
$
\begin{array}{rrrrr}
           -0.053  & \phantom{-}0.000  &  \phantom{-}0.000  & \phantom{-0.000} & \phantom{-0.000} \\
 \phantom{-}0.000  &           -0.053  &  \phantom{-}0.000  & \phantom{-0.000} & \phantom{-0.000} \\
 \phantom{-}0.000  & \phantom{-}0.000  &  \phantom{-}0.107  & \phantom{-0.000} & \phantom{-0.000} \\
 \phantom{-0.000}  &  \phantom{-0.000} &   \phantom{-0.000} & \phantom{-0.000} & \phantom{-0.000} \\
 \phantom{-0.000}  &  \phantom{-0.000} &   \phantom{-0.000} & \phantom{-0.000} & \phantom{-0.000} \\
\end{array}
$ &
$
\begin{array}{rrrrr}
             0.141 & \phantom{-}0.000  & \phantom{-}0.000  & \phantom{-}0.008  & \phantom{-}0.056 \\
  \phantom{-}0.000 & \phantom{-}0.141  & \phantom{-}0.000  &           -0.056  & \phantom{-}0.008 \\
  \phantom{-}0.000 & \phantom{-}0.000  & \phantom{-}0.190  & \phantom{-}0.000  & \phantom{-}0.000 \\
  \phantom{-}0.008 &           -0.056  & \phantom{-}0.000  & \phantom{-}2.326  & \phantom{-}0.000 \\
  \phantom{-}0.056 & \phantom{-}0.008  & \phantom{-}0.000  & \phantom{-}0.000  & \phantom{-}2.326 \\
\end{array}
$ \\
\colrule
1-2 &
$
\begin{array}{rrrrr}
 \phantom{-}0.024  & \phantom{-}0.089  &            -0.103  & \phantom{-0.000} & \phantom{-0.000} \\
 \phantom{-}0.089  &          -0.150  &  \phantom{-}0.207  & \phantom{-0.000} & \phantom{-0.000} \\
          -0.103  & \phantom{-}0.207  &            -0.025  & \phantom{-0.000} & \phantom{-0.000} \\
 \phantom{-0.000} & \phantom{-0.000} &   \phantom{-0.000} & \phantom{-0.000} & \phantom{-0.000} \\
 \phantom{-0.000} & \phantom{-0.000} &   \phantom{-0.000} & \phantom{-0.000} & \phantom{-0.000} \\
\end{array}
$ &
$
\begin{array}{rrrrr}
  \phantom{-}0.009 &  \phantom{-}0.080 &            -0.104 &  \phantom{-}0.026 &            -0.043 \\
  \phantom{-}0.080 &            -0.103 &  \phantom{-}0.202 &            -0.085 &  \phantom{-}0.102 \\
            -0.104 &  \phantom{-}0.202 &            -0.032 &  \phantom{-}0.114 &            -0.068 \\
  \phantom{-}0.026 &            -0.085 &  \phantom{-}0.114 &            -0.067 &            -0.030 \\
            -0.043 &  \phantom{-}0.102 &            -0.068 &            -0.030 &            -0.042 \\
\end{array}
$ \\
\colrule
1-3 &
$
\begin{array}{rrrrr}
          -0.037  &          -0.027  &  \phantom{-}0.000  & \phantom{-0.000} & \phantom{-0.000} \\
          -0.027  & \phantom{-}0.037  & \phantom{-}0.000  & \phantom{-0.000} & \phantom{-0.000} \\
\phantom{-}0.000  & \phantom{-}0.000  &            -0.342  & \phantom{-0.000} & \phantom{-0.000} \\
 \phantom{-0.000} & \phantom{-0.000} &   \phantom{-0.000} & \phantom{-0.000} & \phantom{-0.000} \\
 \phantom{-0.000} & \phantom{-0.000} &   \phantom{-0.000} & \phantom{-0.000} & \phantom{-0.000} \\
\end{array}
$ &
$
\begin{array}{rrrrr}
            -0.032 &           -0.054 &  \phantom{-}0.000 & \phantom{-}0.064 &  \phantom{-}0.163 \\
            -0.054 & \phantom{-}0.032 &  \phantom{-}0.000 & \phantom{-}0.163 &            -0.064 \\
  \phantom{-}0.000 & \phantom{-}0.000 &            -0.331 & \phantom{-}0.000 &  \phantom{-}0.000 \\
  \phantom{-}0.064 & \phantom{-}0.163 &  \phantom{-}0.000 &           -0.014 &            -0.065 \\
  \phantom{-}0.163 &           -0.064 &  \phantom{-}0.000 &           -0.065 &  \phantom{-}0.014 \\
\end{array}
$ \\
\colrule
1-4 &
$
\begin{array}{rrrrr}
          -0.089  & \phantom{-}0.024  &            -0.102  & \phantom{-0.000} & \phantom{-0.000} \\
\phantom{-}0.024  &           -0.018  &  \phantom{-}0.024  & \phantom{-0.000} & \phantom{-0.000} \\
          -0.102  & \phantom{-}0.024  &            -0.125  & \phantom{-0.000} & \phantom{-0.000} \\
 \phantom{-0.000} &  \phantom{-0.000} &   \phantom{-0.000} & \phantom{-0.000} & \phantom{-0.000} \\
 \phantom{-0.000} &  \phantom{-0.000} &   \phantom{-0.000} & \phantom{-0.000} & \phantom{-0.000} \\
\end{array}
$ &
$
\begin{array}{rrrrr}
            -0.112 &  \phantom{-}0.013 &            -0.113 & -0.053 &           -0.002 \\
  \phantom{-}0.013 &  \phantom{-}0.000 &  \phantom{-}0.048 & -0.356 &           -0.029 \\
            -0.113 &  \phantom{-}0.048 &            -0.124 & -0.092 &           -0.004 \\
            -0.053 &            -0.356 &            -0.092 & -0.213 &           -0.031 \\
            -0.002 &            -0.029 &            -0.004 & -0.031 & \phantom{-}0.000 \\
\end{array}
$ \\
\colrule
1-5 &
$
\begin{array}{rrrrr}
 \phantom{-}0.053  &           -0.053  &  \phantom{-}0.019  & \phantom{-0.000} & \phantom{-0.000} \\
 \phantom{-}0.004  &           -0.039  &  \phantom{-}0.034  & \phantom{-0.000} & \phantom{-0.000} \\
           -0.047  & \phantom{-}0.113  &            -0.054  & \phantom{-0.000} & \phantom{-0.000} \\
  \phantom{-0.000} &  \phantom{-0.000} &   \phantom{-0.000} & \phantom{-0.000} & \phantom{-0.000} \\
  \phantom{-0.000} &  \phantom{-0.000} &   \phantom{-0.000} & \phantom{-0.000} & \phantom{-0.000} \\
\end{array}
$ &
$
\begin{array}{rrrrr}
  \phantom{-}0.046 &            -0.060 &  \phantom{-}0.027 &            -0.076 &            -0.175 \\
            -0.002 &            -0.050 &  \phantom{-}0.046 &  \phantom{-}0.069 &  \phantom{-}0.151 \\
            -0.071 &  \phantom{-}0.105 &            -0.050 &            -0.020 &            -0.032 \\
            -0.124 &  \phantom{-}0.017 &  \phantom{-}0.055 &            -0.039 &            -0.078 \\
            -0.259 &  \phantom{-}0.041 &  \phantom{-}0.161 &            -0.133 &            -0.234 \\
\end{array}
$ \\
\end{tabular}
\end{ruledtabular}
\end{table}

  As an example, we shown in Table~\ref{tab.CFsplitting} the crystal-filed splitting between
$e^{\pi}_g$ and $a_{1g}$ levels for the series of corundum-type oxides,
obtained after the diagonalization of the
matrices
$\hat{h}_{11}^{(3)}$ and $\hat{h}_{11}^{(5)}$.
The splitting turns out to be very different in two different models.
Since many properties of TM oxides
are controlled by this crystal-field
splitting,\cite{CFcorundum}
such a model-dependence may be viewed as somewhat unphysical.
However, the crystal-field splitting cannot be considered independently
from other model parameters, such as the Coulomb and transfer interactions, which
should be defined
\textit{on the same footing and for the same type of model}.
For the Coulomb interactions, it is important to follow the concept of
WFs, which we will consider in the next section.
\begin{table}[h!]
\caption{\label{tab.CFsplitting}
Parameters of crystal-field splitting between $e^{\pi}_g$ and $a_{1g}$ levels
obtained for the series of corundum-type oxides in the three- ($\Delta_{\rm CF}^{(3)}$)
and five- ($\Delta_{\rm CF}^{(5)}$) orbital models. The positive value of
$\Delta_{\rm CF}$ means that the $a_{1g}$ level lies
\textit{higher} than $e^{\pi}_g$. The nonsphericity of electron-ion interactions
has been included (the values in parenthesis show the crystal-field splitting
originating from the transfer interactions alone).
}
\begin{ruledtabular}
\begin{tabular}{ccc}
  compound   & $\Delta_{\rm CF}^{(3)}$ (eV) &  $\Delta_{\rm CF}^{(5)}$ (eV)  \\
\colrule
 Ti$_2$O$_3$ & \phantom{-}0.070 (0.254)     &           -0.033 (0.159)       \\
  V$_2$O$_3$ & \phantom{-}0.160 (0.233)     & \phantom{-}0.051 (0.128)       \\
 Cr$_2$O$_3$ & \phantom{-}0.140 (0.216)     & \phantom{-}0.051 (0.129)       \\
\end{tabular}
\end{ruledtabular}
\end{table}

  We also note an appreciable contribution coming
from nonsphericity of the electron-ion interactions. This contribution, which is
ignored in conventional ASA, acts \textit{against}
the crystal-field splitting originating from the transfer interactions and tends
to stabilize the $a_{1g}$ level. This may revise certain conclusions obtained
in the framework of ASA-LMTO method.\cite{CFcorundum}

\begin{figure}[h!]
\begin{center}
\resizebox{3cm}{!}{\includegraphics{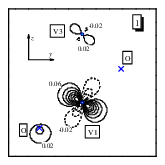}}
\resizebox{3cm}{!}{\includegraphics{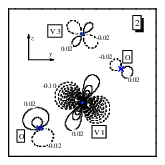}}
\resizebox{3cm}{!}{\includegraphics{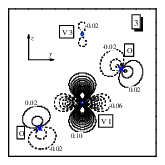}}
\resizebox{3cm}{!}{\makebox{ }}
\resizebox{3cm}{!}{\makebox{ }}
\end{center}
\begin{center}
\resizebox{3cm}{!}{\includegraphics{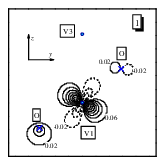}}
\resizebox{3cm}{!}{\includegraphics{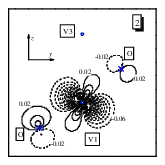}}
\resizebox{3cm}{!}{\includegraphics{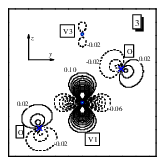}}
\resizebox{3cm}{!}{\includegraphics{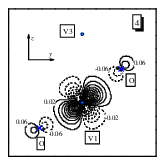}}
\resizebox{3cm}{!}{\includegraphics{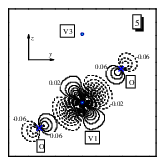}}
\end{center}
\caption{\label{fig.V2O3WF}
Contour plot Wannier orbitals
in the $(100)$ plane of V$_2$O$_3$, obtained in the three-orbital (top)
and five-orbital (bottom) models.
Projections of the vanadium and oxygen sites onto the plane are shown
by circles and crosses, respectively.
The numbering of V atoms (``V 1'' and ``V 3'')
is the same as in Fig.~\protect\ref{fig.V2O3bands}.
First two orbitals ($1$ and $2$) are of the $e_g^\pi$ type,
third orbital ($3$) is of the $a_{1g}$ type, and last
two orbitals ($4$ and $5$) are of the $e_g^\sigma$ type.
Other notations are the same is
in Fig.~\protect\ref{fig.YTiO3WF}.}
\end{figure}

\begin{figure}[h!]
\begin{center}
\resizebox{10cm}{!}{\includegraphics{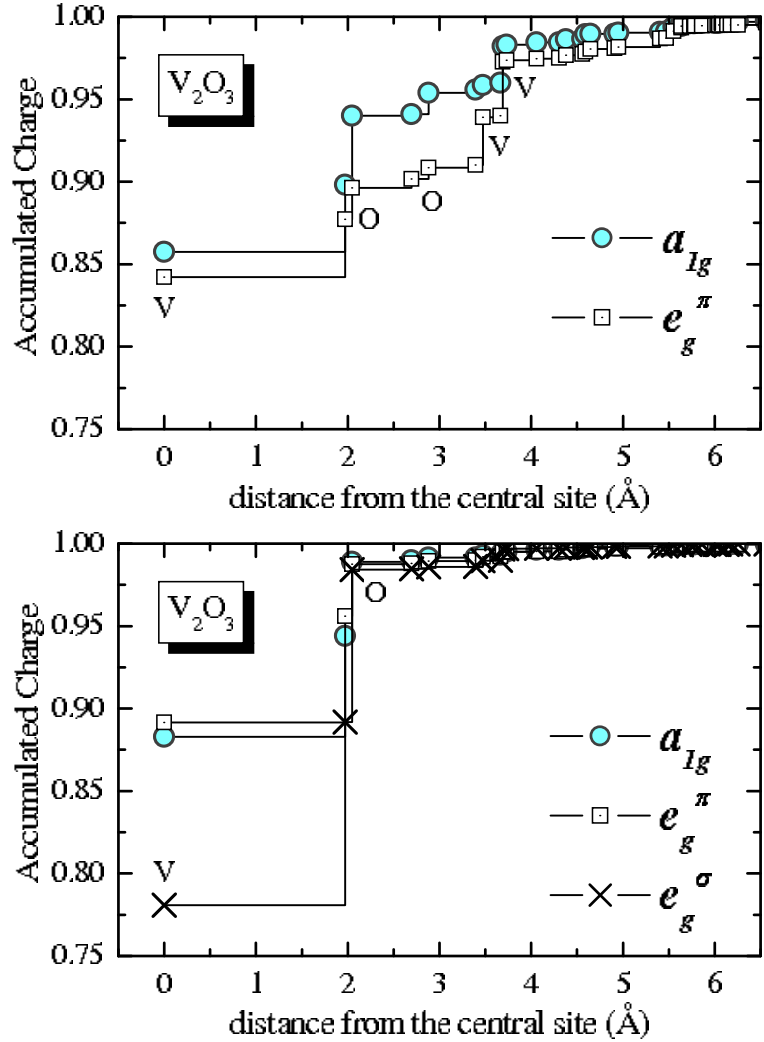}}
\end{center}
\caption{\label{fig.V2O3WFextension}
Spacial extension of Wannier functions in the three- (top)
and five- (bottom) orbital models for V$_2$O$_3$.}
\end{figure}

  Corresponding WFs are shown in Fig.~\ref{fig.V2O3WF}, and their
spacial extension -- in Fig.~\ref{fig.V2O3WFextension}.
All functions have been orthogonalized to the O($2p$) band. In the three-orbital
model, the WFs have been additionally orthogonalized to the
V($e_g$) band.
Similar to Y$_2$Mo$_2$O$_7$, the spacial extension of the WFs strongly
depends on their symmetry.
Generally, the $a_{1g}$ and $e_g^\pi$ orbitals are more localized, while the
$e_g^\sigma$ orbitals have a considerable weight (more than 20\% of the total charge)
at the neighboring oxygen sites. Furthermore, the spacial extension of the WFs
depends on the model for which they are constructed.
Generally, the five-orbital model allows to construct more compact WFs
rather than the three-orbital one.
For example, in the three-orbital model we have
$\langle {\bf r}^2 \rangle$$=$ 1.75 and 2.38 \AA$^2$, correspondingly for the
$a_{1g}$ and $e_g^\pi$ orbitals. For comparison, the five-orbital yields
$\langle {\bf r}^2 \rangle$$=$ 1.04, 1.04, and 1.41 \AA$^2$ for the
$a_{1g}$, $e_g^\pi$, and $e_g^\sigma$ orbitals, respectively.

  This is not surprising. \\
1. The transfer interactions in the three-orbital model are longer-ranged, as they
contain additional contributions mediated by the $e_g^\sigma$ orbitals. \\
2. For the three-orbital model, the WFs should be additionally
orthogonalized to the V($e_g$) band. At the central site, this condition can be easily
satisfied by choosing proper atomic
$t_{2g}$ and $e_g$ orbitals, which diagonalize the density matrix.
However, the WF has a tail spreading to the neighboring sites, which should
be additionally orthogonalized to the V($e_g$) band by including
partial waves of the $e_g^\sigma$
type into the $r$-part of the WF.

  Thus, there is certain compromise with the choice of the suitable model for
compounds like V$_2$O$_3$,
where smaller dimensionality of the Hilbert space in the
three-orbital model is counterbalanced by necessity to deal with more
extended WFs.

\section{\label{sec:screenedU}The Effective Coulomb Interaction}

  The calculation of effective Coulomb interactions for the
first principles is an extremely complicated problem because they are subjected
to different mechanisms of screening which should be taken into consideration
in the process of these calculations. So far, the solution of this
problem has not been fully accomplished by any of the research groups, despite a
vast activity in this
direction.\cite{Dederichs,Norman,McMahan,Gunnarsson,Hybertsen,GunnarssonPostnikov,
AnisimovGunnarsson,PRB94.2,NormanBrooks,PRB96,
Pickett,Springer,Kotani,Ferdi04,PRB05}
It would be probably fair to say from the very beginning that we were not
able to solve this problem either, without additional approximations,
which will be considered in Sec.~\ref{sec:approximations}.
However, we hope to present certain systematics on different points
of view, which can be found in the literature.
We will also summarize several open questions and unresolved problems.

  It is convenient to start with the basic definition of the effective Coulomb
interaction $U$, as it was discussed in many details by Herring.\cite{Herring}
According to this definition, (the spherically averaged part of) $U$ is nothing but the
energy cost for the reaction
$2(d^n)$$\rightleftharpoons$$d^{n+1}$$+$$d^{n-1}$, i.e.
for moving a $d$-electron between two atoms, located at ${\bf R}$
and ${\bf R}'$, and initially populated by $n_{\bf R}$$=$$n_{{\bf R}'}$$\equiv$$n$
electrons:
\begin{equation}
U = E [ n_{{\bf R}}+1,n_{{\bf R}'}-1 ] -
E [ n_{{\bf R}},n_{{\bf R}'}  ].
\label{eqn:HerringU1}
\end{equation}
This $U$ may depend on ${\bf R}$ and ${\bf R}'$, and using several combinations
of ${\bf R}$ and ${\bf R}'$ one can extract the values of both on-site and intersite
interactions. A typical example for SrVO$_3$ will be considered in
Sec.~\ref{sec:cLDAresults}. However, here we drop these atomic indices and consider
more general aspects of calculations of the effective Coulomb interactions.

  It is implied that the electron is transferred between two Wannier orbitals,
and $n_{\bf R}$ and $n_{{\bf R}'}$ are the populations of these Wannier orbitals.
A special precaution should be taken in order to avoid the double counting
of the kinetic energy term.
Indeed,
since the kinetic-energy term is included explicitly
into the
Hubbard model (\ref{eqn:Hmanybody}),
it should not contribute to
the total energy
difference~(\ref{eqn:HerringU1}).
This point was emphasized
by Gunnarsson and co-workers,
in the series of publications.\cite{Gunnarsson,GunnarssonPostnikov,AnisimovGunnarsson}
They proposed to derive $U$ from constraint-LDA (c-LDA)
calculations,\cite{Dederichs} and suppress \textit{all} matrix elements of hybridization
involving the atomic $d$-states. Such a procedure can be easily
implemented in the LMTO method. For the $3d$-compounds, this method typically
yields $U$$\sim$5-12 eV,\cite{AZA,PRB96}, which is too large (if correct,
results of this approach would imply that all nature surrounding us
would be ``strongly correlated'').
Therefore, although the basic strategy is correct, there is an important piece
of physics, which is missing in the method of Gunnarsson \textit{et~al}.

  Similar strategy can be pursued in
our WF method. Our basic idea is to switch off the kinetic-energy term
during the construction of the WFs, and to use these functions
in
calculations of the effective interaction $U$.
Therefore,
instead of regular WFs $\{\tilde{W}(\hat{h})\}$,
which after applying to the KS Hamiltonian generate the matrix
$\hat{h} \equiv \langle \tilde{W}(\hat{h}) | H_{\rm KS} | \tilde{W}(\hat{h}) \rangle$,
we introduce the set of
\textit{auxiliary Wannier functions} $\{\bar{W}(\hat{c})\}$, satisfying the
condition
$\langle \bar{W}(\hat{c}) | H_{\rm KS} | \bar{W}(\hat{c}) \rangle = \hat{c}$.
In the ground-state configuration ($n_{\bf R}$$=$$n_{{\bf R}'}$$\equiv$$n$),
$\hat{c}$ is a constant, which can be dropped.\cite{comment.3}
In the excited state ($n_{\bf R}$$\neq$$n_{{\bf R}'}$), $\hat{c}$ is a diagonal matrix
with respect to the site indices,
$\hat{c} \equiv \| c_{\bf R}~\delta_{{\bf RR}'} \|$, where each
matrix element $c_{\bf R}$ may
depend on occupation numbers $\{ n_{\bf R} \}$.
Since such auxiliary WFs do not interact with each other through the kinetic-energy term,
they can be used as the basis functions for the effective Coulomb interaction $U$.

  The auxiliary WFs can be easily constructed using the method proposed
in Sec.~\ref{sec:Wannier} after the substitution $\hat{h}$$=$$\hat{c}$ in all equations.
Meanwhile, the orthogonality condition to other LDA bands is strictly observed
by including proper solutions of KS equations inside atomic spheres and their
energy derivatives into the $r$-part of auxiliary WFs.
This allows to retain the hybridization between
TM $d$- and oxygen $p$-states, which is an important feature of
TM oxides. As we will see below,
the change of this hybridization, induced by the reaction
$2(d^n)$$\rightleftharpoons$$d^{n+1}$$+$$d^{n-1}$,
represents a very important channel of screening, which
substantially reduces $U$ and explains
many details of its behavior
in solids.
This channel of screening has been overlooked by
Gunnarsson \textit{et~al}.

  A similar idea, although formulated in the very different way, has been
recently proposed by Aryasetiawan \textit{et al.}\cite{Ferdi04}
They proposed to extract the parameter $U$ from the
GW method,\cite{Hedin,FerdiGunnarsson}
and suppressed all contributions
to the GW polarization function associated with the transitions between Hubbard
(in our case -- $t_{2g}$)
bands, in order to avoid the double counting of these effects in the process
of solution of the Hubbard model.
Clearly, since the polarization function in the GW method will vanish without
the kinetic-energy term, this procedure appears to be similar to the setting
$\hat{h}$$=$$\hat{c}$ for the WFs,
which are used as the basis functions for
the effective Coulomb interaction $U$.

\begin{figure}[h!]
\begin{center}
\resizebox{6cm}{!}{\includegraphics{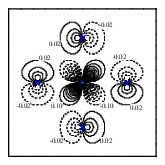}}
\resizebox{6cm}{!}{\includegraphics{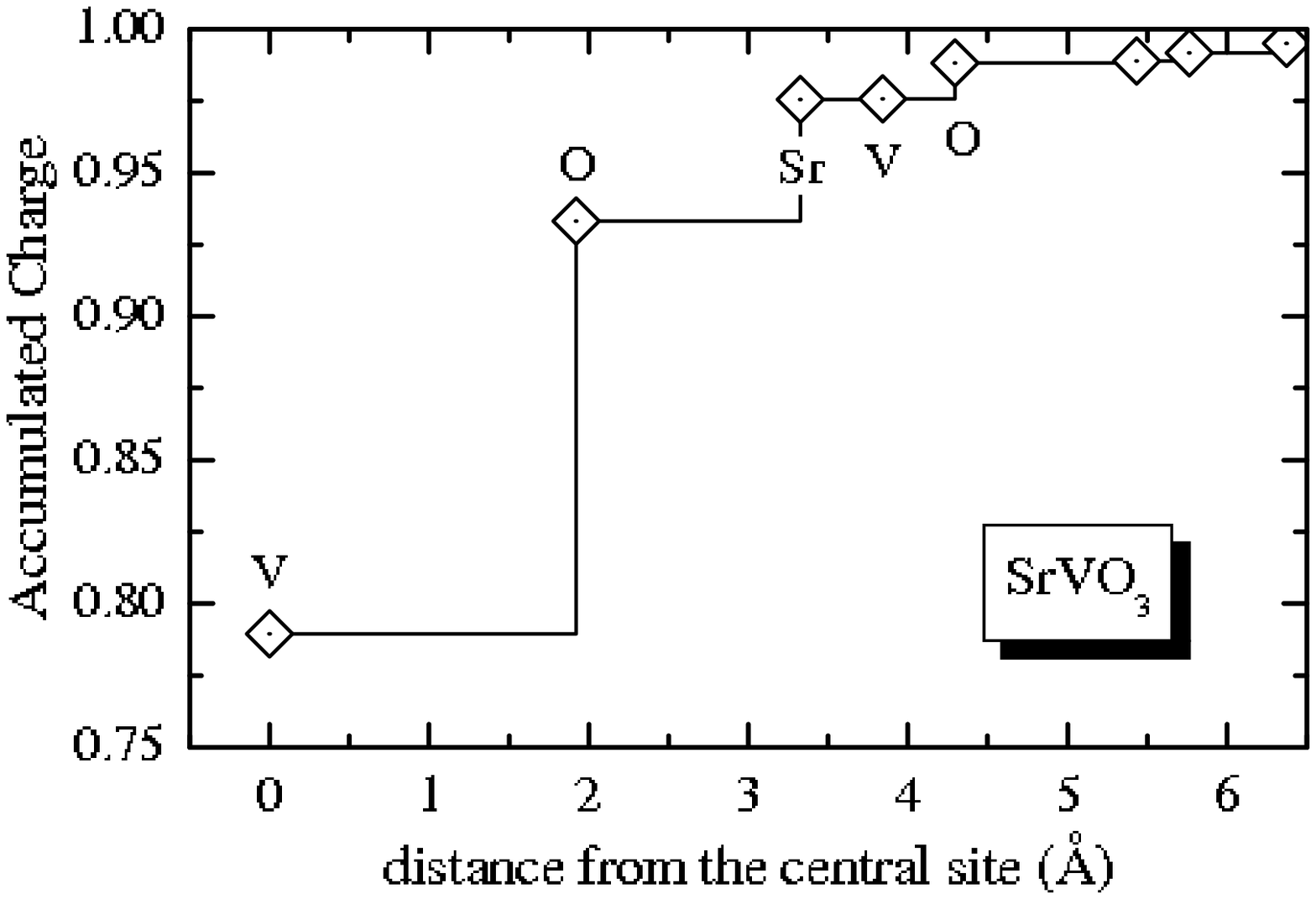}}
\end{center}
\caption{\label{fig.SrVO3WFdummy}
An example of
auxiliary Wannier function $\{ \bar{W}(\hat{c})\}$ for SrVO$_3$
for which all parameters of the kinetic energy are set to be zero.
Left panel shows the contour plot of the $xy$-orbital in the (001) plane.
Right panel shows the spacial extension of this orbital.
Other notations are the same as in Figs.~\protect\ref{fig.SrVO3WF} and
\protect\ref{fig.SrVO3WFextension}.}
\end{figure}

  A characteristic example of the auxiliary WFs
is shown in Fig.~\ref{fig.SrVO3WFdummy} for SrVO$_3$.
We note only a minor difference between auxiliary
WFs and the regular ones shown in Fig.~\ref{fig.SrVO3WF},
meaning that the main details of the WFs for the $t_{2g}$
bands are predetermined by orthogonality condition to other bands.
For example, in the case of auxiliary WFs, 79\% of the total charge
are accumulated at the central V site (instead of 77\% for the regular
WFs). The values of $\langle {\bf r}^2 \rangle$ obtained for
the auxiliary and regular WFs are 2.27 and 2.37 \AA, respectively.
Thus, the auxiliary WFs appear to be more localized. However,
the difference is small. The result is well anticipated for
strongly correlated systems, for which the kinetic-energy term
$\hat{h}$ is expected to be small.

  Since the KS Hamiltonian is diagonal in the basis of auxiliary WFs,
the latter can be regarded as eigestates of this Hamiltonian
corresponding to certain boundary condition.
The corresponding set of KS eigenvalues will be denoted as
$\{ \varepsilon_{\bf R} \}$.
Then, the occupation numbers
$\{ n_{\bf R} \}$ become well defined and one can use the standard
properties of the density-functional theory. Namely,
by using Janak's theorem for the KS eigenvalues
$$
\varepsilon_{\bf R} = \frac{\delta E}{\delta n_{\bf R}}
$$
and Slater's transition-state arguments, Eq.~(\ref{eqn:HerringU1})
can be further rearranged as
$$
U =
\varepsilon_{{\bf R}} [ n_{{\bf R}}+\frac{1}{2},n_{{\bf R}'}-\frac{1}{2} ] -
\varepsilon_{{\bf R}} [ n_{{\bf R}}-\frac{1}{2},n_{{\bf R}'}+\frac{1}{2} ].
$$
The final expression for the parameter $U$ is obtained by considering the
deviations $\pm$$1/2$ from $n_{\bf R}$ and $n_{{\bf R}'}$ as a weak perturbation
and employing the Taylor expansion. Then,
in the first order of $\pm$$1/2$ one obtains:
\begin{equation}
U = \frac{\delta \varepsilon_{{\bf R}}}{\delta n_{{\bf R}}},
\label{eqn:HerringU1TSD}
\end{equation}
where the energy derivative is calculated under the following condition:
\begin{equation}
n_{{\bf R}}+n_{{\bf R}'}={\rm const},
\label{eqn:OccupationsConstraint}
\end{equation}
which guarantees the conservation of the
total number of particles. Strictly speaking, the definition
(\ref{eqn:HerringU1TSD}) corresponds to the
\textit{infinitesimal change} of the occupation numbers
$2(d^n)$$\rightleftharpoons$$d^{n+\delta n}$$+$$d^{n-\delta n}$, which is different
from original Herring's definition (\ref{eqn:HerringU1}). However,
for practical purposes, these definitions can be regarded as equivalent as
they yield very similar values for the parameter $U$.\cite{PRB94.2}

  Then, it is convenient to use the Hellman-Feinman theorem, which allows
to relate $U$ with the change of the Hartree potential
(the change of the exchange-correlation potential in LDA is typically small
and can be neglected):\cite{PRB94.2}
$$
U = \langle \bar{W} | \frac{\delta V_{\rm H}}{\delta n_{\bf R}} | \bar{W} \rangle.
$$
Taking into account that
$V_{\rm H}({\bf r})$$=$$e^2 \int d{\bf r}' \rho({\bf r}')/|{\bf r}-{\bf r}'|$,
and using Eq.~(\ref{eqn:rho}) for the electron density,
the above expression can be rearranged as
$$
U = e^2 \int d{\bf r} \int d{\bf r}'
\frac{|\bar{W}({\bf r}')|^2}{|{\bf r}-{\bf r}'|}
\frac{\delta \rho({\bf r})}{\delta n_{\bf R}},
$$
where
\begin{equation}
\frac{\delta \rho({\bf r})}{\delta n_{\bf R}} = \sum_i
\left\{
\frac{\delta n_i}{\delta n_{\bf R}} |\psi_i({\bf r})|^2 +
n_i \frac{\delta}{\delta n_{\bf R}} |\psi_i({\bf r})|^2
\right\}.
\label{eqn:DensityChange}
\end{equation}
The last expression points out at the existence of two additive
channels of screening. \\
(i)
The first one comes from the change of occupation numbers.
Due to the constraint (\ref{eqn:OccupationsConstraint}) imposed
on the occupation numbers, this channel involves two Wannier orbitals,
centered at different TM sites, and describes the
screening of
on-site Coulomb interactions by intersite interactions.
Other states can be affected by this term only through the change
of wavefunctions in the process of iterative solution of the KS equations.
We also would like
to note that this channel of screening is absent in the GW methods, which
may lead to an error for metallic compounds.\cite{PRB05} \\
(ii)
The second channel describes the relaxation of the wavefunctions.
It affects both the auxiliary WFs and the electronic states
belonging to the rest of the spectrum.

\subsection{\label{sec:approximations}Approximations and Simplifications}

  The usual way in calculating the parameter $U$ is the c-LDA approach,
that is to solve iteratively Eqs.~(\ref{eqn:KS}) and (\ref{eqn:rho})
for a fixed set of occupation numbers $\{ n_i \}$, which does not necessarily
follow the Fermi-Dirac distribution for the ground
state.\cite{Dederichs,Norman,McMahan,Hybertsen,Gunnarsson,GunnarssonPostnikov,AnisimovGunnarsson,PRB94.2}
In practical calculations, these occupation numbers are controlled by
an external potential $\delta V_{\rm ext}({\bf r})$, playing a role of Lagrange multipliers
in the constrained density functional theory.
In spite of many limitations for the strongly correlated systems,
LDA is formulated as the ground-state theory. Therefore, there is a general belief
that it should provide a good estimate for the total energy difference
given by Eq.~(\ref{eqn:HerringU1}) and all other expressions which can be
derived from Eq.~(\ref{eqn:HerringU1}) using usual arguments of DFT.

  From the practical point of view, the basic difficulty of combining c-LDA
with the WF method is the necessity to deal with relaxation
of these WFs. This means that the auxiliary WFs should be
recalculated on each iteration, for every new value of the electron density
and the KS potential.
Taking into account an arbitrariness
with the choice of the WFs, this procedure cannot be easily
implemented in the standard
c-LDA calculations.
Instead, we will employ a hybrid method, which
starts with c-LDA and then takes into account the effects of relaxation
of the WF in an analytical form, using the well-known
expressions for the
screened Coulomb interaction in the
random-phase approximation (RPA).\cite{Springer,Kotani,Ferdi04,Hedin,FerdiGunnarsson}

  In c-LDA calculations, the change of the occupation numbers $\{ \delta n_i \}$
is associated with some change of the total potential
$\delta V$$=$$\delta V_{\rm ext}$$+$$\delta V_{\rm H}$$+$$\delta V_{\rm XC}$. Then, the
change of the electron density in Eq.~(\ref{eqn:DensityChange})
can be identically expressed in terms of the polarization function as
\begin{equation}
\delta \rho({\bf r}) = \int d {\bf r}' P({\bf r},{\bf r}') \delta V ({\bf r}').
\label{eqn:Polarization}
\end{equation}
Using Eq.~(\ref{eqn:DensityChange}), one can identify three main
contributions to the polarization function associated with the
following processes (correspondingly $P^I$, $P^{II}$, and $P^{III}$):
\begin{enumerate}
\item[I.]
the change of the occupation numbers of the auxiliary WFs;
\item[II.]
the relaxation of the auxiliary WFs;
\item[III.]
the relaxation of the rest of the electronic states.
\end{enumerate}
Then, each $\bar{W}$ can be expressed in terms of the basis
functions (or partial waves)
$\{ \phi \}$ and $\{ \dot{\phi} \}$, using Eq.~(\ref{eqn:WF2}).
Therefore, the change of $\bar{W}$ includes the relaxation of these
basis functions as well as the change of hybridization of the
TM $t_{2g}$ states with (mainly) the oxygen states.
The latter is given by the change of coefficients $\{ \Gamma_r \}$ in the right-hand side
of Eq.~(\ref{eqn:WF2}).
The corresponding contributions to the polarization function are denoted as
$P^{IIB}$ and $P^{IIH}$, which stand for the change of basis functions and hybridization,
respectively.
The same arguments are applied to relaxation of the rest of the electronic states.
The corresponding polarization function can be divided accordingly in
$P^{IIIB}$ and $P^{IIIH}$. We also introduce combined notations:
$P^B$$=$$P^{IIB}$$+$$P^{IIIB}$ and $P^H$$=$$P^{IIH}$$+$$P^{IIIH}$.

\begin{figure}[h!]
\begin{center}
\resizebox{10cm}{!}{\includegraphics{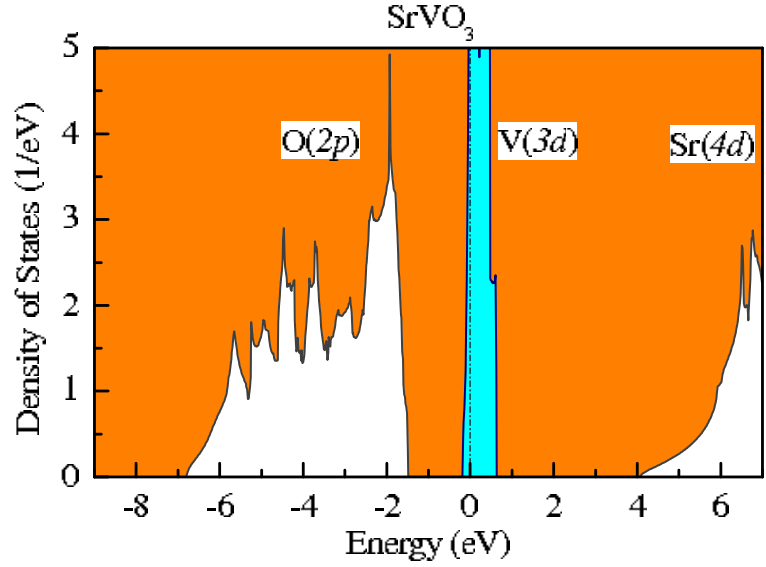}}
\end{center}
\caption{\label{fig.SrVO3CanonicalBands}
An example of model electronic structure for SrVO$_3$, which is
used in constraint-LDA calculations. The electronic structure corresponds to the
canonical-bands approximation.\protect\cite{LMTO} The Fermi level is at zero energy.
}
\end{figure}

  We would like to point out here that, conceptually, the
RPA approach for treating the relaxation effects is similar to c-LDA.
The main difference is that RPA is based on an analytical
expression for the change of the wavefunctions, formulated in terms of the
perturbation theory expansion, while c-LDA treats the same effects numerically.\cite{PRB05}
Therefore, we use a hybrid c-LDA+RPA scheme, which was originally considered in Ref.~\onlinecite{PRB05}.
It consists of two steps. \\
(i) First, we take into account the screening associated with
$P^I$, and $P^B$ in the framework of conventional c-LDA method,
and neglect all kinds of hybridization effects involving the
TM $d$-orbitals.
An example of such a model electronic structure is shown in Fig.~\ref{fig.SrVO3CanonicalBands}.
This part is totally equivalent to the method
of Gunnarsson and co-workers.\cite{Gunnarsson,GunnarssonPostnikov,AnisimovGunnarsson}
It allows to calculate the Coulomb repulsion $u$ and the
intra-atomic exchange (Hund's rule) coupling
$j$$=$$-$$2\delta^2E/\delta {\bf m}^2$ in the atomic limit (${\bf m}$ being the spin
magnetization).
By using these $u$ and $j$
one can construct the full $5$$\times$$5$$\times$$5$$\times$$5$
matrix $\hat{u}$ of Coulomb interactions between
atomic $d$ electrons, as it is typically done in the LDA$+$$U$ method.\cite{PRB94,comment.4}
This matrix will be used as the starting point in RPA calculations.

  c-LDA is supplemented with additional approximations, such as the
atomic-spheres approximation. It also disregards some hybridization effects.
However, in Sec.~\ref{sec:RPAresultsSrVO3} we will see that
at least for the static Coulomb interaction,
the
RPA results are close
to the strong-coupling regime. In such a situation, the precise value of the parameter $u$,
which is used as the starting point for these calculations,
appears to
be less
important, and
it is sufficient to have an ``order of magnitude'' estimate, which can be obtained
from c-LDA.
(ii)
We turn on the hybridization, and evaluate the screening associated with
the last portion of the polarization function, $P^H$, in RPA:
\begin{equation}
\hat{U} = \left[1-\hat{u}\hat{P}^H \right]^{-1} \hat{u},
\label{eqn:URPA}
\end{equation}
where $\hat{P}^H$ is the $5$$\times$$5$$\times$$5$$\times$$5$ matrix
$\hat{P}^H$$\equiv$$\| P^H_{\alpha \beta \gamma \delta }\|$, which will
be specified below.

Since total $P$ is an additive function of $P^I$, $P^B$, and $P^H$,
this procedure can be justified within RPA,
where each new contribution to the polarization function ($P^H$)
can be included consequently by
starting with the Coulomb interaction $\hat{u}$, which already incorporates
the effects of other terms ($P^I$ and $P^B$).\cite{Ferdi04}

\begin{figure}[h!]
\begin{center}
\resizebox{10cm}{!}{\includegraphics{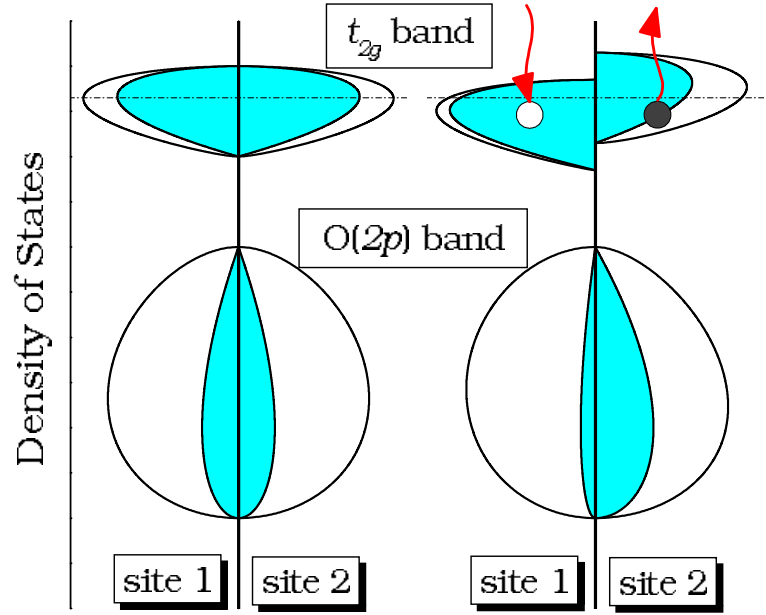}}
\end{center}
\caption{\label{fig.pdhybridization}
A schematic view on the change of the $p$-$d$ hybridization in the
O($2p$) and $t_{2g}$ bands of the TM oxides associated
with the repopulation of the Wannier orbitals at the neighboring
transition-metal sites: $2d^n$$\rightleftharpoons$$d^{n-1}$$+$$d^{n+1}$.
Left panel corresponds to the ground-state configuration ($2d^n$).
In the right panel, the removal (addition) of an electron from (to)
the Wannier orbital in the $t_{2g}$ part of the spectrum is simulated
by the shift of these orbitals relative to the Fermi level
(shown by dot-dashed line). Around
each transition-metal site, it changes the Coulomb potential,
which controls the distribution of the $d$-states
as well as the degree of their hybridization between TM($d$) and O($2p$) states.
Generally, the removal of an electron from (or the addition of an electron to)
the Wannier orbital is partially compensated by the change of the
amount of $d$-states, which is admixed
into the O($2p$) band. This transfer of the spectral weight
works as a very efficient channel of
screening of local Coulomb interactions in the transition-metal oxides.
}
\end{figure}

The physical meaning of processes associated with the change of the hybridization
and their role in the screening of local Coulomb interactions
is illustrated schematically in Fig.~\ref{fig.pdhybridization}.

\begin{figure}[h!]
\begin{center}
\resizebox{10cm}{!}{\includegraphics{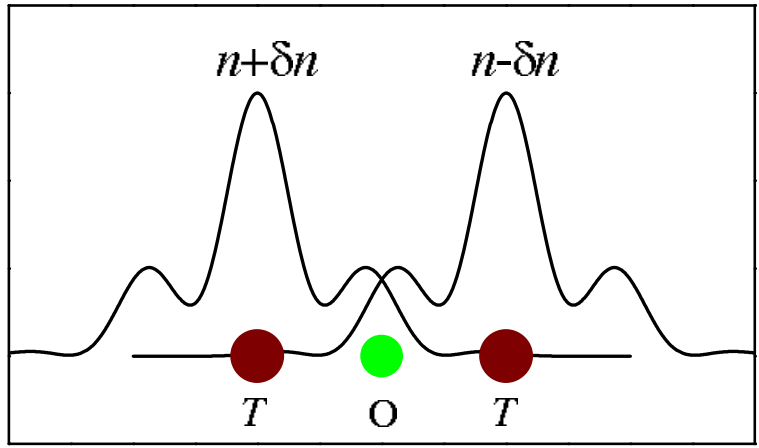}}
\end{center}
\caption{\label{fig.WOverlapCartoon}
A cartoon picture showing two Wannier orbitals centered at
neighboring transition-metal sites ($T$) and having tails
spreading to the intermediate oxygen (O) site. The charge
disproportionation associated with the reaction
$2(d^n)$$\rightleftharpoons$$d^{n+\delta n}$$+$$d^{n-\delta n}$
will affect mainly the transition-metal sites. The change of the electron
density at the intermediate oxygen site will be largely cancelled out.
}
\end{figure}

Since the creation and the annihilation of an electron in RPA are treated as two
independent processes,
the screening of Coulomb interactions
will be generally different from that associated with the true
reaction $2(d^n)$$\rightleftharpoons$$d^{n+1}$$+$$d^{n-1}$.
To some extent, the true screening can be simulated by
imposing certain constraints on the form of RPA polarization function $P^H$.
Namely, one can expect certain cancellation of contributions coming from
Wannier orbitals centered at different TM
sites at the intermediate (i.e. oxygen) sites
(see Fig.~\ref{fig.WOverlapCartoon}).
Therefore, we believe that it is more
physical to take into account only those contributions to the polarization
function which are associated with the TM sites, and to suppress
contributions associated with intermediate sites.
This makes some difference from the conventional RPA,\cite{Springer,Kotani,Ferdi04}
which constitutes the basis of the GW method.\cite{Hedin,FerdiGunnarsson}
We expect this scheme to work well for the on-site Coulomb interactions.
However, the effect of hybridization on the intersite Coulomb interactions
remain an open and so far unresolved problem.
Some estimates of these effects will be given in Sec.~\ref{sec:RPAvscLDA}, using the
c-LDA method. The calculations suggest that the intersite interactions are
screened very efficiently by the change of hybridization.
Therefore we speculate that for the considered compounds, the effective Coulomb
interactions between different TM sites are small and can be neglected.

The analytical expression for $P^H$ can be obtained from Eq.~(\ref{eqn:Polarization})
by considering the perturbation-theory expansion for the wavefunctions
with the fixed occupation numbers.\cite{PRB05}
The time-dependent perturbation theory, corresponding to the external
perturbation $\delta V_{\rm ext} e^{-i\omega t}$, yields in the
first order:
\begin{equation}
P^H_{\alpha \beta \gamma \delta}(\omega) = \sum_{ij}
\frac{(n_i-n_j)
d^\dagger_{\alpha j} d_{\beta i}
d^\dagger_{\gamma i} d_{\delta j}}
{\omega - \varepsilon_j + \varepsilon_i + i\delta (n_i-n_j)},
\label{eqn:PolarizationPT}
\end{equation}
where $d_{\gamma i}$$=$$\langle \phi_\gamma | \psi_i \rangle$ is the projection
of LDA eigenstate $\psi_i$ onto one of partial $d$-waves $\phi_\gamma$, belonging to the TM
site, and $i$ and $j$ are the joint index, incorporating spin and band indices as well as the
position of ${\bf k}$-point in the first Brillouin zone.
In this notations, the matrix multiplication in Eq.~\ref{eqn:URPA} implies the convolution over two
indices. Namely, the matrix element of the product $\hat{u}\hat{P}^H$ is given by
$(\hat{u}\hat{P}^H)_{\alpha \beta \gamma \delta}$$=$$\sum_{\mu \nu}
u_{\alpha \beta \mu \nu} P^H_{\mu \nu \gamma \delta}$.
Note that all transitions in Eq.~(\ref{eqn:PolarizationPT}) are allowed only between
occupied and empty bands.

  In the next Sections, we will discuss different contributions to the screening
of Coulomb interactions more in details.

\subsection{\label{sec:cLDAresults}Results of Constraint-LDA Calculations}

  The results of conventional constraint-LDA calculations for TM
oxides have been widely discussed in the
literature.\cite{AZA,Norman,McMahan,Gunnarsson,GunnarssonPostnikov} Here we only illustrate the main idea
and show some basic results using SrVO$_3$ as an example.

  The calculations are performed in the supercell geometry, in which the number of
atomic $3d$ electrons is modulated around the ``ground-state'' value $n$$=$$1$
according to the formula:
$$
n_{\bf R} = n + \delta n \cos ({\bf k} {\bf R}).
$$
Corresponding values of interaction parameter $u_{\bf k}$,
calculated in several different points of the Brillouin zone, are listed in
Table~\ref{tab:SrVO3Uqall}.
\begin{table}[h!]
\caption{The values of $u_{\bf k}$ (in eV) obtained in
constraint-LDA calculations for SrVO$_3$
in different points
of the Brillouin zone:
$\Gamma$$=$$(0,0,0)$,
${\rm X}$$=$$(\pi/a)(1,0,0)$,
${\rm M}$$=$$(\pi/a)(1,1,0)$,
and
${\rm R}$$=$$(\pi/a)(1,1,1)$. The right column shows the weights of these
${\bf k}$-points used in the process of integration over the
Brillouin zone.} \label{tab:SrVO3Uqall}
\begin{ruledtabular}
\begin{tabular}{ccc}
${\bf k}$  & $u_{\bf k}$    & $w_{\bf k}$    \\
\hline
 $\Gamma$  & $-$            & $0$            \\
 ${\rm X}$ & $10.6$         & $3/7$          \\
 ${\rm M}$ & $9.9$          & $3/7$          \\
 ${\rm R}$ & $9.5$          & $1/7$          \\
\end{tabular}
\end{ruledtabular}
\end{table}
Then, we map $u_{\bf k}$ onto the model
$$
u_{\bf k}=u-v \sum_{\bf R} \cos ({\bf k} {\bf R}),
$$
and extract parameters of on-site ($u$) and NN ($v$)
interactions after integration over the Brillouin zone.
We note that in the present context the $\Gamma$-point result has no
physical meaning, as it corresponds to the transfer of an electron to the same
atomic site. Therefore, we exclude it in the process of integration, and recalculate
the weights of other ${\bf k}$-points using the symmetry arguments.
The new weights are shown in Table~\ref{tab:SrVO3Uqall}.
This yields the following parameters:
$u$$=$$10.1$ eV and $v$$=$$1.2$ eV.
Similar calculations for intra-atomic exchange coupling yield $j$$=$$1.0$ eV.

  For comparison, the values of \textit{bare} Coulomb and exchange integrals,
calculated on the atomic V($3d$) wavefunctions are 21.7 and 1.2 eV, respectively.
Thus, in the c-LDA scheme
the effective Coulomb interaction is reduced by factor two.
The intra-atomic exchange interaction is
reduced by
20\%.
As we will see in the next section, the Coulomb interaction
will be further reduced by
relaxation of the WFs due
the change of hybridization.

\subsection{\label{sec:RPAresultsSrVO3}The Role of Hybridization}

  Because of
hybridization, the $d$-states of the TM sites may have a significant
weight in other bands. A typical situation
for the series of TM oxides is shown in
Fig.~\ref{fig.DOSsummary}, where besides the $t_{2g}$-band,
the $d$-states contribute
to the TM $e_g$ as well as the O($2p$) bands.
On both sides
of the reaction $2(d^n)$$\rightleftharpoons$$d^{n+1}$$+$$d^{n-1}$,
the WFs constructed for the $t_{2g}$ bands should be orthogonal to
other bands. As it was already pointed out in Sec.~\ref{sec:approximations},
this mechanism is responsible for an additional
channel of screening of the on-site Coulomb interaction associated
with the change of this hybridization. The corresponding contribution can be
evaluated
using the Dyson equation~(\ref{eqn:URPA}) and taking the matrix of
Coulomb interactions $\hat{u}$ obtained in c-LDA
as the starting interaction.
Then, the relevant expression for the polarization matrix is given by
Eq.~(\ref{eqn:PolarizationPT}).

  According to the electronic structure of the TM oxides,
one can identify three main contributions to the polarization matrix
$\hat{P}^H$, associated with the following inter-band transitions:
O($2p$)$\rightarrow$TM($e_g$), O($2p$)$\rightarrow$TM($t_{2g}$), and
TM($t_{2g}$)$\rightarrow$TM($e_g$).
Meanwhile, all transitions between $t_{2g}$ bands should be switched off,
in order to avoid the double counting of these effects in the process of solution
of the Hubbard model.\cite{Ferdi04}
As it was already pointed out in Sec.~\ref{sec:screenedU}, this procedure
is similar to the setting $\hat{h}$$=$$\hat{c}$ for the auxiliary WFs.

\begin{figure}[h!]
\begin{center}
\resizebox{14cm}{!}{\includegraphics{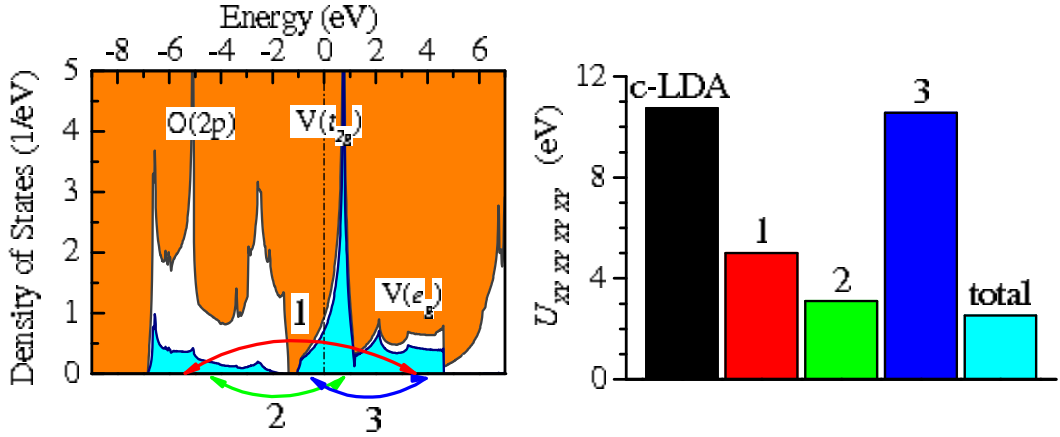}}
\end{center}
\caption{\label{fig.SrVO3DOSURPA}
Left panel shows the local density of states of SrVO$_3$ with the notation
of the main
inter-band
transitions which contribute to the polarization function in RPA:
O($2p$)$\rightarrow$V($e_g$) (1), O$(2p$)$\rightarrow$V($t_{2g}$) (2), and
V($t_{2g}$)$\rightarrow$V($e_g$) (3). Right panel shows the intra-orbital
interaction $\mathcal{U}$ in c-LDA ($\mathcal{U}$$=$$U$$+$$\frac{8}{7}J$),\protect\cite{comment.4}
the effect of screening
corresponding to each type
of transitions in the polarization function, and the final value of $\mathcal{U}$
in RPA which incorporates all three transitions.
All RPA results are for $\omega$$=$$0$.
}
\end{figure}

  Details of static screening, corresponding to
$\omega$$=$$0$, are explained in Fig.~\ref{fig.SrVO3DOSURPA} for SrVO$_3$.
It is convenient to introduce three Kanamori parameters:\cite{Kanamori}
the intra-orbital Coulomb interaction $\mathcal{U}$$=$$U_{xy~xy~xy~xy}$,
the inter-orbital interaction $\mathcal{U}'$$=$$U_{xy~xy~yz~yz}$,
and the off-diagonal (exchange-type) interaction
$\mathcal{J}$$=$$U_{xy~yz~xy~yz}$.
In addition to the total value of $U$,
we calculate intermediate
interactions corresponding to each type of transitions
in the polarization matrix (and neglecting the other two). The
screening
caused by the change of the hybridization appears to be very efficient.
So, by going from c-LDA to RPA the intra-orbital interaction $\mathcal{U}$
is reduced from 11.2 to 2.5 eV (i.e., by factor four and even more).
The main contribution to this screening comes from the
O($2p$)$\rightarrow$V($e_g$) and O$(2p$)$\rightarrow$V($t_{2g}$)
transitions in the polarization functions.
In the cubic perovskites,
the direct interaction between V($t_{2g}$) and V($e_g$) bands plays a
minor role and can be neglected.

\begin{figure}[h!]
\begin{center}
\resizebox{8cm}{!}{\includegraphics{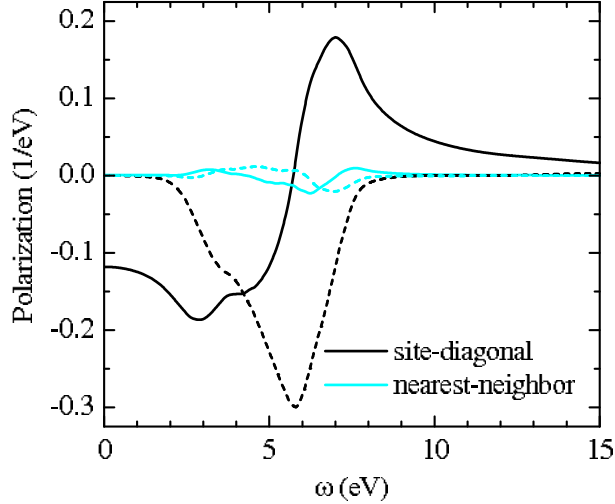}}
\end{center}
\caption{\label{fig.SrVO3P1111}
The diagonal matrix element $P^H_{xy~xy~xy~xy}$ of the polarization
function
for SrVO$_3$ in the real space: the black curve shows the site-diagonal part,
the light blue curve shows nearest-neighbor elements in the $xy$-plane.
The real and imaginary parts are shown by solid and dashed curves, respectively.
}
\end{figure}

  Matrix elements of the (total) polarization function are displayed in Fig.~\ref{fig.SrVO3P1111}.
The largest contribution comes from the site-diagonal elements
of the type
$P(\omega)$$\equiv$$P^H_{\alpha \alpha \alpha \alpha}(\omega)$$=$$
P^H_{\alpha \beta \beta \alpha}(\omega)$.
Other contributions are considerably smaller.
The static polarization $P$$\equiv$$P(0)$ is about $-$$0.12$ eV$^{-1}$.
This is the large value because the renormalization of the on-site
Coulomb interaction in the multi-orbital systems is controlled by the
parameter $MP$, rather than $P$ (see Appendix).
The prefactor $M$ stands for the total number of orbitals per one TM site
($M$$=$$3$ for $t_{2g}$ systems). Therefore,
$- uMP$ can be estimates as $3.6$, and
the situation appears to be close to the strong-coupling regime.
Then, the effective interaction
is \textit{not sensitive to the exact value of the parameter} $u$,
which is used as the starting point in RPA calculations.
For example, had we started with the bare Coulomb interaction,
which exceed
the c-LDA value by factor two and even more,
we would have obtained
$\mathcal{U}$$=$$2.7$ eV, which is close to $2.5$ eV derived by starting with
c-LDA.
This justifies some approximations discussed in Sec.~\ref{sec:approximations},
particularly the use of fast but not extremely accurate c-LDA for some
channels of screening.

\subsection{\label{sec:RPAvscLDA}RPA versus Constraint-LDA for $t_{2g}$ Electrons}

  In this section we briefly return to the problem considered in
Ref.~\onlinecite{PRB96} and reinterpret some results obtained in that
work in the light of present RPA approach.
The basic idea of Ref.~\onlinecite{PRB96} was to evaluate the effective
Coulomb interaction for the series of TM perovskite oxides in the
framework of c-LDA,
which would incorporate the screening by itinerant
TM($e_g$) electrons. Since for the considered type of screening,
RPA has many similarities with c-LDA, the present section
can be also regarded as a test for these two approaches.

\begin{figure}[h!]
\begin{center}
\resizebox{10cm}{!}{\includegraphics{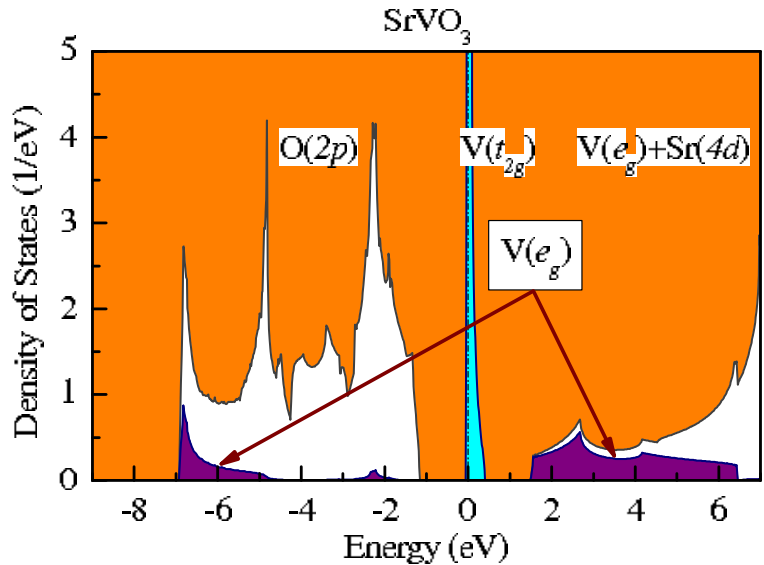}}
\end{center}
\caption{\label{fig.SrVO3Canonicalt2gBands}
Electronic structure of SrVO$_3$ in
canonical-bands approximation for the
V($t_{2g}$) states.\protect\cite{LMTO}
the contributions of V($e_g$) states are shown by arrows.
Other notations are the same as in Fig.~\protect\ref{fig.DOSsummary}.
}
\end{figure}

  However, it is important to remember that several basic assumptions
of Ref.~\onlinecite{PRB96} were different from the present work.
So, the TM($t_{2g}$) states in Ref.~\onlinecite{PRB96} were totally decoupled
from the rest of the electronic states by switching off the matrix elements
of hybridization in the LMTO method, and only the TM($e_g$) states were
allowed to hybridize.
The corresponding electronic structure
of SrVO$_3$ is shown in Fig.~\ref{fig.SrVO3Canonicalt2gBands}.
In terms of RPA polarization function, this means that the only allowed
contributions to the screening in Ref.~\onlinecite{PRB96} were due to the
O($2p$)$\rightarrow$V($e_g$) inter-band transitions
(type 1 in Fig.~\ref{fig.SrVO3DOSURPA}).

  Results of such c-LDA calculations for SrVO$_3$,
which have been
performed along the same line as in Sec.~\ref{sec:cLDAresults},
are summarized in Table~\ref{tab:SrVO3Uqt2g}.
\begin{table}[h!]
\caption{Results of constraint-LDA calculations for $U_{\bf k}$ between
$t_{2g}$ states
in SrVO$_3$ obtained after including the $e_g$ states to the screening of $t_{2g}$ interactions.
Other notations are the same as in Table~\protect\ref{tab:SrVO3Uqall}.}
\label{tab:SrVO3Uqt2g}
\begin{ruledtabular}
\begin{tabular}{ccc}
${\bf k}$  & $U_{\bf k}$    & $w_{\bf k}$    \\
\hline
 $\Gamma$  & $-$            & $0$            \\
 ${\rm X}$ & $3.9$          & $3/7$          \\
 ${\rm M}$ & $3.2$          & $3/7$          \\
 ${\rm R}$ & $2.9$          & $1/7$          \\
\end{tabular}
\end{ruledtabular}
\end{table}
After the Fourier transformation to the real space, we obtain the following parameters
of on-site and NN interactions:
$U$$=$$3.4$ eV and $V$$=$$0.3$ eV.
This value of $U$ appears to be in a reasonable agreement with the final $U$$=$$2.5$ eV, extracted
from RPA (Fig.~\ref{fig.SrVO3DOSURPA}).
However, c-LDA employs an
additional
atomic-spheres approximation. Therefore, for a proper comparison with RPA,
one should use the same level of approximation and suppress all nonspherical
interactions in the matrix of Coulomb interactions $\hat{u}$, which is used as the
starting point in RPA.
In this approximation,
and considering only the
O($2p$)$\rightarrow$V($e_g$) transitions in the polarization function, we obtain
$U$$=$$3.6$ eV, which is close to the c-LDA value obtained using the method
proposed in Ref.~\onlinecite{PRB96}.
The small difference is caused
by different approximations used for treating the intersite Coulomb interactions, which
were neglected
in RPA and taken into account in c-LDA.
For comparison, the total value of $U$ obtained in RPA after neglecting the
nonsphericity effects is only $1.6$ eV.

  In summarizing this section, there is a reasonable agreement between
results of RPA calculations and the c-LDA approach proposed in Ref.~\onlinecite{PRB96}.
However, the agreement is somewhat fortuitous because this c-LDA takes into account
only one part of the total screening, corresponding to the
O($2p$)$\rightarrow$TM($e_g$) transitions in the polarization function.
The error caused by this approximation is partially compensated by the
atomic-spheres approximation supplementing the c-LDA scheme.

  The c-LDA calculations give some idea about the effect of
hybridization on the screening of intersite Coulomb interactions. The screening appears
to be very efficient. So, by taking into account only the
O($2p$)$\rightarrow$V($e_g$) transitions,
the NN interaction is reduced from 1.2 to 0.3 eV. We expect this value
to be further reduced by including other types of transitions in
the polarization function.
Thus, for the considered compounds, the effective Coulomb interaction
between different TM sites
seems to be small and can be neglected.

\subsection{\label{sec:Kanamori}Doping-Dependence and Kanamori Rules for Cubic Compounds}

\begin{figure}[h!]
\begin{center}
\resizebox{10cm}{!}{\includegraphics{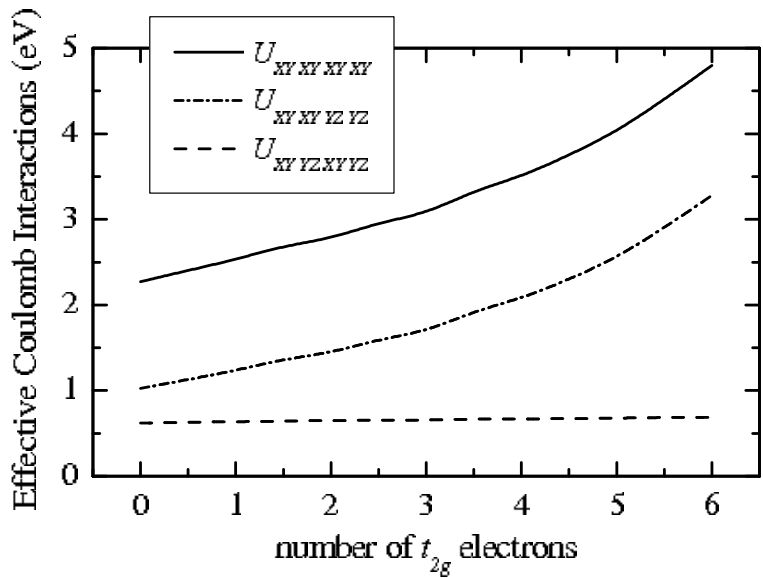}}
\end{center}
\caption{\label{fig.SrVO3RPADoping}
Doping-dependence of the effective Coulomb interactions in SrVO$_3$.
}
\end{figure}

  In this section we discuss the effects of electron/hole doping on the
static Coulomb interactions in SrVO$_3$ using the rigid-band
approximation. Results of such c-LDA+RPA calculations are shown in Fig.~\ref{fig.SrVO3RPADoping}
versus the total number of electrons in the TM($t_{2g}$) band, $n_{t_{2g}}$.
We monitor the behavior of three Kanamori parameters:\cite{Kanamori}
$\mathcal{U}$,
$\mathcal{U}'$,
and
$\mathcal{J}$.

  The Coulomb interactions reveal a monotonic behavior as the function
of doping. The screening is the most efficient when the $t_{2g}$ band is empty
($n_{t_{2g}}$$=$$0$). The situation corresponds to SrTiO$_3$. In this case all
O($2p$)$\rightarrow$TM($t_{2g}$) transitions contribute to the screening in RPA
(see Fig.~\ref{fig.SrVO3DOSURPA}).
This channel of screening vanishes when the $t_{2g}$ band becomes occupied ($n_{t_{2g}}$$=$$6$).
Then, the only possible screening is associated with the O($2p$)$\rightarrow$TM($e_g$) transitions
in the polarization function, and the effective Coulomb interaction becomes large.

  The screening of off-diagonal matrix element $\mathcal{J}$ practically does not depend on
doping.
Therefore, the well know Kanamori rule, $\mathcal{U}$$=$$\mathcal{U}'$$+$$2 \mathcal{J}$,
which was originally established for atoms, works well in the cubic compounds, even after
the screening of $t_{2g}$ interactions by other electrons.

   The present result also supports an old empirical rule suggesting that only
the Coulomb integral $U$ is sensitive to the crystal environment in solids.
The nonspherical
interactions, which are responsible for Hund's first and second rules,
appears to be much closer to their atomic values.\cite{NormanBrooks,MarelSawatzky}

\subsection{\label{sec:Womega}Frequency-Dependence}

\begin{figure}[h!]
\begin{center}
\resizebox{8cm}{!}{\includegraphics{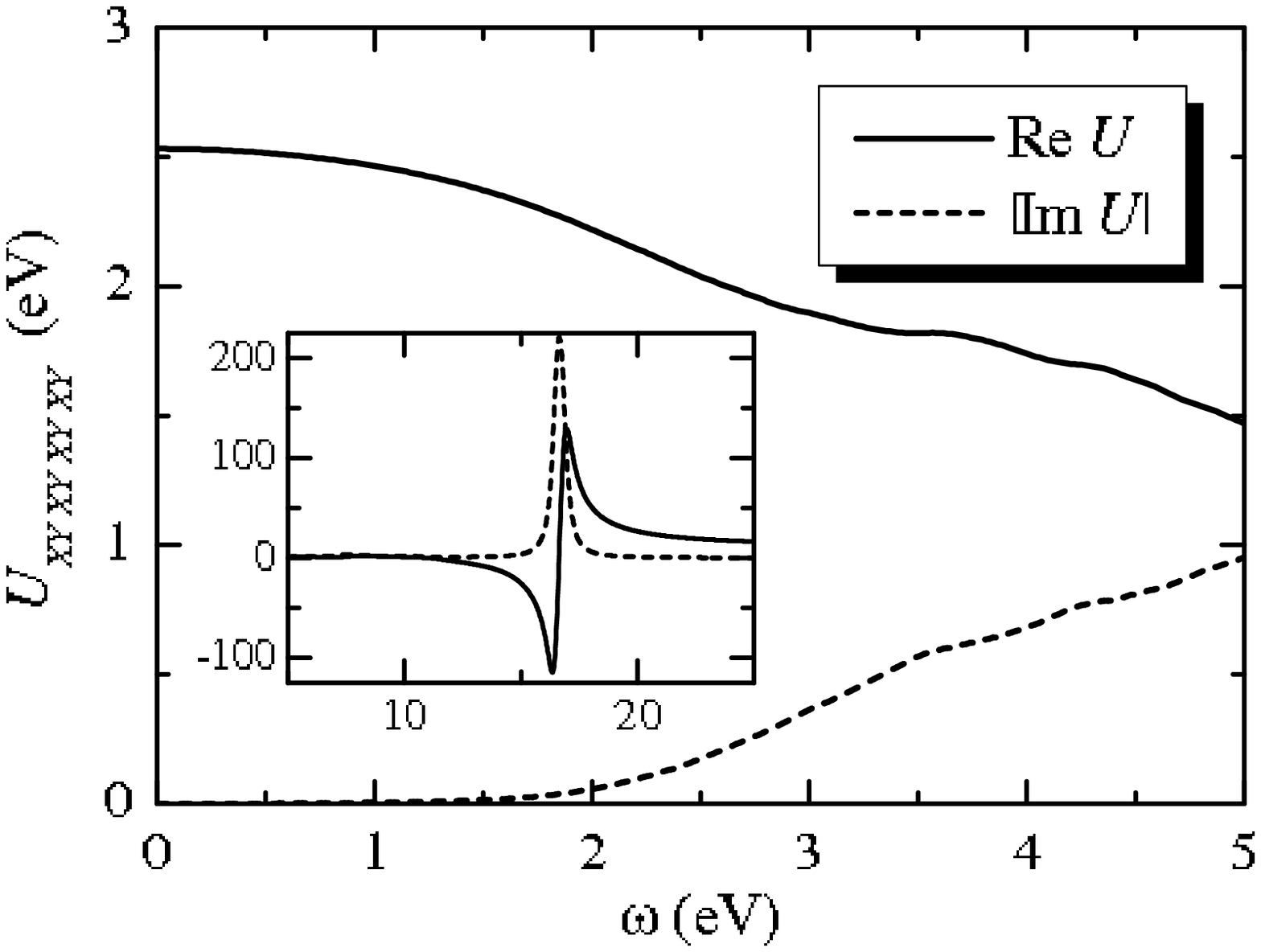}}
\resizebox{8cm}{!}{\includegraphics{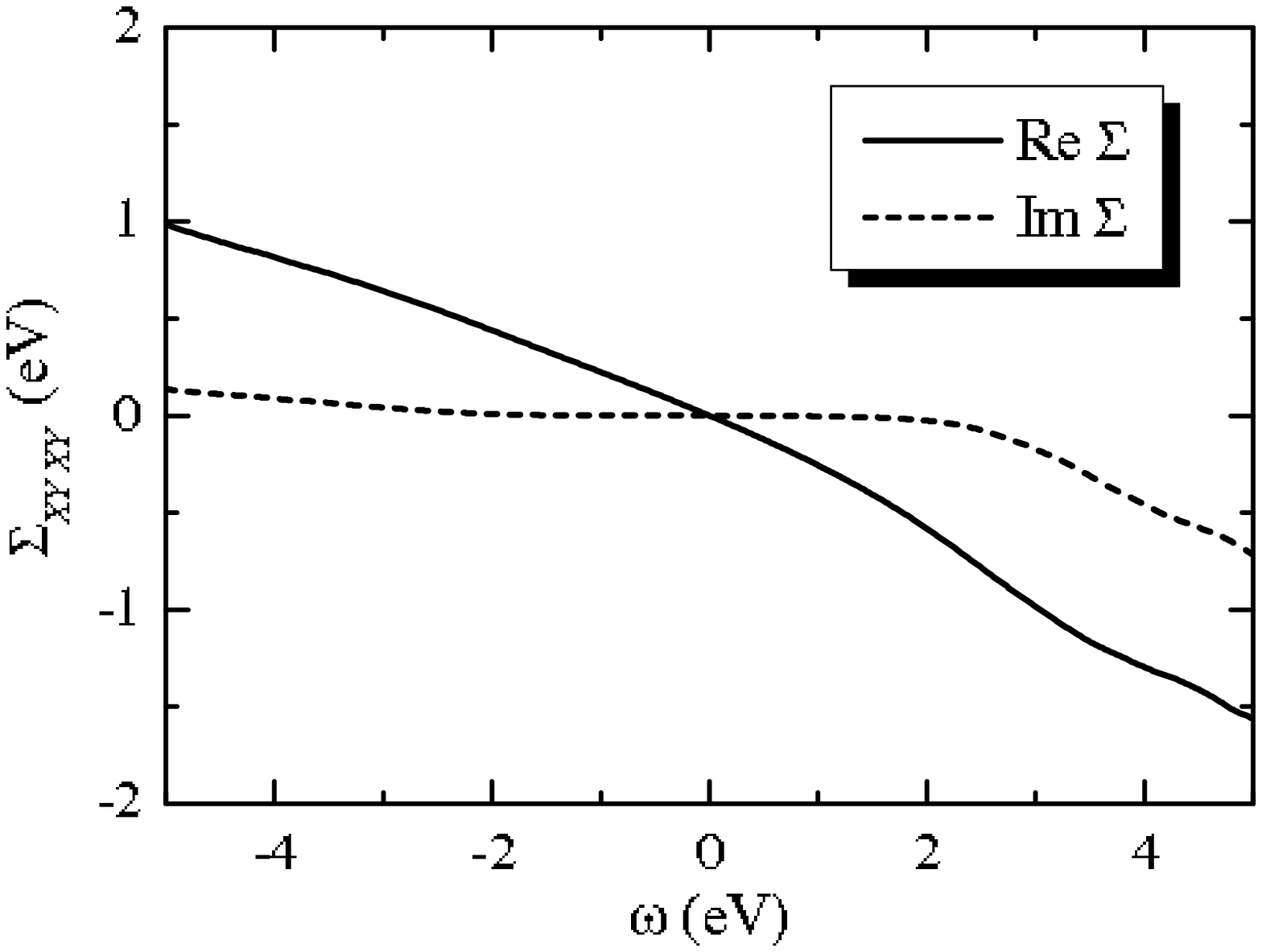}}
\end{center}
\caption{\label{fig.SrVO3WSomega}
Frequency dependence of diagonal matrix elements of screened Coulomb interaction (left) and self-energy
obtained in the GW scheme for SrVO$_3$ The inset shows the high-frequency part of $\mathcal{U}$.
}
\end{figure}

  We have shown that the change of hybridization
plays a very important role and
strongly reduces the static value of $U$. However, this is only one part of the story
because the same effect
implies the strong frequency-dependence of the effective interaction,
as it
immediately follows from the Kramers-Kronig transformation in RPA:\cite{FerdiGunnarsson}
\begin{equation}
{\rm Re}\hat{U}(\omega) = \hat{u} - \frac{2}{\pi}
{\cal P} \int_0^\infty d \omega' \frac{\omega' | {\rm Im} \hat{U}(\omega') |}
{\omega^2-\omega'^2}.
\label{eqn:KramersKronig}
\end{equation}
Indeed, the difference $[\hat{u}$$-$${\rm Re}\hat{U}(\omega)]$ at $\omega$$=$$0$,
which for the diagonal matrix elements of SrVO$_3$ is about 8.7 eV,
should be related with
the existence of the finite spectral weight of $|{\rm Im} \hat{U}(\omega)|$ at finite $\omega$.
These dependencies are shown in Fig.~\ref{fig.SrVO3WSomega} for SrVO$_3$.

  The high-frequency part of $\hat{U}$ can also contribute to the
low-energy part of the
spectrum through the self-energy effect.\cite{Ferdi04}
The latter can be evaluated in the GW approximation,\cite{Hedin,FerdiGunnarsson}
where
the self-energy is given by the convolution of $\hat{U}(\omega)$ with the
one-particle Green function $\hat{G}(\omega)$ for the $t_{2g}$ band:
$$
\hat{\Sigma}(\omega)=\frac{i}{2 \pi} \int d\omega'
\hat{G}(\omega + \omega') \hat{U}(\omega').
$$
For SrVO$_3$,
the diagonal matrix element of
$\hat{\Sigma}(\omega)$ is also shown in Fig.~\ref{fig.SrVO3WSomega}.
The low-frequency part of ${\rm Im} \hat{\Sigma}$ is small and can be neglected,
while ${\rm Re} \hat{\Sigma}$ mainly contributes to the renormalization factor:
$$
Z_{\alpha \beta}=
\left[ \left . 1 - \partial {\rm Re} \Sigma_{\alpha \beta} / \partial \omega \right|_{\omega=0} \right]^{-1}.
$$
The latter is estimated as $0.8$, for the diagonal matrix elements.

\subsection{\label{sec:Wlattice}Lattice Distortion, Formal Valency and Screening}

  In Sec.~\ref{sec:tYTiO3} we already pointed out that the degree of hybridization
between atomic TM($t_{2g}$) and O($2p$) states can differ substantially for
different TM oxides.
A typical example is two isoelectronic perovskites: SrVO$_3$ and YTiO$_3$.
The TM($t_{2g}$)-O($2p$) hybridization is stronger in SrVO$_3$, because of
two reasons: \\
(i)
A direct proximity of O($2p$) and V($t_{2g}$) bands in SrVO$_3$, which is expected for
tetra-valent compounds; \\
(ii)
A strong orthorhombic distortion observed in YTiO$_3$, which generally deteriorates
the Ti($t_{2g}$)-O($2p$) hybridization.

\begin{figure}[h!]
\begin{center}
\resizebox{8cm}{!}{\includegraphics{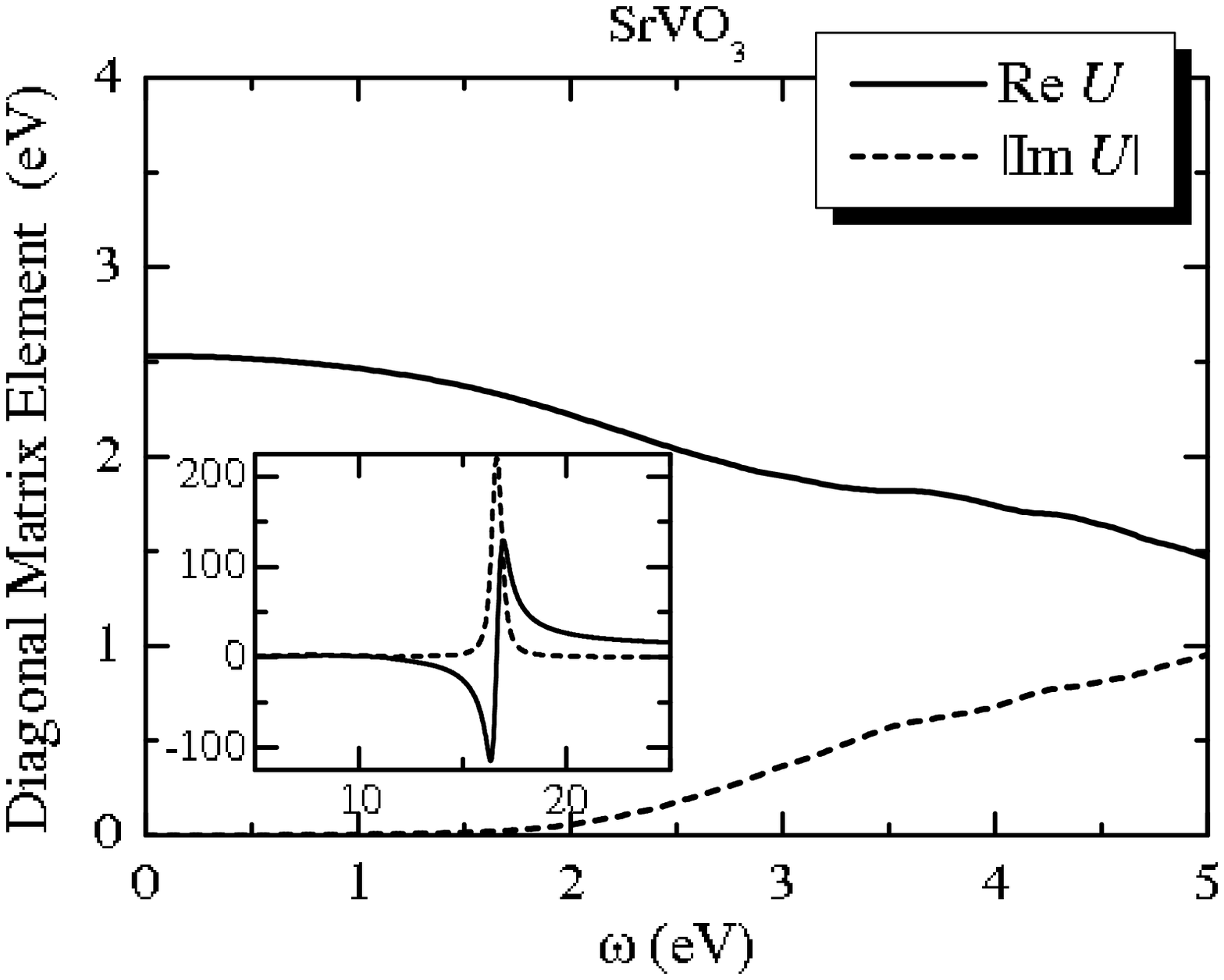}}
\resizebox{8cm}{!}{\includegraphics{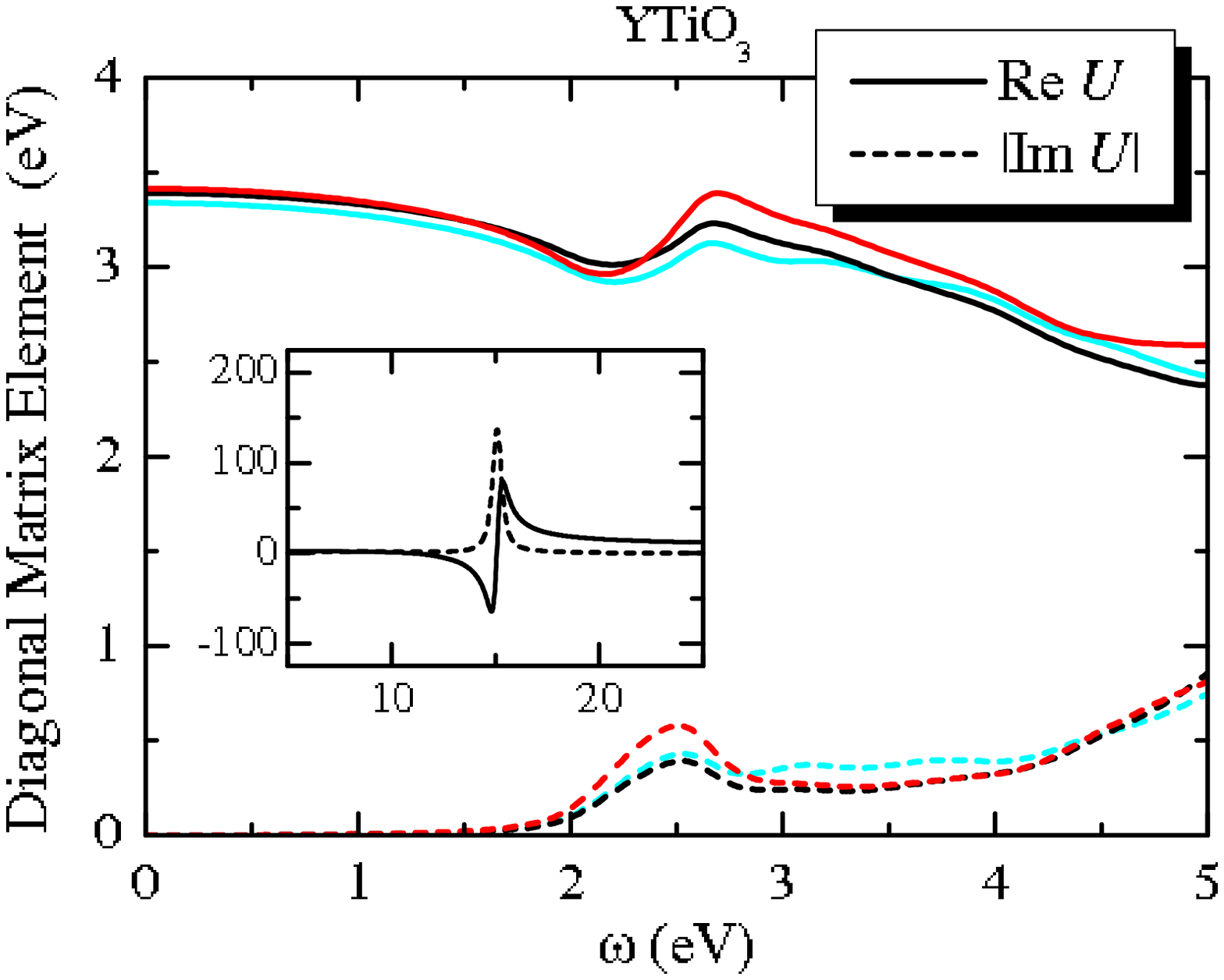}}
\end{center}
\caption{\label{fig.Wdistortion}
Diagonal matrix elements of screened Coulomb interactions for SrVO$_3$ (left)
and YTiO$_3$ (right).
Three different lines in the case of YTiO$_3$ show the behavior of three
diagonal matrix elements in the local coordinate frame
given by Eq.~(\protect\ref{eqn:YTiO3localframe}).
The insets show the high-frequency part of
$\hat{U}$.
The lines corresponding to different $t_{2g}$ orbitals in the case of YTiO$_3$
are indistinguishably close.
}
\end{figure}

  Therefore, it is reasonable to expect a very different screening of on-site Coulomb
interactions in these two compounds. This idea is nicely supported by results of
RPA calculations shown in Fig.~\ref{fig.Wdistortion}.
The static $U$ is larger YTiO$_3$. For example, the diagonal matrix element
of $\hat{U}$
is about 3.4 eV,
against 2.5 eV in SrVO$_3$. This is despite the fact that c-LDA has an
opposite tendency.
Generally, the
value of $u$ in c-LDA is expected to be larger for SrVO$_3$
rather than for YTiO$_3$, due to
different shape of the atomic $3d$-wavefunctions in the
three- and tetra-valent compounds.\cite{PRB94.2}
So, c-LDA yields $u$$=$ 8.9 and 10.1 eV, correspondingly for YTiO$_3$ and SrVO$_3$.
These parameters have been used as the starting point in RPA, which
results in an opposite trend for the static $U$.

  Therefore, the RPA screening is more efficient in SrVO$_3$, which is consistent with
stronger TM($t_{2g}$)-O($2p$) hybridization in this compound.
On the other hand, the frequency-dependence of $\hat{U}$ is weaker
in YTiO$_3$, as it immediately follows from the Kramers-Kronig
transformation (\ref{eqn:KramersKronig}).

  Finally, because of different hybridization of the $t_{2g}$ orbitals in YTiO$_3$,
the diagonal matrix elements of $\hat{U}$ are also different (see Fig.~\ref{fig.Wdistortion}).
In this case, there is some deviation from the Kanamori rules.
The difference is small (about 0.07 eV for diagonal matrix elements
of $\hat{U}$ at $\omega$$=$$0$).
However, it may play some role in more delicate applications, such as
the orbital magnetism in solids, for example.\cite{NormanBrooks}

\subsection{\label{sec:WV2O3}Two Models for V$_2$O$_3$}

  In Sec.~\ref{sec:tV2O3} we introduced two possible models for the kinetic-energy part
of V$_2$O$_3$: the ``five-orbital model''
treats all V($3d$) bands on an equal footing, while the ``three-orbital model'' is limited
by twelve V($t_{2g}$) bands, close to the Fermi level. We argued that parameters of the kinetic energy can be
different
for these two models. The same is true for the effective Coulomb interactions.
The RPA provides a very transparent explanation for this difference, which is
based on the following arguments. Recall that RPA incorporates the screening of
on-site Coulomb
interactions
caused by relaxation of the wavefunctions.
This relaxation is treated analytically, using regular perturbation theory
for the wavefunctions,
which results in Eq.~(\ref{eqn:PolarizationPT}) for the polarization function.\cite{PRB05}

  Since the V($e_g$) band is eliminated
in the three-orbital model, the effective $\hat{U}$ should include the screening
caused by
that change of the wavefunctions, which
is formulated in terms of transitions between V($t_{2g}$) and V($e_g$) bands
in the perturbation-theory expansion.
In the five-orbital model, the V($e_g$) band is included explicitly.
Therefore, the relaxation caused by possible interactions between
V($t_{2g}$) and V($e_g$) bands will be automatically taken into
account in the process of
solution of the five-orbital model, and we should get
rid of this parasitic screening at the stage the construction of the
model Hamiltonian.
Thus, the tree-orbital model will include an additional screening of
on-site interactions, caused by the V($t_{2g}$)$\rightarrow$V($e_g$)
transitions in the polarization function, which does not appear in the
five-orbital model.

\begin{figure}[h!]
\begin{center}
\resizebox{8cm}{!}{\includegraphics{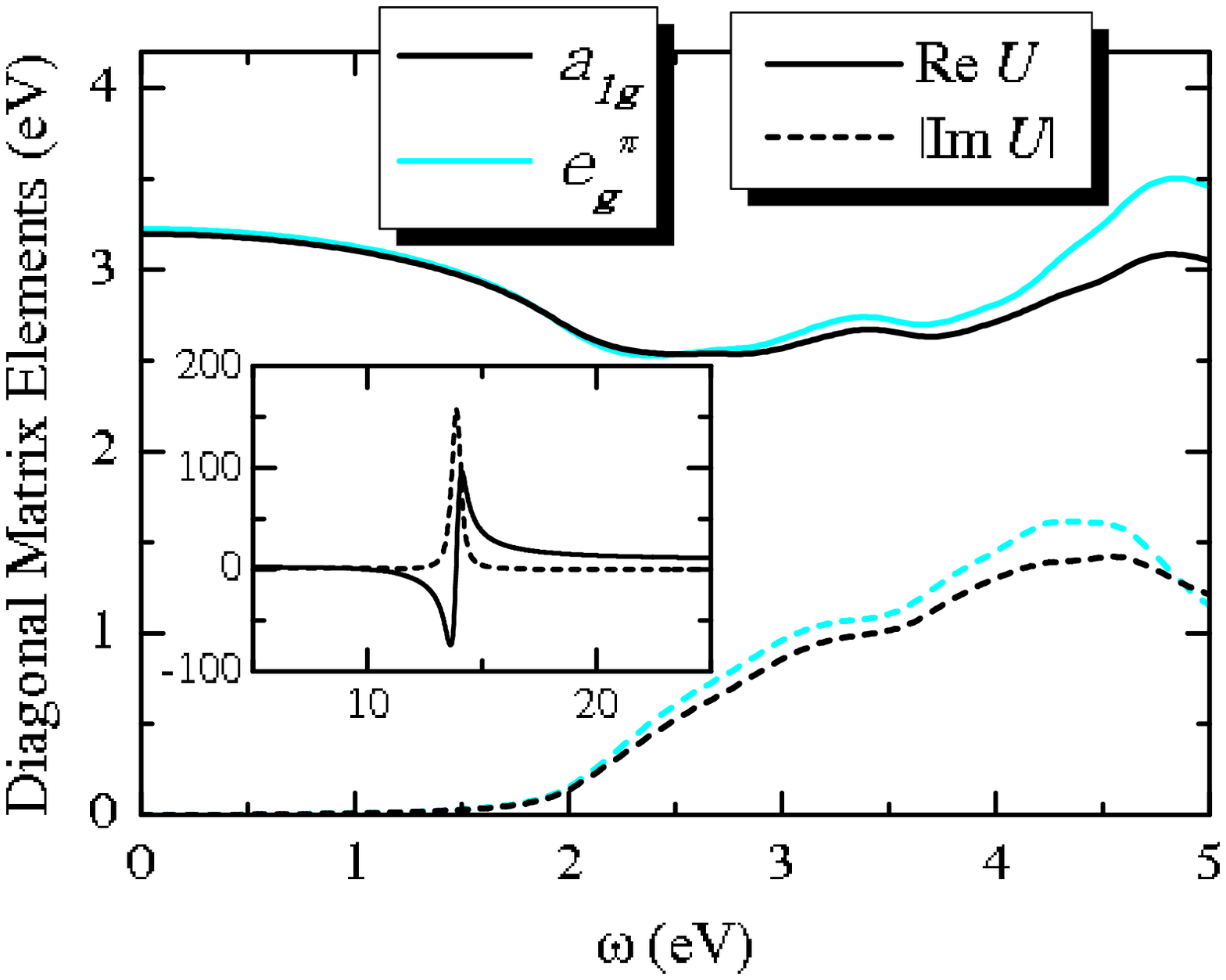}}
\resizebox{8cm}{!}{\includegraphics{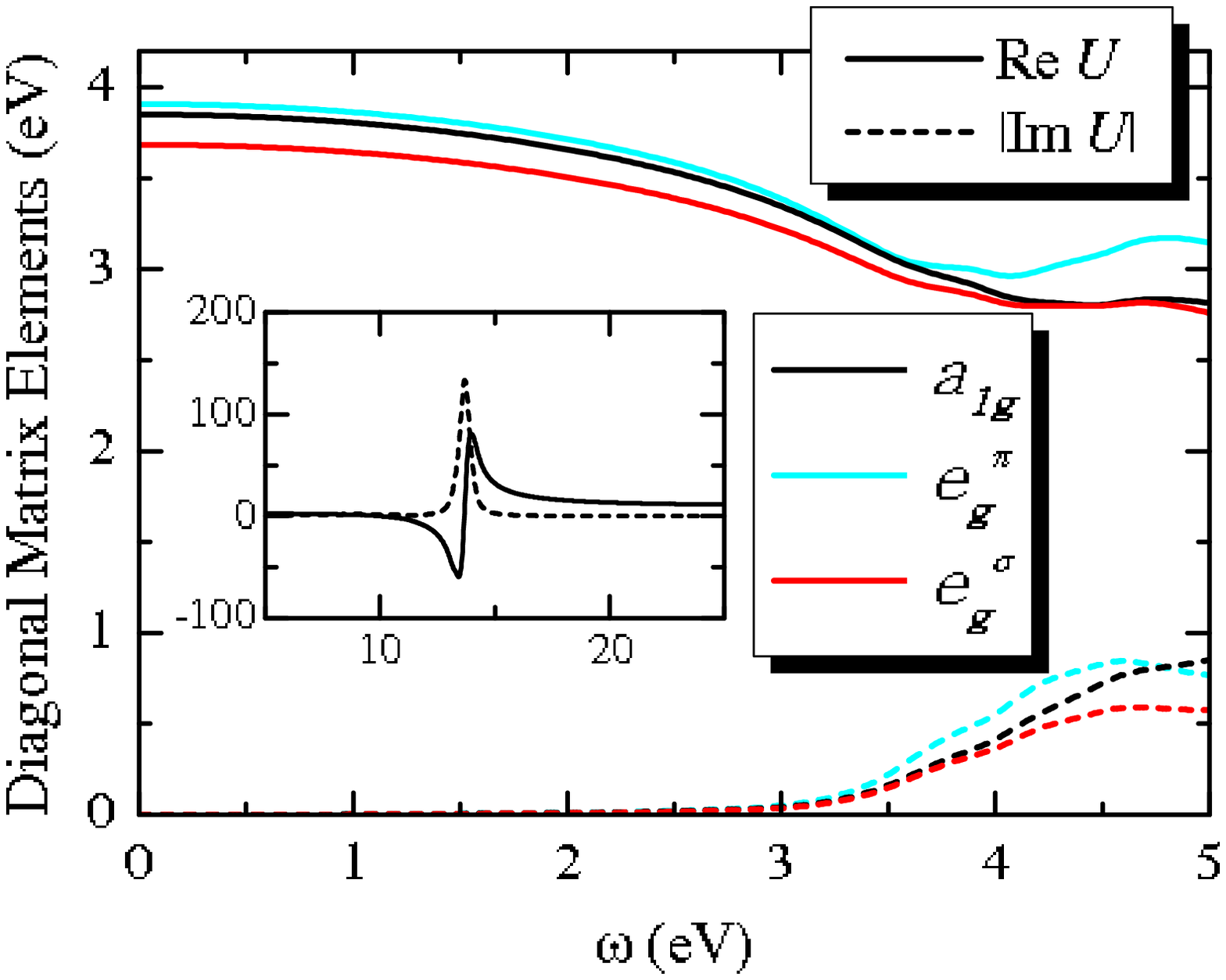}}
\end{center}
\caption{\label{fig.WV2O3}
Diagonal matrix elements of screened Coulomb interactions obtained in the three- (left)
and five-(right) orbitals models for V$_2$O$_3$.
The notations of orbitals are given in Sec.~\protect\ref{sec:tV2O3}.
Other notations are the same as in Fig.~\protect\ref{fig.Wdistortion}.
}
\end{figure}

  This screening appears to be rather efficient (unlike in cubic
perovskites considered in Sec~\ref{sec:RPAresultsSrVO3}).
So, the diagonal matrix element of static $t_{2g}$ interactions
is about 3.2 and 3.9 eV,
for the three- and five-orbital model, respectively (Fig.~\ref{fig.WV2O3}).
On the other hand, according to the
Kramers-Kronig
transformation (\ref{eqn:KramersKronig}),
the frequency-dependence of the effective interaction is
more important in the three-orbital model.
Because of different hybridization of $t_{2g}$ and $e_g^\sigma$
states in the five-orbital model,
the $e_g^\sigma$ interactions are screened more efficiently (the static
interaction between $e_g^\sigma$ orbitals is about 3.7 eV).

\begin{center}
\section{\label{sec:Summary}Summary, open questions, and comparison with other methods}
\end{center}

  The ultimate goal of this work was to make a bridge between
first-principle electronic structure calculations and
the universe of Hubbard parameters for strongly-correlated systems.
We have presented a comprehensive analysis of the problem, by starting with
the brief description of the ASA-LMTO method for electronic structure calculations
and ending up with realistic parameters of the kinetic-energy
and the Coulomb interactions for the series
of TM oxides obtained on the basis of this LMTO method.
A particular attention has been paid to the analysis of
microscopic processes responsible for the
screening of on-site Coulomb interactions in oxide compounds.

  Our strategy consists of three steps:\\
(i)
Derivation of the kinetic-energy part of the Hubbard model
from the single-particle electronic structure in LDA,
using the downfolding method.
We have also considered corrections to the crystal-field splitting
caused by nonsphericity of electron-ion interactions, beyond the
conventional atomic-spheres-approximation. \\
(ii)
Construction of the Wannier functions using results of the downfolding method.
At this stage we closely follow the idea of LMTO method, and construct the
WFs as the LMTO basis functions, which after applying to the
Kohn-Sham Hamiltonian in the real space
generate the matrix elements of the kinetic energy
obtained in the downfolding method. \\
(iii)
Calculation of screened Coulomb interactions using the concept of
\textit{auxiliary} WFs.
The latter are defined as the Wannier orbitals
for which the kinetic-energy part is set to be zero.
This construction allows to avoid the double counting of the
kinetic-energy term, which is included explicitly in the Hubbard
model.
The screened Coulomb interactions are calculated on the basis of
a hybrid approach, combining the conventional constraint-LDA with
the random-phase approximation for treating the hybridization
effects between atomic TM($3d$) and O($2p$) orbitals.
The latter play a very important role and
yields a strong renormalization of the
effective Coulomb interaction
for
isolated $t_{2g}$ band.
It also explains a strong material-dependence of this interaction, which is sensitive
to
the crystal environment in solids, the number of
$t_{2g}$ electrons, the valent state of the TM ions, etc.

  Taking into account a wide interest to the construction
of effective lattice fermion models from the first principles, we would like to make a
brief comparison with other works on a similar subject.

  Majority of methods start with the construction of the WFs, which
are then used as the basis for calculations of the parameters of the kinetic
energy and the Coulomb interactions. This is different from our approach, where we
start with the kinetic-energy part, and only after that construct the WFs
\textit{for a given set of parameters of the kinetic energy}.
We believe that such an order is extremely important, as it allows us to
control the contributions of the kinetic energy to the WFs and
the Coulomb interactions.

  Among recent works, a considerable attention is paid to the method of
Marzari and Vanderbilt, because it allows to control the spacial extension of
the WFs. Very recently, Schnell~\textit{et~al.} applied this method
to calculations of the parameters of the Hubbard Hamiltonian for the series
of $3d$ transition metals.\cite{Schnell}
In each ${\bf k}$-point of the Brillouin zone,
they constructed the WFs from all 16 bands
of the LDA Hamiltonian, corresponding to the
$4s4p3d4f$ LMTO basis. Therefore, the total number of Wannier orbitals
was also 16.
The coefficients of this expansion
has been chosen so to minimize
the square of the
position operator, $\langle {\bf r}^2 \rangle$$=$$\langle W | {\bf r}^2 | W \rangle$.
The obtained WFs were indeed well localized, and the parameter
$U$ estimated for the $3d$ bands was very close to the atomic value (about 25 eV).
However, the corresponding Wannier basis set is too large, that does
not make a big difference from the original LMTO basis set, for which one can also introduce
a localized (tight-binding) representation.\cite{LMTO}
From the view point of numerical solution of the Hubbard model
using modern many-body techniques,
it is still hardly feasible to work in the basis of 16 Wannier orbitals per one TM site,
while the simplest Hartree-Fock approximation is definitely not sufficient for
the transition metals.\cite{Schnell}
It would be interesting to see how this method will work for the TM oxides,
considered in the present work,
where the physical basis set is limited by three Wannier orbitals per
one TM site.
For example, is it possible to construct the localized Wannier orbitals
for \textit{isolated $t_{2g}$ bands} in the TM oxides, which would be as good as
the Wannier orbitals derived for \textit{all bands}?
The problem is that when the number of bands decreases,
the number of variational parameters for the
optimization of the Wannier orbitals will also decreases. Therefore,
the Wannier orbitals will generally become less localized, as it was demonstrated in our
work for V$_2$O$_3$.
Also, when the problem is formulated in a reduced Hilbert space of bands closest to
the Fermi level, it is very important to consider the screening of Coulomb interactions
by other bands, which comes from relaxation of the wavefunctions.
These effects are beyond the scopes of the work of Schnell~\textit{et~al.}, who only
considered the bare Coulomb interactions.

  In order to construct the WFs, Ku~\textit{et~al.} employed the
projector-operator scheme,\cite{WeiKu} which is basically the initial step
of the method of Marzari and Vanderbilt, prior the optimization.\cite{MarzariVanderbilt}
In this scheme, each Wannier orbital is generated by
projecting
a trial wavefunction, $| g \rangle$, onto a chosen subset of bands (in our case,
$t_{2g}$ bands):
$$
|W \rangle = \sum_{i \in t_{2g}} | \psi_i \rangle \langle \psi_i | g \rangle.
$$
Since such orbitals are not orthonormal, the procedure is followed by the
numerical orthonormalization, similar to the one described in Sec.~\ref{sec:LMTO}.
These WFs have been used as the basis for the construction of the low-energy
Hamiltonian for the series of cuprates, like La$_4$Ba$_2$Cu$_2$O$_{10}$, where
the Wannier basis consisted of a single orbital centered around each Cu site.
A weak point of this approach is that
it is difficult to assess
the spacial extension of the WF, which crucially
depends on the trial wavefunction,
and can be affected by the orthonormalization.
In some sense, the result strongly depends
on authors' intuition on how they choose the trial wavefunction.
For example, the \textit{bare} Coulomb interaction obtained in
Ref.~\onlinecite{WeiKu}
in the basis of their WFs was only 7.5 eV, which is much smaller
than the atomic Coulomb integral for the Cu($3d$) orbitals.
This means that the WFs are not well localized.
It is not clear at present, whether this is a result of the bad choice
of the trial wavefunction, or there is a more fundamental problem
related with the fact that
a more compact representation for the WFs simply
may not exist in this case.
Note, that apart from the Berry phase, there is no further parameters
available for the optimization of the WFs in the
single-orbital case.

  The delocalization of the WFs gives rise to
appreciable
direct exchange interactions operating between different Cu sites.\cite{WeiKu}
This result, however, rises additional questions. Note, that the kinetic part
of the Hubbard model and the WFs are evaluated in LDA,
where the exchange-correlation potential is set to be \textit{local}.
In principle, non-local effects can be already incorporated in LDA,
through the renormalization of parameters of the kinetic energy
and the local interactions.\cite{PRB99}
Therefore, it is not clear whether the nonlocal exchange interactions should be
regarded as independent parameters of the Hubbard Hamiltonian or not.

  Finally, Ku~\textit{et~al.} calculated only \textit{bare} Coulomb interactions.
They did not consider the screening of these
interactions caused by relaxation effects,
which are extremely important.

  Anisimov~\textit{et.~al.} employed a similar approach
for the analysis of spectroscopic properties
of TM oxides.\cite{Anisimov2005}
They extracted only the kinetic-energy part of the Hubbard Hamiltonian,
using the WFs constructed in the LMTO basis,
and treated the Coulomb interaction $U$ as a parameter.

  A completely different strategy has been proposed by Andersen~\textit{et.~al.},
on the basis of their order-$N$ muffin-tin orbital (NMTO) method, which is an
extension of the LMTO method.\cite{AndersenDasgupta}
From the very beginning, they construct the NMTO basis functions
in certain energy interval as the WFs of the original KS Hamiltonian.
Pavarini~\textit{et al.} applied this method to the series of $d^1$ perovskites.\cite{Pavarini}
At present, it is not clear how the nonuniqueness of the WFs is
reflected in the construction of the NMTO basis set. Obviously, such a
basis set is also not unique, and there is some freedom left
for the localization of the
Wannier orbitals, which does not seem to be well controlled.
Generally, our transfer integrals for the $d^1$ perovskites seems to be
more localized and our crystal-field splitting is smaller.
For example, had we relaxed the constraint condition for the construction of the ``heads''
of the WFs, based on the diagonalization of the density matrix
(\ref{eqn:DensityMatrix}), our conclusion would have been also different:
the crystal-field splitting would increase, but the transfer integrals would become
less localized.

  Pavarini~\textit{et al.} did not calculate the Coulomb
interactions. Instead, they used $U$$\sim$$5$ eV as a parameter, with the
reference to the photoemission data.\cite{MizokawaFujimori}
However, the photoemission data are typically interpreted in the cluster
model, which treats the O($2p$) band \textit{explicitly}. For the isolated
$t_{2g}$ band, the effective interaction should include an additional
renormalization coming from the relaxation of the O($2p$) band, which is
\textit{eliminated} in the $t_{2g}$-model.
Therefore, the value of the effective $U$ should be smaller.

  Finally, due to unknown for us reason, there is a substantial difference
of the parameters of $t_{2g}$ bandwidth ($W_{t_2g}$) between our work and
Ref.~\onlinecite{Pavarini}, even for cubic SrVO$_3$.
The parameters reported by Pavarini~\textit{et al.}
are generally overestimated by about 30\%.\cite{comment.5}
Our conclusions about magnetic properties
of YTiO$_3$ and LaTiO$_3$ are also different.\cite{PRB04}
Details will be presented in a separate
paper.

\begin{acknowledgments}
I thank Professor Masatoshi Imada for valuable suggestions,
especially for drawing my attention to the problem of Wannier functions,
nonsphericity of electron-ion interactions and its effect on the crystal-field
splitting,\cite{MochizukiImada} and the random-phase approximation for
calculating the effective interaction parameters in the Hubbard model.\cite{Ferdi04}
\end{acknowledgments}

\appendix*
\section{\label{sec:Appendix}
Orbital degeneracy and screening in RPA}

  In this Appendix we consider the screening of on-site Coulomb interactions in RPA for an
$M$-orbital system. All orbitals are supposed to be equivalent. For simplicity we neglect
small nonsphericity of bare Coulomb interactions. Then, the nonvanishing matrix elements of
the Coulomb
interactions are $u_{\alpha \alpha \beta \beta} \equiv u$, where
$\alpha$ ($\beta$)$=$$1,\dots,M$.
They can be presented in the form $\hat{u}$$=$$u\hat{I}$, where $\hat{I}$ is the
$M$$\times$$M$ matrix, consisting of only the units:
$$
\hat{I} =
\left(
\begin{array}{cccc}
1      & 1      & \cdots & 1      \\
1      & 1      &        & \vdots \\
\vdots &        & \ddots & \vdots \\
1      & \cdots & \cdots & 1      \\
\end{array}
\right).
$$
The part of the polarization polarization matrix (\ref{eqn:PolarizationPT}),
which can interact with the matrix $\hat{u}$,
is assumed to be diagonal:
$P_{\alpha \alpha \beta \beta} \equiv P \delta_{\alpha \beta}$.
The assumption is justified for
cubic perovskites, where different $t_{2g}$ orbitals belong to
different bands.
For other compounds it can be regarded as an approximation, which
does not change our qualitative conclusion.
Hence, for the screened Coulomb interaction (\ref{eqn:URPA}) we
have:
\begin{equation}
\hat{U} = [1 - \hat{u}\hat{P}]^{-1}\hat{u} = \sum_{n=0}^\infty
(uP)^n \hat{I}^n u \hat{I}.
\label{eqn:Appendix1}
\end{equation}
Since $\hat{I}^{n+1}$$=$$M^n\hat{I}$, Eq.~(\ref{eqn:Appendix1}) can be
converted to
$$
\hat{U} = \sum_{n=0}^\infty (uMP)^n u \hat{I} = \frac{u}{1-uMP} \hat{I}.
$$
This means that in the multi-orbital systems, the renormalization of the
Coulomb repulsion is more efficient as it is controlled by the quantity $(MP)$,
where the prefactor $M$ stands for the number of orbitals.

  In the strong coupling limit, $-uMP$$\gg$$1$,
the effective interaction is $\hat{U}$$=$$-$$(MP)^{-1}\hat{I}$,
which does not depend
on the value of bare Coulomb interaction $u$.


\begin{thebibliography}{10}
\bibitem{comment.1}
Note that this splitting is different from
the ``two-down, one-up'' scheme of the atomic
$t_{2g}$ levels ($-$$0.069$, $-$$0.042$, and $0.112$ eV)
reported in Ref.~\onlinecite{PRB04}.
The difference is related with
nonsphericity of the electron-ion interaction which was not considered
in Ref.~\onlinecite{PRB04}.
The new scheme of the crystal-field splitting tends to stabilize
a single $t_{2g}$ orbital and thereby suppress the orbital fluctuations.
On the Hartree-Fock level, the magnetic ground state of
YTiO$_3$ and LaTiO$_3$ is expected to ferromagnetic and A-type antiferromagnetic,
respectively, in agreement with Ref.~\onlinecite{PRB04}.
Other details will be reported in a separate paper.
\bibitem{comment.3}
In the present work we employ the setting:
$\langle \tilde{W}(\hat{c}) | H_{\rm KS} | \tilde{W}(\hat{c}) \rangle$$=$$E_\nu$
\bibitem{comment.4}
In the atomic limit, the $5$$\times$$5$$\times$$5$$\times$$5$
matrix $\hat{u}$ of the Coulomb interactions between $d$ electrons
is totally specified by three radial Slater's integral: $F^0$,
$F^2$, and $F^4$. Then, we use the following definitions for the
parameters $u$ and $j$ in terms of
$F^0$, $F^2$, and $F^4$:
$u$$=$$F^0$ and $j$$=$$\frac{1}{14}(F^2$$+$$F^4)$.
In LDA$+$$U$, one can use
the ratio:\cite{PRB94} $F^4/F^2$$\simeq$$0.63$, which holds
in the atomic limit.
It allows to obtain all three Slater's integrals,
using the values of only two parameters, $u$ and $j$, extracted from
c-LDA.
We also list here several useful identities which hold for Coulomb matrix
elements between $t_{2g}$ orbitals:
$u_{xy~xy~xy~xy}$$=$$F^0$$+$$\frac{4}{49}F^2$$+$$\frac{36}{441}F^4$
($\equiv u$$+$$\frac{8}{7}j$),
$u_{xy~xy~yz~yz}$$=$$F^0$$-$$\frac{2}{49}F^2$$-$$\frac{4}{441}F^4$,
and
$u_{xy~yz~xy~yz}$$=$$\frac{3}{49}F^2$$+$$\frac{20}{441}F^4$.
\end{thebibliography}
\end{document}